\input amstex
\input amssym.tex
\input amssym

%
%
%
%
\def\temp{1.31}
\let\tempp=\relax
\expandafter\ifx\csname psboxversion\endcsname\relax
  \message{version: \temp}
\else
    \ifdim\temp cm>\psboxversion cm
      \message{version: \temp}
    \else
      \message{psbox(\psboxversion) is already loaded: I won't load
        psbox(\temp)!}
      \let\temp=\psboxversion
      \let\tempp= 
    \fi
\fi
\tempp
\let\psboxversion=\temp
\catcode`\@=11
%
%
\def\execute#1{#1}
\def\psm@keother#1{\catcode`#112\relax}
\def\executeinspecs#1{%
\execute{\begingroup\let\do\psm@keother\dospecials\catcode`\^^M=9#1\endgroup}}
%
%
\def\psfortextures{
\def\PSspeci@l##1##2{%
\special{illustration ##1\space scaled ##2}%
}}
\def\psfordvitops{
\def\PSspeci@l##1##2{%
\special{dvitops: import ##1\space \the\drawingwd \the\drawinght}%
}}
\def\psfordvips{
\def\PSspeci@l##1##2{%
\d@my=0.1bp \d@mx=\drawingwd \divide\d@mx by\d@my%
\includegraphics{##1\space}%
}}
\def\psforoztex{
\def\PSspeci@l##1##2{%
\special{##1 \space
      ##2 1000 div dup scale
      \putsp@ce{\number-\psllx} \putsp@ce{\number-\pslly} translate
}%
}}
\def\putsp@ce#1{#1 }
\def\psfordvitps{
\def\psdimt@n@sp##1{\d@mx=##1\relax\edef\psn@sp{\number\d@mx}}
\def\PSspeci@l##1##2{%
\special{dvitps: Include0 "psfig.psr"}
\psdimt@n@sp{\drawingwd}
\special{dvitps: Literal "\psn@sp\space"}
\psdimt@n@sp{\drawinght}
\special{dvitps: Literal "\psn@sp\space"}
\psdimt@n@sp{\psllx bp}
\special{dvitps: Literal "\psn@sp\space"}
\psdimt@n@sp{\pslly bp}
\special{dvitps: Literal "\psn@sp\space"}
\psdimt@n@sp{\psurx bp}
\special{dvitps: Literal "\psn@sp\space"}
\psdimt@n@sp{\psury bp}
\special{dvitps: Literal "\psn@sp\space startTexFig\space"}
\special{dvitps: Include1 "##1"}
\special{dvitps: Literal "endTexFig\space"}
}}
\def\psforDVIALW{
\def\PSspeci@l##1##2{
\special{language "PS"
literal "##2 1000 div dup scale"
include "##1"}}}
\def\psonlyboxes{
\def\PSspeci@l##1##2{%
\at(0cm;0cm){\boxit{\vbox to\drawinght
  {\vss
  \hbox to\drawingwd{\at(0cm;0cm){\hbox{(##1)}}\hss}
  }}}
}%
}
\def\psloc@lerr#1{%
\let\savedPSspeci@l=\PSspeci@l%
\def\PSspeci@l##1##2{%
\at(0cm;0cm){\boxit{\vbox to\drawinght
  {\vss
  \hbox to\drawingwd{\at(0cm;0cm){\hbox{(##1) #1}}\hss}
  }}}
\let\PSspeci@l=\savedPSspeci@l
}%
}
%
%
\newread\pst@mpin
\newdimen\drawinght\newdimen\drawingwd
\newdimen\psxoffset\newdimen\psyoffset
\newbox\drawingBox
\newif\ifNotB@undingBox
\newhelp\PShelp{Proceed: you'll have a 5cm square blank box instead of
your graphics (Jean Orloff).}
\def\@mpty{}
\def\s@tsize#1 #2 #3 #4\@ndsize{
  \def\psllx{#1}\def\pslly{#2}%
  \def\psurx{#3}\def\psury{#4}
  \ifx\psurx\@mpty\NotB@undingBoxtrue
  \else
    \drawinght=#4bp\advance\drawinght by-#2bp
    \drawingwd=#3bp\advance\drawingwd by-#1bp
  \fi
  }
\def\sc@nline#1:#2\@ndline{\edef\p@rameter{#1}\edef\v@lue{#2}}
\def\g@bblefirstblank#1#2:{\ifx#1 \else#1\fi#2}
\def\psm@keother#1{\catcode`#112\relax}
\def\execute#1{#1}
{\catcode`\%=12
\xdef\B@undingBox{
}   
\def\ReadPSize#1{
 \edef\PSfilename{#1}
 \openin\pst@mpin=#1\relax
 \ifeof\pst@mpin \errhelp=\PShelp
   \errmessage{I haven't found your postscript file (\PSfilename)}
   \psloc@lerr{was not found}
   \s@tsize 0 0 142 142\@ndsize
   \closein\pst@mpin
 \else
   \immediate\write\psbj@inaux{#1,}
   \loop
     \executeinspecs{\catcode`\ =10\global\read\pst@mpin to\n@xtline}
     \ifeof\pst@mpin
       \errhelp=\PShelp
       \errmessage{(\PSfilename) is not an Encapsulated PostScript File:
           I could not find any \B@undingBox: line.}
       \edef\v@lue{0 0 142 142:}
       \psloc@lerr{is not an EPSFile}
       \NotB@undingBoxfalse
     \else
       \expandafter\sc@nline\n@xtline:\@ndline
       \ifx\p@rameter\B@undingBox\NotB@undingBoxfalse
         \edef\t@mp{%
           \expandafter\g@bblefirstblank\v@lue\space\space\space}
         \expandafter\s@tsize\t@mp\@ndsize
       \else\NotB@undingBoxtrue
       \fi
     \fi
   \ifNotB@undingBox\repeat
   \closein\pst@mpin
 \fi
\message{#1}
}
%
%
\newcount\xscale \newcount\yscale \newdimen\pscm\pscm=1cm
\newdimen\d@mx \newdimen\d@my
\let\ps@nnotation=\relax
\def\psboxto(#1;#2)#3{\vbox{
   \ReadPSize{#3}
   \divide\drawingwd by 1000
   \divide\drawinght by 1000
   \d@mx=#1
   \ifdim\d@mx=0pt\xscale=1000
         \else \xscale=\d@mx \divide \xscale by \drawingwd\fi
   \d@my=#2
   \ifdim\d@my=0pt\yscale=1000
         \else \yscale=\d@my \divide \yscale by \drawinght\fi
   \ifnum\yscale=1000
         \else\ifnum\xscale=1000\xscale=\yscale
                    \else\ifnum\yscale<\xscale\xscale=\yscale\fi
              \fi
   \fi
   \divide \psxoffset by 1000\multiply\psxoffset by \xscale
   \divide \psyoffset by 1000\multiply\psyoffset by \xscale
   \global\divide\pscm by 1000
   \global\multiply\pscm by\xscale
   \multiply\drawingwd by\xscale \multiply\drawinght by\xscale
   \ifdim\d@mx=0pt\d@mx=\drawingwd\fi
   \ifdim\d@my=0pt\d@my=\drawinght\fi
   \message{scaled \the\xscale}
 \hbox to\d@mx{\hss\vbox to\d@my{\vss
   \global\setbox\drawingBox=\hbox to 0pt{\kern\psxoffset\vbox to 0pt{
      \kern-\psyoffset
      \PSspeci@l{\PSfilename}{\the\xscale}
      \vss}\hss\ps@nnotation}
   \global\ht\drawingBox=\the\drawinght
   \global\wd\drawingBox=\the\drawingwd
   \baselineskip=0pt
   \copy\drawingBox
 \vss}\hss}
  \global\psxoffset=0pt
  \global\psyoffset=0pt
  \global\pscm=1cm
  \global\drawingwd=\drawingwd
  \global\drawinght=\drawinght
}}
%
%
\def\psboxscaled#1#2{\vbox{
  \ReadPSize{#2}
  \xscale=#1
  \message{scaled \the\xscale}
  \divide\drawingwd by 1000\multiply\drawingwd by\xscale
  \divide\drawinght by 1000\multiply\drawinght by\xscale
  \divide \psxoffset by 1000\multiply\psxoffset by \xscale
  \divide \psyoffset by 1000\multiply\psyoffset by \xscale
  \global\divide\pscm by 1000
  \global\multiply\pscm by\xscale
  \global\setbox\drawingBox=\hbox to 0pt{\kern\psxoffset\vbox to 0pt{
     \kern-\psyoffset
     \PSspeci@l{\PSfilename}{\the\xscale}
     \vss}\hss\ps@nnotation}
  \global\ht\drawingBox=\the\drawinght
  \global\wd\drawingBox=\the\drawingwd
  \baselineskip=0pt
  \copy\drawingBox
  \global\psxoffset=0pt
  \global\psyoffset=0pt
  \global\pscm=1cm
  \global\drawingwd=\drawingwd
  \global\drawinght=\drawinght
}}
%
\def\psbox#1{\psboxscaled{1000}{#1}}
%
%
%
\newif\ifn@teof\n@teoftrue
\newif\ifc@ntrolline
\newif\ifmatch
\newread\j@insplitin
\newwrite\j@insplitout
\newwrite\psbj@inaux
\immediate\openout\psbj@inaux=psbjoin.aux
\immediate\write\psbj@inaux{\string\joinfiles}
\immediate\write\psbj@inaux{\jobname,}
%
%
\immediate\let\oldinput=\input
\def\input#1 {
 \immediate\write\psbj@inaux{#1,}
 \oldinput #1 }
\def\empty{}
\def\setmatchif#1\contains#2{
  \def\match##1#2##2\endmatch{
    \def\tmp{##2}
    \ifx\empty\tmp
      \matchfalse
    \else
      \matchtrue
    \fi}
  \match#1#2\endmatch}
\def\warnopenout#1#2{
 \setmatchif{TrashMe,psbjoin.aux,psbjoin.all}\contains{#2}
 \ifmatch
 \else
   \immediate\openin\pst@mpin=#2
   \ifeof\pst@mpin
     \else
     \errhelp{If the content of this file is so precious to you, abort (ie
press x or e) and rename it before retrying.}
     \errmessage{I'm just about to replace your file named #2}
   \fi
   \immediate\closein\pst@mpin
 \fi
 \message{#2}
 \immediate\openout#1=#2}
{
\catcode`\%=12
\gdef\splitfile#1 {
 \immediate\openin\j@insplitin=#1
 \message{Splitting file #1 into:}
 \warnopenout\j@insplitout{TrashMe}
 \loop
   \ifeof
     \j@insplitin\immediate\closein\j@insplitin\n@teoffalse
   \else
     \n@teoftrue
     \executeinspecs{\global\read\j@insplitin to\spl@tinline\expandafter
       \ch@ckbeginnewfile\spl@tinline
     \ifc@ntrolline
     \else
       \toks0=\expandafter{\spl@tinline}
       \immediate\write\j@insplitout{\the\toks0}
     \fi
   \fi
 \ifn@teof\repeat
 \immediate\closeout\j@insplitout}
\gdef\ch@ckbeginnewfile#1
 \def\t@mp{#1}
 \ifx\empty\t@mp
   \def\t@mp{#3}
   \ifx\empty\t@mp
     \global\c@ntrollinefalse
   \else
     \immediate\closeout\j@insplitout
     \warnopenout\j@insplitout{#2}
     \global\c@ntrollinetrue
   \fi
 \else
   \global\c@ntrollinefalse
 \fi}
\gdef\joinfiles#1\into#2 {
 \message{Joining following files into}
 \warnopenout\j@insplitout{#2}
 \message{:}
 {
 \edef\w@##1{\immediate\write\j@insplitout{##1}}
 \w@{
 \w@{
 \w@{
 \w@{
 \w@{
 \w@{
 \w@{
 \w@{
 \w@{\string\input\space psbox.tex}
 \w@{\string\splitfile{\string\jobname}}
 }
 \tre@tfilelist#1, \endtre@t
 \immediate\closeout\j@insplitout}
\gdef\tre@tfilelist#1, #2\endtre@t{
 \def\t@mp{#1}
 \ifx\empty\t@mp
   \else
   \llj@in{#1}
   \tre@tfilelist#2, \endtre@t
 \fi}
\gdef\llj@in#1{
 \immediate\openin\j@insplitin=#1
 \ifeof\j@insplitin
   \errmessage{I couldn't find file #1.}
   \else
   \message{#1}
   \toks0={
   \immediate\write\j@insplitout{\the\toks0}
   \executeinspecs{\global\read\j@insplitin to\oldj@ininline}
   \loop
     \ifeof\j@insplitin\immediate\closein\j@insplitin\n@teoffalse
       \else\n@teoftrue
       \executeinspecs{\global\read\j@insplitin to\j@ininline}
       \toks0=\expandafter{\oldj@ininline}
       \let\oldj@ininline=\j@ininline
       \immediate\write\j@insplitout{\the\toks0}
     \fi
   \ifn@teof
   \repeat
   \immediate\closein\j@insplitin
 \fi}
}
\def\autojoin{
 \immediate\write\psbj@inaux{\string\into\space psbjoin.all}
 \immediate\closeout\psbj@inaux
 \input psbjoin.aux
}
%
%
%
%
\def\centinsert#1{\midinsert\line{\hss#1\hss}\endinsert}
\def\psannotate#1#2{\def\ps@nnotation{#2\global\let\ps@nnotation=\relax}#1}
\def\pscaption#1#2{\vbox{
   \setbox\drawingBox=#1
   \copy\drawingBox
   \vskip\baselineskip
   \vbox{\hsize=\wd\drawingBox\setbox0=\hbox{#2}
     \ifdim\wd0>\hsize
       \noindent\unhbox0\tolerance=5000
    \else\centerline{\box0}
    \fi
}}}
\def\psfig#1#2#3{\pscaption{\psannotate{#1}{#2}}{#3}}
\def\psfigurebox#1#2#3{\pscaption{\psannotate{\psbox{#1}}{#2}}{#3}}
%
\def\at(#1;#2)#3{\setbox0=\hbox{#3}\ht0=0pt\dp0=0pt
  \rlap{\kern#1\vbox to0pt{\kern-#2\box0\vss}}}
%
\newdimen\gridht \newdimen\gridwd
\def\gridfill(#1;#2){
  \setbox0=\hbox to 1\pscm
  {\vrule height1\pscm width.4pt\leaders\hrule\hfill}
  \gridht=#1
  \divide\gridht by \ht0
  \multiply\gridht by \ht0
  \gridwd=#2
  \divide\gridwd by \wd0
  \multiply\gridwd by \wd0
  \advance \gridwd by \wd0
  \vbox to \gridht{\leaders\hbox to\gridwd{\leaders\box0\hfill}\vfill}}
%
\def\fillinggrid{\at(0cm;0cm){\vbox{
  \gridfill(\drawinght;\drawingwd)}}}
%
%
\def\textleftof#1:{
  \setbox1=#1
  \setbox0=\vbox\bgroup
    \advance\hsize by -\wd1 \advance\hsize by -2em}
\def\textrightof#1:{
  \setbox0=#1
  \setbox1=\vbox\bgroup
    \advance\hsize by -\wd0 \advance\hsize by -2em}
\def\endtext{
  \egroup
  \hbox to \hsize{\valign{\vfil##\vfil\cr%
\box0\cr%
\noalign{\hss}\box1\cr}}}
%
\def\frameit#1#2#3{\hbox{\vrule width#1\vbox{
  \hrule height#1\vskip#2\hbox{\hskip#2\vbox{#3}\hskip#2}%
        \vskip#2\hrule height#1}\vrule width#1}}
\def\boxit#1{\frameit{0.4pt}{0pt}{#1}}
\catcode`\@=12 
%
 \psfordvips   



\magnification=\magstephalf

\font \srm = cmr8 scaled \magstephalf
\font \brm = cmr10 scaled \magstep 2
\font \bbrm = cmr10 scaled \magstep 3
\font \bbbrm = cmr10 scaled \magstep 4
\font \bbf = cmbx10 scaled \magstep 2

\def \hb{\hfill\break}
\def \vo{\vskip 5mm}
\def \vsm{\vskip 1cm}
\def \vsss{\vskip 2.5cm}
\def \hs {\hskip 0.5cm}
\def \ve{\vfill\eject}
\def \ce{\centerline}
\def \di{\displaystyle}
\def \d{\partial}
\def \ti{\widetilde}
\def \({\left(}
\def \){\right)}
\def \[{\left[}
\def \]{\right]}
\def \wt{\widetilde}
\def \Tr{\text{\rm Tr}\,}
\def \Re{\text{\rm Re}\,}
\def \Im{\text{\rm Im}\,}
\def \dist{\text{\rm dist}\,}
\def \diag{\text{\rm diag}\,}
\def \inn{\text{\rm in}\,}
\def \term{\text{\rm term}\,}
\def \TP{\text{\rm TP}\,}
\def \CP{\text{\rm CP}\,}
\def \const{\text{\rm const}\,}
\def \formal{\text{\rm formal}}
\def \ds{\text{\rm ds}}
        \def \G{\Gamma}
        \def \g{\gamma}
        \def \a{\alpha}
        \def \b{\beta}
        \def \de{\delta}
        \def \De{\Delta}
        \def \ep{\varepsilon}
        \def \kappa{\varkappa}
        \def \la{\lambda}
        \def \La{\Lambda}
        \def \r{\rho}
        \def \t{\theta}
        \def \z{\zeta}
        \def \Sg{\Sigma}
        \def \sg{\sigma}
        \def \Om{\Omega}
        \def \U{\Cal Q}
        \def \om{\omega}
        \def \Var{\text{\rm Var}\;}
        \def \tr{\,\text{\rm tr}\,}
        \def \sign{\,\text{\rm sign}\,}
        \def \f{\varphi}
        \def \N{\Bbb N}
        \def \Q{\Bbb Q}
        \def \R{\Bbb R}
        \def \C{\Bbb C}
        \def \T{\Bbb T}
        \def \Z{\Bbb Z}
        \def \rf{\root 4 \of}
        \def \Ai{\,\text{\rm Ai}\,}
        \def \iff{\quad\text{\rm if}\quad}
        \def \ON{O(N^{-2})}
        \def \iz{\int_{z_1}^{z_2}}
        \def \ix{\int_{x_1}^{x_2}}
        \def \A{\text{\rm A}}
        \def \E{\text{\rm E}\,}
        \def \WKB{\text{\rm WKB}}
        \def \lacr{\la_c}

\null
\vskip 4cm

\ce{\bbbrm Double Scaling Limit in the Random Matrix}

\vskip 5mm

\ce{\bbbrm Model: the Riemann-Hilbert Approach}

\vskip 2cm     

\ce {\bbf Pavel Bleher and Alexander Its}

\vskip 1cm

\ce{\brm  Department of Mathematical Sciences}

\vskip 2mm 

\ce{\brm Indiana University-Purdue University Indianapolis}

\vskip 2cm

\vskip 1cm

{\bf Abstract.} We prove the existence of 
the double scaling limit in the unitary matrix
model with quartic interaction, and we show that 
the correlation functions
in the double scaling limit
are expressed in terms of the integrable kernel determined
by the psi-function for the Hastings-McLeod solution
to the Painlev\'e II equation. The proof is based on 
the Riemann-Hilbert approach.

\vskip 3cm

\vfill\eject


\beginsection 1. Introduction \par

\vskip 2mm

In this paper we will concern with the double scaling limit in the 
unitary random matrix model with quartic interaction. 
The unitary random matrix  model, or the unitary ensemble of random
matrices, is defined by the probability distribution
$$
\mu_N(dM)=Z_N^{-1}\exp\(-N\Tr V(M)\) dM,\quad Z_N=\int_{\Cal H_N}
\exp\(-N\Tr V(M)\) dM,
\eqno (1.1)  
$$
on the space $\Cal H_N$
of Hermitian $N\times N$ matrices $M=\(M_{ij}\)_{1\le
i,j\le N}$, where in general $V(M)$ is a polynomial of even degree
with a 
positive leading coefficient, or even more generally, a real analytic
function with some conditions at infinity.
 The basic case for the double scaling
limit is the quartic matrix model when 
$$
V(M)={t\over 2}M^2+{g\over 4}M^4,\qquad g>0,
\eqno (1.2)
$$ 
and we will consider this case only. By a change of variable one can
reduce the general case to the one with $g=1$ but
we prefer to keep $g$ because it is useful in some questions.
  The double scaling limit describes the  
asymptotics of correlation functions between eigenvalues in the limit
when simultaneously $N\to\infty$ and $t$ approaches the critical value
$$
t_c=-2 \sqrt g,
$$
with an appropriate relation between $N$ and $t-t_c$ (see below).
The critical value $t_c$ is a bifurcation point: for $t\ge t_c$
the support of the limiting distribution of eigenvalues consists
of one interval, while for $t<t_c$ it consists of two intervals.
This can be described as follows.

Let $d\nu(x)$ be the limiting eigenvalue
distribution (for the proof of the existence of the limiting
eigenvalue distribution 
and formulae for it  see [BPS],
 [DKM] and [DKMVZ1]; see also earlier
physical works [BIPZ], [BIZ], and others).
The distribution $d\nu(x)$ is a unique solution to a
variational problem and for  
 a general analytic function $V(M)$ satisfying certain
conditions at infinity (see [DKM] and [KM] for more
detail) the distribution
$d\nu(x)$ is
supported by a finite number of segments $[a_1,b_1],\ldots,[a_q,b_q]$,
it is absolutely continuous with respect to $dx$,
and on  $J=\cup_{j=1}^q[a_j,b_j]$ its density function is of the form
$$
p(x)=\frac{1}{2\pi i} h(x)R^{1/2}_+(x),\quad
R(x)=\prod_{j=1}^q(x-a_j)(x-b_j), 
$$
where $h(x)$ is a polynomial and $R^{1/2}_+(x)$ means the
value on the upper cut of
the principal sheet  of the function
$R^{1/2}(z)$ with cuts on $J$. The variational problem for $d\nu(x)$
implies that the polynomial $h(z)$ satisfies
the equation
$$
\om(z)=\frac{V'(z)}{2}-\frac{h(z)R^{1/2}(z)}{2},
\eqno (1.3)
$$
where
$$
\om(z)\equiv\int_{J} \frac {p(x)\,dx}{z-x}=z^{-1}+O(z^{-2}),
\quad z\to\infty
$$
(see e.g. [DKMVZ1]). 
Equation (1.3) enables one (see e.g. [BPS])
to determine uniquely the limiting
eigenvalue distribution for quartic polynomial (1.2).
For $t\ge t_c$ the support of the distribution consists of
one interval $[-a,a]$
and on $[-a,a]$,
$$
p(x)={1\over \pi}(b_0+b_2x^2)\sqrt{a^2-x^2},\quad |x|\le a;
\qquad t>t_c.
\eqno (1.4)
$$
Here
$$
a=\(\frac{-2t+\(4t^2+48g\)^{1/2}}{3g}\)^{1/2};\qquad
b_2=\frac{g}{2},\quad
b_0=\frac{t+\((t^2/4)+3g\)^{1/2}}{3}.
$$
For $t=t_c$, (1.4) reduces to
$$
p(x)=\frac{gx^2\sqrt{a^2-x^2}}{2\pi}\,,\quad |x|\le a\,;
\qquad t=t_c,
\eqno (1.5)
$$
where $a=2/g^{1/4}$. Observe 
that $p(0)=0$ in this case. For $t<t_c$ the support consists of
two intervals,
$[-a,-b]$ and $[b,a]$, and on these intervals,
$$
p(x)={1\over \pi}b_0 |x|\sqrt{(a^2-x^2)(x^2-b^2)},
\quad 0<b\le |x|\le a;\qquad t<t_c.
\eqno (1.6)
$$
Here
$$
a=\(\frac{2\sqrt g-t} {g}\)^{1/2}\,,\quad  b=\(\frac{-2\sqrt g-t} 
{g}\)^{1/2}\,
;\qquad 
b_0={g\over 2}\,.
\eqno (1.7)
$$
Fig.1 depicts the density function $p(x)$ for $g=1$ and
$t=-1,-2,-3$. In this case $t_c=-2$.

\centinsert{\pscaption{\psboxto (6in;1.2in){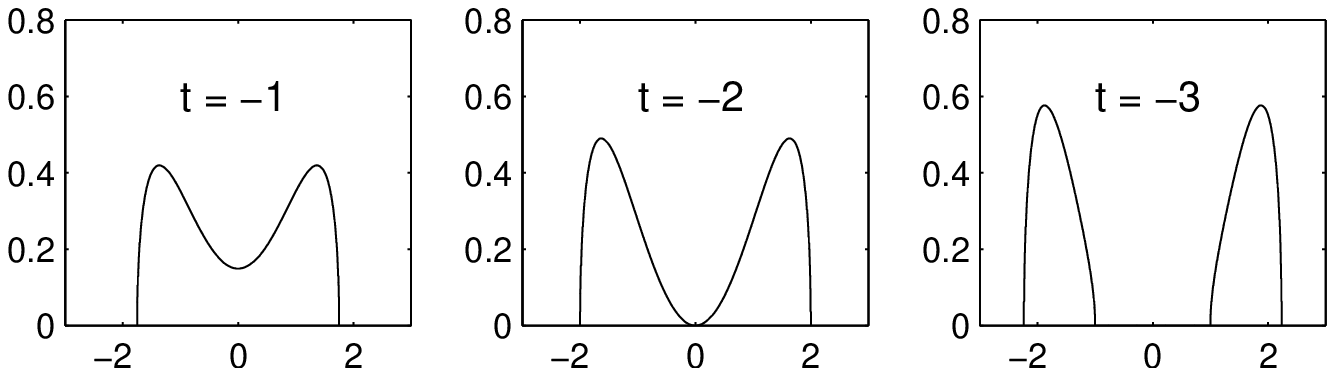}}
{\srm Fig 1:  The density function $p(x)$ for $g=1$ and
$t=-1,-2,-3$.}}

\vskip 3mm

{\it Correlation Functions and Orthogonal Polynomials.} 
Our basic object of interest is correlations between eigenvalues. 
The $m$-point correlation function
$K_{Nm}(z_1,\dots,z_m)$, $m=1,2,\dots$,
 is a distribution over the space $\Cal
D(\R^m)$ of 
$C^\infty$-functions $\f(z_1,\dots,z_m)$ on $\R^m$ with compact
support such that for the
product functions $\f(z_1,\dots,z_m)=\f_1(z_1)\dots \f_m(z_m)$,
$$
K_{Nm}\(\f_1(z_1)\dots \f_m(z_m)\)
=\int_{\Cal H_N}\[\chi_{\f_1}(M)\dots\chi_{\f_m}(M)\]\mu_N(dM)
\eqno (1.8)
$$
where 
$$
\chi_{\f}(M)=\sum_{j=1}^N \f(\la_j),\quad Me_j=\la_j e_j.
$$
By linearity $K_{Nm}(\f)$ is extended then from the set of product
functions $\f_1(z_1)\dots \f_m(z_m)$ to the whole space $\Cal
D(\R^m)$.

The $m$-point correlation function
$K_{Nm}(z_1,\dots,z_n)$ turns out to be
 a regular function on the set $\{ z_i\not=
z_j\}$ and on this set it can be expressed in terms of orthogonal
polynomials. Namely, let
$$
P_n(z)=z^n+\dots,\qquad n=0,1,2,\dots,
\eqno (1.9)
$$
be monic orthogonal polynomials on the line with respect to
the weight $e^{-NV(z)}$, so that
$$
\int_{-\infty}^\infty P_n(z)P_m(z)\,e^{-NV(z)}dz=h_n\de_{mn}.
\eqno (1.10)
$$
We normalize $P_n(z)$ by the condition that the leading coefficient
of $P_n(z)$ is equal to 1. 
The polynomials $P_n(z)$ satisfy the recursive equation [Sze]
$$
zP_n(z)=P_{n+1}(z)+R_nP_{n-1}(z),
\eqno (1.11)
$$
where
$$
R_n={h_n\over h_{n-1}}>0.
\eqno (1.12)
$$
We associate with $P_n(z)$ the functions
$$
\psi_n(z)={1\over \sqrt{h_n}}\, P_n(z) e^{-NV(z)/2}\,,
\eqno (1.13)
$$
which form an orthonormal basis in $L^2(\R^1)$,
$$
\int_{-\infty}^\infty \psi_n(z)\psi_m(z)\, dz=\de_{nm}.
\eqno (1.14)
$$
From (1.11),
$$
z\psi_n(z)=R_{n+1}^{1/2}\psi_{n+1}(z)+R_n^{1/2}\psi_{n-1}(z),
\eqno (1.15)
$$
The $m$-point correlation function $K_{Nm}(z_1,\dots,z_n)$
is expressed in terms of the functions $\psi_n(z)$  as 
$$
K_{Nm}(z_1,\dots,z_m)=\det \( Q_N(z_i,z_j)\)_{i,j=1,\dots,m},\quad
z_i\not=z_j, 
\eqno (1.16)
$$
where
$$
Q_N(z,w)=\sum_{k=0}^{N-1}\psi_k(z)\psi_k(w)
\eqno (1.17)
$$
([Dys]; see also [Meh], [TW1]).
The Christoffel-Darboux identity reduces $Q_N(z,w)$ to
$$
Q_N(z,w)=
R_N^{1/2}\,{\psi_N(z)\psi_{N-1}(w)-\psi_{N-1}(z)\psi_N(w)\over z-w}\,.
\eqno (1.18)
$$
We are interested in the asymptotics of the correlation functions 
$K_{Nm}(z_1,\dots,z_n)$ in the limit when $N\to\infty$. Formula 
(1.16) reduces the problem to the asymptotics of the kernel $Q_N(z,w)$
and (1.18) further to the asymptotics
of the functions $\psi_N(z)$ and $\psi_{N-1}(z)$. In noncritical case
$t<t_c$ this problem was solved in [BI]. Noncritical
asymptotics for the general $V(M)$ were found in [DKMVZ1].

{\it Double scaling limit for correlation functions.} We will
consider the correlation functions
$K_{Nm}(z_1,\dots,z_m;t)$ in the situation  when the parameter $t$
approaches 
the critical value $t_c$ and $z_1,\dots,z_m$ are near 0.  
The problem is how to scale eigenvalues and $t-t_c$
to get a nontrivial limit for the correlation functions.
More precisely, we want
to find numbers $\xi$ and $\eta$, critical exponents,  such that the following
``double scaling'' limit exists:
$$
\lim_{N\to\infty} \frac{1}{N^{(m-1)\eta}}
K_{Nm}\(\frac{z_1}{N^{\eta}},\dots,\frac{z_m}{N^{\eta}};t_c+yN^{-\xi}\)
=K_m(z_1,\dots,z_m;y),
\eqno (1.19)
$$
and it is a nonconstant function of the parameter $y,\; -\infty<y<\infty$.
We also want to find the limiting correlation functions
$K_m(z_1,\dots,z_m;y)$. 
We will derive the double scaling limit for correlation functions from
corresponding asymptotics for orthogonal polynomials. 
The relation between the double scaling limit for
correlation functions and the one for orthogonal
polynomials can be described as follows.

{\it Double scaling limit for orthogonal polynomials.} Let us fix
the parameters $t<0$ and $g>0$ in quartic polynomial (1.2).
Since $t<0$, $V(z)$ is a double-well function.
Consider orthogonal polynomials $P_n(z)$ with respect to the 
weight $e^{-NV(z)}$.
The bifurcation of $p(x)$
in the quartic matrix model is closely
related  to the bifurcation of the distribution of zeros of the
orthogonal polynomials $P_n(z)$. This relation is
motivated by the Heine formula
for orthogonal polynomials as follows.
Let $x_1,\ldots,x_n$
be zeros of $P_n(z)$ which are all real. Consider the 
corresponding probability measure 
$$
d\mu_n(x)=n^{-1}\sum_{j=1}^n\de(x-x_j)\,dx.
$$
We can write the weight  as
$$
e^{-NV(x)}=e^{-nV_\la(x)},\quad V_\la(x)\equiv \la^{-1}V(x),
\quad \la\equiv \frac{n}{N}.
$$
By the Heine formula for orthogonal polynomials
(see e.g. [BIZ]),
$$
P_n(z)=\left\langle \det (z-M) \right\rangle_{V_\la}
\equiv
Z_n^{-1}(V_\la)\int \det (z-M) e^{-n\Tr V_\la(M)}dM,
$$
or equivalently, in the ensemble of eigenvalues,
$$
P_n(z)=\left\langle \prod_{j=1}^n (z-\la_j) \right\rangle_{V_\la}.
$$
Due to the variational principle, we expect
that if $n$ is large then the probability measure
$$
d\nu(x;\{\la_j\})\equiv n^{-1}\sum_{j=1}^n\delta(x-\la_j)d\la
$$ 
is close to the
equilibrium measure $p(x;V_\la)dx$ for typical $\{\la_j\}$.
Therefore,
$$
n^{-1}\log\left\langle \prod_{j=1}^n (z-\la_j) \right\rangle_{V_\la}
\approx
\int_{J_\la} \log(z-x) p(x;V_\la)dx,
$$
hence by Heine,
$$
\int_{-\infty}^\infty \log(z-x) d\mu_n(x)=
n^{-1}\log P_n(z)\approx
\int_{J_\la} \log(z-x) p(x;V_\la)dx,\quad z\in\C\setminus \R.
$$
That is, as $n\to\infty$ the logarithmic potential of $d\mu_n(x)$
converges to the logarithmic potential of $p(x;V_\la)dx$.
Thus, we can expect that as $n,N\to \infty$ in such a way
that $\frac{n}{N}\to\la$, the measure $d\mu_n(x)$
converges to $d\mu_\infty(x;\la)=p(x;V_\la)dx$. 
This convergence has been rigorously
established for the general $V(M)$ in [DKMVZ1] (see also [BPS],
[BI] and works on the theory of orthogonal polynomials
[LS], [Lub], [Nev], [ST], and references therein).

For $\la$ small, the zeros of $P_n(z)$ are located near the minima
of the function $V(z)$ and the support of $d\mu_{\infty}(x;\la)$
consists of two intervals. For $\la$ big, the support of
$d\mu_{\infty}(x;\la)$ 
consists of one interval, and 
there exists a critical value,
 $$
\la_c=\frac{t^2}{4g}\,,
\eqno (1.20)
$$
 which separates the two-interval
and one-interval regimes (see [BI]).
The problem of double scaling limit for orthogonal polynomials is
to prove the existence of   a nontrivial scaling limit 
for the $\psi$-function,  
$$
\lim_{n,N\to\infty:\;(n/N)
=\la_c+yN^{-\xi}}C_n^{-1}\psi_n\(\frac{z}{N^{\eta}}\)= 
\psi_{\infty}(z;y),
$$
with some critical exponents $\xi$ and $\eta$,
where $C_n\not=0$ are some normalizing constants. Our main goal in this
paper is to derive uniform asymptotics for the functions $\psi_n(z)$
on the whole complex plane as $n,N\to\infty$ in the double scaling 
limit regime,
$(n/N)=\la_c+yN^{-\xi}$. When applied to the quartic matrix model,
these asymptotics will give us the double scaling limit for the correlation
functions at the origin as well as scaling limits in the bulk of
the spectrum and at the edges.
There are three basic ingredients in our approach: the string (Freud)
equation, the Lax pair for the string equation, and the
Riemann-Hilbert problem.

{\it String Equation}. Let $Q=(Q_{mn})_{m,n=0,1,\dots}$ be the matrix
of the operator of multiplication by $z$ in the basis $\psi_n(z)$.
By (1.15),
$$
Q=\pmatrix
0 & \sqrt {R_1} & 0 & \ldots \\
\sqrt {R_1} & 0 & \sqrt{R_2} & \ldots \\
0 & \sqrt{R_2} & 0 & \ldots \\
\vdots & \vdots & \vdots & \ddots
\endpmatrix.
$$
Let $P$ be the matrix of the operator $d/dz$. Then $P^t=-P$, where
$P^t$ is the transposed matrix. 
Observe that
$$
\psi_n'=-(NV'/2)\psi_n+\frac{P_n'e^{-NV/2}}{\sqrt{h_n}}
=-(NV'/2)\psi_n+\frac{n}{\sqrt {R_n}}\psi_{n-1}+\dots,
$$
so that 
$$
\[P+NV'(Q)/2\]_{n,n-1}=\frac{n}{\sqrt{R_n}},
\eqno (1.21)
$$
and
$$
\[P+NV'(Q)/2\]_{n,n+1}=0,
$$
hence
$$
0=\[P+NV'(Q)/2\]_{n-1,n}=\[-P+NV'(Q)/2\]_{n,n-1}.
\eqno (1.22)
$$
Combining (1.21) with (1.22) we obtain that
$$
\frac{n}{N\sqrt {R_n}}=[V'(Q)]_{n,n-1},
\eqno (1.23)
$$
the discrete string equation
(see [BK], [DS], and [GM]). 
When $V$ is as in (1.2), this reduces to 
\vskip 1mm
$$
\frac{n}{N}=R_n\(t+gR_{n-1}+gR_n+gR_{n+1}\).
\eqno (1.24)
$$
\vskip 1mm
\noindent
(cf. [Fre], [BIPZ], [IZ], and [BIZ]). 
Note that from (1.12) and (1.24) it follows that 
$$
\frac{n}{N}<tR_n+gR_n^2,
$$
hence
$$
0<R_n<{-t+\sqrt{t^2+4gn/N}\over 2g}\,.
\eqno (1.25)
$$

{\it Lax Pair.} Indroduce the vector valued function 
$$
\vec\Psi_n(z)=
\pmatrix
\psi_n(z) \\
\psi_{n-1}(z)
\endpmatrix\,.
\eqno (1.26)
$$
It satisfies  the following $2\times 2$ matrix 
differential equation: 
$$
\vec\Psi_n'(z)=N A_n(z)\vec\Psi_n(z),
\eqno (1.27) 
$$
where
$$
A_n(z)=
\pmatrix
-({tz\over 2}+{gz^3\over 2}+gzR_n) 
& \sqrt{R_n}(t+gz^2+gR_n+gR_{n+1})\\
-\sqrt{R_n}(t+gz^2+gR_{n-1}+gR_{n}) 
& {tz\over 2}+{gz^3\over 2}+gzR_n
\endpmatrix
\eqno (1.28)
$$
([FIK2]; see also [BI]). For the sake of completeness, let us remark,
that for a general even polynomial 
$$
V(z)=\sum_{j=1}^k\frac{t_{2j}z^{2j}}{2j}
$$
the matrix $A_n(z)$ has the form
$$
A_n(z)=
\pmatrix
-a_n(z) & \sqrt{R_n}\,\di\frac{a_n(z)+a_{n+1}(z)}{z} \\
-\sqrt{R_n}\,\di\frac{a_{n-1}(z)+a_n(z)}{z} & a_n(z)
\endpmatrix
$$
where
$$
a_n(z)=\frac{V'(z)}{2}+\sqrt{R_n}\sum_{j=2}^k t_{2j}
\sum_{l=1}^{j-1}z^{2l-1}[Q^{2j-2l-1}]_{n,n-1}
$$
(see [FIK4]).
This reduces to (1.28) when $V$ is as in (1.2).

In addition, we have the recurrence equation
$$
\vec\Psi_{n+1}=U_n(z)\vec\Psi_n(z),
\eqno (1.29)
$$
where
$$
U_n(z)=
\pmatrix
\frac{z}{R_{n+1}^{1/2}} & -\frac{R_n^{1/2}}{R_{n+1}^{1/2}} \\
1 & 0
\endpmatrix.
\eqno (1.30)
$$
The compatibility condition of equations (1.27) and (1.29) is
$$
U_n'(z)=NA_{n+1}(z)U_n(z)-NU_n(z)A_n(z).
\eqno (1.31)
$$
Restricting this matrix equation to the element 11 one can rederive
string equation (1.24) (see [FIK]). And vice versa, (1.24) implies
(1.31). Thus, system (1.27,29) is the Lax pair for
equation (1.24).

{\it Remark.} The Lax pair (1.27,29)
can be alternatively written (see [FIK]) as the following linear 
differential-difference  system,
$$\left\{
\eqalign{
&{d\psi \over {dz}} =NV'_-(Q)\psi,\cr
&Q\psi=z\psi,}\right.,\quad
\psi \equiv (\psi_{1}, \psi_{2}, ..., \psi_{n}, ...).
$$
Accordingly, string equation (1.42) takes the ``quantum''
(cf. [Moo], [Nov]) commutator form,
$$
[Q, V_{-}'(Q)] = {1\over N}
$$
(see [FIK1]).
Here, as usual, the low triangular part, $M_{-}$, of
a matrix $M$ is defined
by the equations, $[M_{-}]_{nm} = [M]_{nm}$ if $n > m$
and $[M_{-}]_{nm} = 0$ if $n \leq m$. 

To describe the Riemann-Hilbert problem
consider the adjoint functions
$$
\phi_n(z)= e^{NV(z)/2}\int_{-\infty}^\infty
{e^{-NV(u)/2}\psi_n(u)\over u-z}\,du,\quad z\in \C\setminus\R.
\eqno (1.32)
$$
They share the following properties:
\item {(1)} As $z\to\infty$,
$$
\phi_n(z)\sim \sqrt{h_n} z^{-n-1}e^{NV(z)/2}
\( 1+\sum_{k=1}^\infty \frac{\g_{2k}}{z^{2k}}\).
$$
\item{(2)} The function $\phi_n(z)$ has limits 
$\phi_{n\pm}(z)$ as $z$ approaches the real axis from above
and below, and the limits  are related as
$$
\phi_{n+}(z)=\phi_{n-}(z)-2\pi i\psi_n(z).
$$
\item{(3)} The vector function
$$
\vec\Phi_n(z)=\pmatrix
\phi_n(z) \\ \phi_{n-1}(z)
\endpmatrix
$$
solves Lax pair equations (1.27,29).

To prove (1) expand $\frac{1}{u-z}$ into geometric series and use
orthogonality. Property (2) follows from the jump condition
for Cauchy integral. Property (3) will follow from the Riemann-Hilbert 
problem (see Proposition 1.1 below). 
Define the matrix-valued function $\Psi_n(z)$ on the
complex plane as
$$
\Psi_n(z)=\pmatrix
\psi_n(z) & \phi_n(z) \\
\psi_{n-1}(z) & \phi_{n-1}(z)
\endpmatrix.
\eqno (1.33)
$$
It solves the Lax pair equations,
$$\left\{
\eqalign{
&\Psi_n'(z)=NA_n(z)\Psi_n(z),\cr
&\Psi_{n+1}(z)=U_n(z)\Psi_n(z),}\right.
\eqno (1.34)
$$
and the Riemann-Hilbert problem for orthogonal polynomials.

{\it Riemann-Hilbert Problem.} The Riemann-Hilbert problem
for orthogonal polynomials
is formulated as follows ([FIK]; see also [BI],
and [DKMVZ1]).
 One has to find a
$2\times 2$ matrix-valued  function $\Psi_n(z)$ 
on the
complex plane which is analytic outside of the real line, 
and which has continuous limits from
above and below of the real line,
$$
\Psi_{n\pm}(z)=\lim_{u\to z,\;\pm\Im u>0} \Psi_n(u),
$$
so that  

(i) $\Psi_n(z)$ satisfies the jump condition on the real
line,
$$
\Psi_{n+}(z)=\Psi_{n-}(z)S,\quad \Re z=0,
\eqno (1.35)
$$
where
$$
S=
\pmatrix
1 & -2\pi i \\
0 & 1
\endpmatrix\,;
\eqno (1.36)
$$
and

(ii) As $z\to\infty$,
the function $\Psi_n(z)$ has the following uniform asymptotic expansion:
$$
\Psi_n(z)\sim
\(\sum_{k=0}^\infty {\G_k\over z^k}\)
e^{-\({NV(z)\over 2}-n\ln z+\la_n\)\sg_3},\quad z\to\infty,
\eqno (1.37)
$$
where $\G_k,\; k=0,1,2,\dots$, are some constant $2\times 2$ matrices,
with
$$
\G_0=
\pmatrix
1 & 0 \\
0 & R_n^{-1/2}
\endpmatrix,\quad
\G_1=
\pmatrix
0 & 1 \\
R_n^{1/2} & 0
\endpmatrix,
\eqno (1.38)
$$
$\la_n$ is a constant, and $\sg_3$ is the Pauli matrix,  
$$
\sg_3=
\pmatrix
1 & 0 \\
0 & -1
\endpmatrix.
\eqno (1.39)
$$

To solve the Riemann-Hilbert problem we have to find the function
$\Psi_{n}(z)$ and real numbers $\la_n$ and $R_n>0$ such that (i)
and (ii) hold. 

{\bf Proposition 1.1.} {\it There exists a unique solution to the
Riemann-Hilbert problem (1.35-38), and it is given by (1.33).
The number $R_n$ in this solution coincides with the recursive
coefficient in (1.11) and 
$$
\la_n=\frac{\ln h_n}{2}\,,
\eqno (1.40)
$$ 
where $h_n$ is defined 
in (1.10). The functions $\Psi_n(z),\; n=0,1,2,\dots,$ satisfy the Lax
pair equations (1.34).}

For the proof of Proposition 1.1 see [FIK], [BI], and [DKMVZ2].
From now on, we shall ``forget'' (cf. [BI]) about explicit equation (1.33) 
for the solution
$\Psi_{n}(z)$ of the Riemann-Hilbert problem (1.35-38).  It is the
Riemann-Hilbert problem itself
which becomes now the principal characteristic of the function
$\Psi_{n}(z)$. In other words, we shift the focus from
equation (1.33), which represents
the function $\Psi_{n}(z)$ in terms of the
orthogonal polynomials $P_{n}(z)$, to the
equation
$$
\psi_{n}(z) = [\Psi_{n}(z)]_{11},
$$
which represents the orthogonal polynomials
$P_{n}(z)$ in terms of the function 
$\Psi_{n}(z)$ (which in turn is
uniquely determined as a solution
of the Riemann-Hilbert problem (1.35-38)).
It also worth to emphasize that (see [BI]) in the setting
of the  Riemann-Hilbert problem  the quantities
$R_{n}$ and $\lambda_{n}$ {\it are not the given data}. They
are evaluated via the solution $\Psi_{n}(z)$, which
is determined by conditions  (1.35-38) uniquely
without any prior specification of $R_{n}$ and $\lambda_{n}$.

Since $V(-z)=V(z)$, we obtain that
$$
\psi_n(-z)=(-1)^n\psi_n(z).
\eqno (1.41)
$$
This leads to the following equation on $\Psi_n(z)$:
$$
\Psi_n(-z)=(-1)^n\sg_3\Psi_n(z)\sg_3.
\eqno (1.42)
$$
We turn now to formulation of the main result concerning
the semiclassical asymptotics of the functions $\psi_n(z)$ in the  
double scaling limit.

{\it Formulation of the main result.} We will consider $n$ such
that $n/N$ is close to the critical value $\la_c=t^2/(4g)$. We
introduce the  
variable $y$  as
$$
y=c_0^{-1}N^{2/3}\(\frac{n}{N}-\la_c\),\qquad
c_0=\(\frac{t^2}{2g}\)^{1/3}.
\eqno (1.43)
$$
We will assume that
$|y|$ is bounded, that is we fix an arbitrary large
number $T_0$ and we will consider such $n$ that 
$$
|y|\le T_0.
\eqno (1.44)
$$
Our results will concern asymptotics of the recurrence coefficients $R_n$
and the functions $\psi_n(z)$. For $R_n$ we will prove the following
asymptotics:
$$\eqalign{
R_n&={|t|\over
2g}+N^{-1/3}c_1(-1)^{n+1}u(y)+N^{-2/3}c_2v(y)+O(N^{-1}),\cr
c_1&=\({2|t|\over g^2}\)^{1/3},\qquad
c_2={1\over 2}\({1\over 2|t|g}\)^{1/3}\,,\cr}
\eqno (1.45)
$$
where $u(y)$ is the Hastings-McLeod solution to the Painlev\'e II equation
$$
u''(y)=yu(y)+2u^3(y),
\eqno (1.46)
$$
which is characterized by the conditions at infinity,
$$
\lim_{y\to-\infty}{u(y)\over (-y/2)^{1/2}}=1,\qquad
\lim_{y\to\infty} u(y)=0,
\eqno (1.47)
$$
and
$$
v(y)=y+2u^2(y).
\eqno (1.48)
$$

{\it Remark.} Asymptotics
(1.45) has been suggested in physical papers 
by Douglas, Seiberg, Shenker [DSS], Crnkovi\'c, Moore [CM],
 and Periwal, Shevitz [PeS].
 The existence and uniqueness of the solution to (1.46,47) 
was first established in [HM] (see also the
later works  [Kap1,2] and [DZ2]). It is also worth noticing
that the asymptotics
of $u(y)$ at $+\infty$ can be specified as
$$
u(y)=\Ai(y)\(1+O(|y|^{-1})\),
\eqno (1.47')
$$
where $\Ai(y)$ is the Airy function. Moreover, 
the asymptotic condition (1.47$'$) characterizes
solution $u(x)$ uniquely, so that the first equation in
(1.47) and equation (1.47$'$) constitute an example of the
so-called connection formulae for Painlev\'e equations
(see [IN], [FI], and [I] for more on this matter). 
We should also point out that the analytic and 
asymptotic properties of the
Hastings-McLeod solution for complex values of $y$
can be extracted from the results of [Kap1,2], [Novok], [Kit2],
and [IK].

We will prove asymptotics (1.45) simultaneously with 
semiclassical asymptotics for the function
$\psi_n(z)$ on the whole complex plane. To that end we will 
 substitute asymptotics (1.45) into
the coefficients of differential equation (1.27) and solve the resulting
equation  in the semiclassical approximation. Realization
of this idea is not trivial and
in comparison to the standard semiclassical
analysis, it involves
a new type of special functions generated by the monodromy
problem associated with the second Painlev\'e equation. The appearance
of the latter is closely related to the fact that ansatz (1.45) leads
to the  coalescence of four turning points in system  (1.27)
(cf. [Kit1]).
To see this let us analyze (1.27).

 System (1.27) can be reduced to
one equation of the second order. Namely, if we denote by 
$a_{ij}\,$, $i,j=1,2$, the matrix elements of
the matrix $A_n(z)$, solve $\psi_{n-1}$ in terms of $\psi_n$
from the first equation in (1.27),
$$
\psi_{n-1}=\frac{1}{a_{12}}\( N^{-1}\psi_n'-a_{11}\psi_n\),
$$
and substitute
$$
\psi_n=\sqrt{a_{12}}\,\eta,
\eqno (1.49)
$$
then we obtain from the second equation in (1.27) 
the Schr\"odinger equation on $\eta$,
$$
-\eta''+N^2U\eta=0,
\eqno (1.50)
$$
where 
$$
U=-\det A_n+N^{-1}\[(a_{11})'
-a_{11}{(a_{12})'\over a_{12}}\]
-N^{-2}\[ {(a_{12})''\over 2a_{12}}-{3((a_{12})')^2\over 4(a_{12})^2}\].
\eqno (1.51)
$$
The zeroth order term in the potential $U$ is $-\det A_n$
so let us consider properties of $-\det A_n$. 

From (1.28) we obtain that $-\det A_n(z)$ is
a polynomial of the sixth degree,
$$
-\det A_n(z)=\frac{g^2z^4}{4}(z^2-z_0^2)+\a_n z^2+\b_n,
\eqno (1.52)
$$
where 
$$
z_0=\sqrt{-2t/g}
\eqno (1.53)
$$
and
$$\eqalign{
&\a_n=\frac{t^2}{4}-gR_n(t+gR_{n-1}+gR_n+gR_{n+1}),\cr
&\b_n=-R_n\t_{n-1}\t_n,\qquad \t_n=t+gR_n+gR_{n+1}.}
\eqno (1.54)
$$
Due to string equation (1.24), $\a_n$ simplifies to
$$
\a_n=\frac{t^2}{4}-\frac{gn}{N}=-g\(\frac{n}{N}-\la_c\),
\eqno (1.55)
$$
or, according to scaling (1.43),
$$
\a_n=-gc_0yN^{-2/3}.
$$
The substitution of asymptotics (1.45) into $\b_n$ gives
$$
\b_n=c_3N^{-4/3}+c_4N^{-5/3}+O(N^{-2}),
\eqno (1.56)
$$
where
$$  
c_3=-\({g^{1/3}|t|^{1/3}\over 2^{5/3}}\)\[v^2(y)-4w^2(y))\],\quad
c_4=(-1)^n\({g^{2/3}\over 2^{1/3}|t|^{1/3}}\)w(y)
\eqno (1.57)
$$
and $w(y)\equiv u'(y)$.
Thus,
$$
-\det A_n(z)=\frac{g^2z^4}{4}(z^2-z_0^2)-g\(\frac{n}{N}-\la_c\) z^2+
\[c_3N^{-4/3}+c_4N^{-5/3}+O(N^{-2})\].
\eqno (1.58)
$$
As $N\to\infty$, $-\det A_n(z)$ approaches the polynomial
$$
a_{\infty}(z)=\frac{g^2z^4}{4}(z^2-z_0^2),
\eqno (1.59)
$$
which has
two simple roots at $\pm z_0$ and a quadruple root at the
origin. Therefore, equation (1.50) has two simple turning points
approaching $\pm z_0$ and four turning points coalescing at the
origin.   

In accord with this analysis,  we divide the
complex plane into several regions. The form of the semiclassical asymptotics
will be different in different regions.
Let $d_1$ and $d_2$ be arbitrary fixed (i.e. independent of $N$)
numbers such that 
$$
0<d_2\le d_1\le \frac{z_0}{4}\,.
\eqno (1.60)
$$
Introduce the rectangular region $\Om$ as
$$
\Om=\{ z\,:\; |\Re z|\le z_0+d_1,\;\;|\Im z|\le d_2\}.
\eqno (1.61)
$$

\centinsert{\pscaption{\psboxto (6in;1.5in){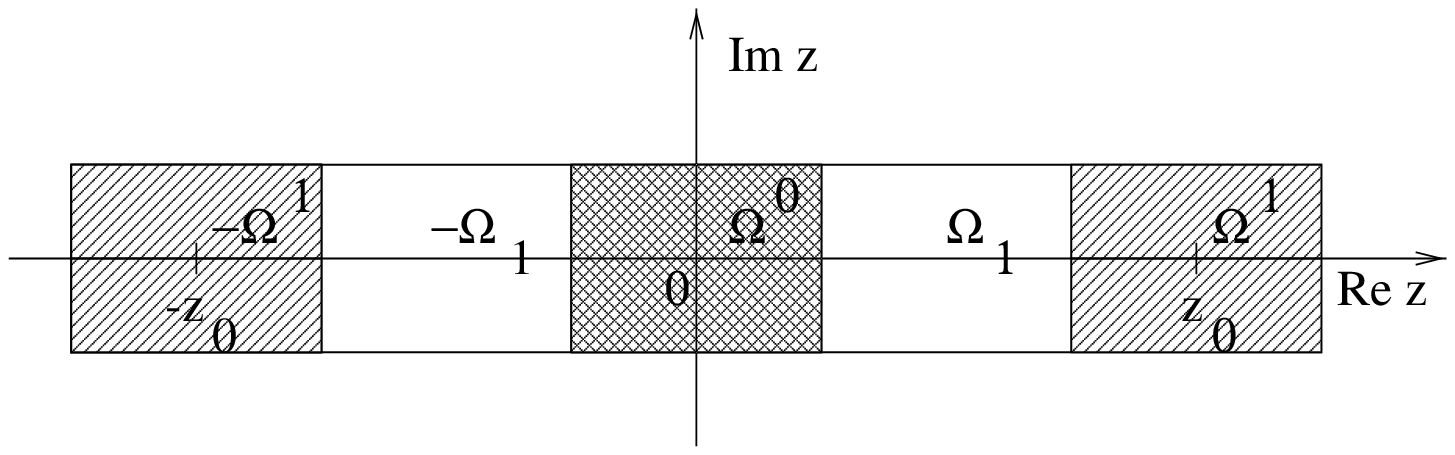}}
{\srm Fig 2: The region $\Om$.  }}

\vskip 3mm

\noindent
Observe that the segment $[-z_0,z_0]$ lies in $\Om$.
Introduce furthemore the rectangular subregions 
$\Om^0$ and $\Om^1$ of the region $\Om$ as
$$
\Om^0=\{ z\,:\; |\Re z|\le d_1,\;\;|\Im z|\le d_2\},\qquad
\Om^1=\{ z\,:\; |\Re z-z_0|\le d_1,\;\;|\Im z|\le d_2\},
\eqno (1.62)
$$
so that $\Om^0$ is a rectangular region centered at 0 and $\Om^1$ at $z_0$.
Finally, let $\Om_1$ be the rectangular region between $\Om^0$ and $\Om^1$,
that is
$$
\Om_1=\{ z\,:\; d_1\le \Re z\le z_0-d_1,\;\;|\Im z|\le d_2\} 
\eqno (1.63)
$$
(see Fig.2).  We will prove the following semiclassical
asymptotics for the vector function  $\vec\Psi_n(z)$:

\item{1.} {In $\Om^c\equiv
\overline{\C\setminus\Om}$: WKB asymptotics of exponential
type.}
\item{2.} {In $\Om_1$: WKB asymptotics of cosine type.}
\item{3.} {In $\Om^1$: turning point (TP) asymptotics (in terms of the 
Airy function).}
\item{4.} {In $\Om^0$: critical point (CP) asymptotics of Painlev\'e II
type.}

\noindent
In the regions $-\Om^1$ and $-\Om_1$ the semiclassical asymptotics
will follow by symmetry equation (1.42). As a matter of fact,
both the TP and CP asymptotics will be extended to the region
$\Om_1$ and we will use  this extension to prove the WKB
asymptotics of the cosine type in $\Om_1$.

{\it Region $\Om^c=\overline{\C\setminus\Om}$, WKB asymptotics.}
In $\Om^c$ 
 we will prove the following asymptotics:
$$
\vec\Psi_n(z)=\(1+O\({1\over
{N(1+|z|)}}\)\)\vec\Psi_{\WKB}(z),\qquad z\in\Om^c,
\eqno (1.64)
$$ 
where 
$$
\vec\Psi_{\WKB}(z)\equiv
 \frac{1}{2\pi^{1/2}}\,(R_n^0)^{-1/4}
T^c(z)
\pmatrix
e^{-N\xi^c(z)} \\
-e^{-N\xi^c(z)}
\endpmatrix,\qquad z\in\Om^c,
\eqno (1.65)
$$
with
the quantities on the right defined as follows. We define
$$
R_n^0\equiv{|t|\over
2g}+N^{-1/3}c_1(-1)^{n+1}u(y)+N^{-2/3}c_2v(y)
\eqno (1.66)
$$
[cf. (1.45)], where $y$ is defined as in (1.43), and we put  
$$
 A_n^0(z)\equiv
\pmatrix
-\di\({tz\over 2}+{gz^3\over 2}+gz R_n^0\) 
&  \sqrt{R_n^0}(t+gz^2+g R_n^0+g R_{n+1}^0)\\
-\sqrt {R_n^0}(t+gz^2+g R_{n-1}^0+g R_n^0) 
& \di{tz\over 2}+{gz^3\over 2}+gz R_n^0
\endpmatrix
\eqno (1.67)
$$
[cf. (1.28)]. We denote the matrix elements of $ A_n^0(z)$ by
$a_{ij}^0(z)$. We define the function $\mu^c(z)$ as
$$
\mu^c(z)\equiv \sqrt{ U^0(z)},
$$
where $U^0(z)$ is a suitable approximation of
the function $U(z)$ in (1.51), namely,
$$\eqalign{
U^0(z)&\equiv\frac{g^2z^4}{4}(z^2-z_0^2)-\(\frac{n}{N}-
\frac{t^2}{4g}\) z^2+\(c_3N^{-4/3}+c_4N^{-5/3}\)\cr
&+N^{-1}\[{a_{11}^0}'(z)
-a_{11}^0(z){{a_{12}^0}'(z)\over a_{12}^0(z)}\],}
\eqno(1.68)
$$
and we put
$$
\xi^c(z)\equiv\int_{ z^N_0}^z \mu^c(u)\,du,\qquad z\in\Om^c,
\eqno (1.69)
$$
where $z^N_0$ is the root of $U^0(z)$ that approaches $z_0$ 
as $N\to\infty$. The contour of integration in (1.69) 
is taken as follows:
first from $z^N_0$ to $z_0+d_1$ and then from $z_0+d_1$ to $z$ around
the region $\Om$ in the counterclockwise direction.
The matrix $T^c(z)$ is defined as
$$ 
T^c(z)\equiv
\(\frac{ a_{12}^0(z)}{ \mu^c(z)}\)^{1/2}
\pmatrix
1 & 0 \\
&\\
-\di\frac{a^0_{11}(z)}{a^0_{12}(z)} &
\di\frac{\mu^c(z)}{a^0_{12}(z)}
\endpmatrix.
\eqno (1.70)
$$
Observe that $\det T^c(z)\equiv 1$.
The branches for $(a_{12}^0(z))^{1/2}$, $(U^0(z))^{1/2}$,
and $(\mu^c(z))^{1/2}$ are fixed by the condition
that they are positive for large positive $z$. We will check
below that the function $\vec\Psi_{\WKB}(z)$ is analytic in $\Om^c$.
The meaning of formula (1.64) and similar formulae to follow
is that there exists a 
$2\times 2$ matrix valued function $\ep_N(z)$ such
that 
$$
\vec\Psi_n(z)=\(1+\ep_N(z)\)\vec\Psi_{\WKB}(z),\qquad z\in\Om^c,
$$ 
and
$$
\sup_{z\in\Om^c}|(1+|z|)\ep_N(z)|\le cN^{-1},\qquad c>0.
$$
For concrete calculations the function
$U^0(z)$ can be simplified as follows. The function
$$
U^1(z)\equiv {a_{11}^0}'(z)
-a_{11}^0(z){{a_{12}^0}'(z)\over a_{12}^0(z)}
$$ 
is analytic in $\Om^c$ and as $N\to \infty$,
$$\eqalign{
U^1(z)&=-{gz^2\over 2}
-N^{-1/3}\wt c_3-N^{-2/3}\wt c_4+O(N^{-1}),\qquad z\in\Om^c;\cr
\wt c_3&=(-1)^n2^{1/3}g^{1/3}|t|^{1/3},\quad 
\wt c_4={2^{2/3}g^{2/3}\[v(y)+4(-1)^nw(y)\]
\over 4|t|^{1/3}},}
\eqno (1.71)
$$
hence
$$\eqalign{
U^0(z)&=\frac{g^2z^4}{4}(z^2-z_0^2)-\(\frac{n}{N}-
\la_c+\frac{1}{2N}\) z^2+\b_N+O(N^{-2}),\cr
\b_N&=(c_3-\wt c_3)N^{-4/3}+(c_4-\wt c_4)N^{-5/3},}
$$
and the error term $O(N^{-2})$ in $U^0(z)$ can be neglected
in concrete calculations, reducing $U^0(z)$ to a polynomial
of the sixth degree.

For the normalizing constant $h_n$ we will prove the asymptotics
$$
h_n=e^{2N\int_{z^N_0}^\infty\mu^c(u)\,du}\(1+O(N^{-1})\).
\eqno (1.72)
$$
The integral in the exponent is regularized at infinity
as in formula (5.3) below.

{\it Region $\Om_1$, WKB asymptotics of cosine type.}
In the region $\Om_1$ we define $\vec\Psi_{\WKB}(z)$ as
$$
\vec\Psi_{\WKB}(z)\equiv
\frac{1}{\pi^{1/2}} (R_n^0)^{-1/4}
\,
T_1(z)
\pmatrix
\cos\(N\xi_1(z)+\di\frac{\pi}{4}\) \\
\\
-\sin\(N\xi_1(z)+\di\frac{\pi}{4}\)
\endpmatrix,
\qquad z\in\Om_1,
\eqno (1.73)
$$
where
$$
\mu_1(z)\equiv  \(-U^0(z)\)^{1/2},\qquad
\xi_1(z)\equiv \int_{ z^N_0}^z\mu_1(u)\,du; \qquad z\in\Om_1,
\eqno (1.74)
$$
and
$$
T_1(z)=
\(\frac{ a_{12}^0(z)}{ \mu_1(z)}\)^{1/2}
\pmatrix
1 & 0 \\
&\\
-\di\frac{a^0_{11}(z)}{a^0_{12}(z)} &
\di\frac{\mu_1(z)}{a^0_{12}(z)}
\endpmatrix,
\eqno (1.75)
$$
so that $\det T_1(z)\equiv 1$.
The functions  $-U^0(z)$ and $a_{12}^0(z)$ are positive on the interval
$d_1\le z\le z_0-d_1$ and the branches for the square roots in
(1.74,75) are 
chosen to be positive on this interval. 
We will prove that
$$
\vec\Psi_n(z)=\(1+O(N^{-1})\)\vec\Psi_{\WKB}(z),\qquad z\in\Om_1.
\eqno (1.76)
$$ 

{\it Region $\Om^1$, TP asymptotics.}
In the region $\Om^1$ we define the turning point vector valued function
$\vec\Psi_{\TP}(z)$ as
$$
\vec\Psi_{\TP}(z)\equiv
 (R_n^0)^{-1/4}\,
W(z)\pmatrix
N^{1/6}\Ai\(N^{2/3}w(z)\)  \\
N^{-1/6}\Ai'\(N^{2/3}w(z)\)
\endpmatrix,\qquad z\in\Om^1,
\eqno (1.77)
$$
where
$$
w(z)=\(\frac{3}{2}\int_{z_0^N}^z \sqrt {U^0(u)}\,du\)^{2/3}. 
\eqno (1.78)
$$
and
$$
W(z)=\({a_{12}^0(z)\over w'(z)}\)^{1/2}
\pmatrix
1 & 0 \\
-\di{a_{11}^0(z)\over a_{12}^0(z)} & \di{w'(z)\over a_{12}^0(z)}
\endpmatrix,
\eqno (1.79)
$$
so that $\det W(z)\equiv 1$.
The branches for fractional powers in (1.78,79) are chosen to be positive
for $z>z^N_0$. Observe that the function $w(z)$ is analytic and has
no critical points in $\Om^1$. We will prove that
$$
\vec\Psi_n(z)=\(1+O(N^{-1})\)\vec\Psi_{\TP}(z),
\qquad z\in\Om^1.
\eqno (1.80)
$$

{\it Region $\Om^0$, CP asymptotics.} The normal form 
for  system (1.27)
at the critical point $z=0$ is the system
$$
\Psi'(z)=A(z)\Psi(z),
\eqno (1.81)
$$
where
$$
A(z)=
\pmatrix
(-1)^n4u(y)z & 4z^2+(-1)^n2w(y)+v(y) \\
-4z^2+(-1)^n2w(y)-v(y)) & -(-1)^n4u(y)z \\
\endpmatrix,
\eqno (1.82)
$$
and
$$
v(y)=y+2u^2(y),\qquad w(y)=u'(y),
\eqno (1.83)
$$
where $u(y)$ is the Hastings-McLeod solution of the Painlev\'e II
equation (1.46) (see Section 3 below). We will consider a special 
solution to (1.81),
$$
\vec\Phi(z)=
\pmatrix
\Phi^1(z) \\
\Phi^2(z)
\endpmatrix,
\eqno (1.84)
$$
which is characterized by the following properties:
\item{1)} $\vec\Phi(z)$ is real, i.e.,
$$
\vec\Phi(\overline{z})=\overline{\vec\Phi(z)}.
\eqno (1.85)
$$
\item{2)} It satisfies the parity equation
$$
\vec\Phi(-z)=(-1)^n\sg_3\vec\Phi(z).
\eqno (1.86)
$$
\item{3)} On the real axis the functions $\Phi^j(z)$ have the
asymptotics
$$\eqalign{
\Phi^1(z)&=\cos\(\frac{4z^3}{3}+yz-\frac{\pi n}{2}\)+O\(z^{-1}\),\cr
\Phi^2(z)&=-\sin\(\frac{4z^3}{3}+yz-\frac{\pi n}{2}\)+O\(z^{-1}\),
\qquad z\to\pm\infty.}
\eqno (1.87)
$$

\noindent
The existence of the solution $\vec\Phi(z)$ is a nontrivial
fact of modern theory of Painlev\'e equations (see e.g. [IN],
[DZ2]). It should be also noticed that, in addition to
properties (1.85-87), the function  $\vec\Phi(z)$ is
an entire function on the complex $z$ plane, and its asymptotic
behavior is known in the whole neighborhood
of $z = \infty$ (see Proposition 3.2 below). 

{\it Remark.} As a function of the parameter $y$, 
the vector  $\vec\Phi(z)$
is a meromorphic function which
satisfies the linear differential  equation (cf. (1.81)),
$$
{\partial \Psi(z) \over{\partial y}}=B(z)\Psi(z),
\eqno (1.81')
$$
where
$$
B(z)=
\pmatrix
(-1)^nu(y) &  z \\
-z & -(-1)^nu(y) \\
\endpmatrix.
\eqno (1.82')
$$
Moreover, the asymptotic distributions
of the complex poles (which coincide with the
poles of the Painlev\'e  function $u(y)$) and the
large $|y|$ asymptotics of $\vec\Phi(z)$ can be extracted
from the general asymptotic results concerning the 
oscillatory Riemann-Hilbert problem associated with
the second Painlev\'e equation (see [IN], [Kap1],
[Kit2], [DZ2], and [IK]). 

In the region $\Om^0$ we define the critical point
function as
$$
\vec\Psi_{\CP}(z)\equiv
\frac{1}{2\pi^{1/2}}\, (R_n^0)^{-1/4}\,V(z)\,\vec\Phi\(N^{1/3}\z(z)\),
\eqno (1.88)
$$
where the matrix valued function $V(z)$ and 
the function $\z(z)$ will be defined
below in Section 4 [see (4.65) and (4.82)].  Both $V(z)$ and $\z(z)$ 
are analytic in $\Om^0$, $\det V(z)\equiv 1$,
and $\z(z)$ has no critical points in $\Om^0$.
It is worth noticing that 
the functions $V(z)$, $\vec\Phi(z)$, and $\z(z)$
depend on the parameter $y$.
We will prove that in $\Om^0$,
$$
\vec\Psi_n(z)=\(1+O(N^{-1})\)\vec\Psi_{\CP}(z),\qquad z\in\Om^0.
\eqno (1.89)
$$
We can now formulate the main result concerning the
double scaling limit for orthogonal polynomials.

{\bf Theorem 1.2.} {\it There exists $d_2^0>0$ such that
for all $d_2$ in the interval $0<d_2\le d_2^0$ and all $d_1$ satisfying
inequalities (1.60), the following holds. Let $\Om$, $\Om^0$, $\Om^1$, 
and $\Om_1$
be the regions defined in (1.61-63) (see Fig.2). Let $T_0>0$ be an
arbitrary number and the variable $y$, defined
in (1.43), satisfies bound (1.44). Then the recurrence
coefficients $R_n$ obey asymptotic formula (1.45),
and for the vector valued function
$\vec\Psi_n(z)$, asymptotic relations (1.64), (1.72), (1.76),
(1.80), and (1.89) hold. } 

{\it Remark.} It is interesting to compare Theorem 1.2
with the results
of  Deift-Kriecherbauer-McLaughlin-Venakides-Zhou [DKMVZ1]  
and  Baik-Deift-Johansson [BDJ],
which are both based on the Riemann-Hilbert approach as well. 
The general theorem 
of [DKMVZ1] can be applied to our problem at the critical
point, $y=0$, and for this case it gives 
asymptotics formulae for the function $\vec\Psi_n(z)$ with
an error term of the order of $N^{-1/3}$.
In the work of Baik-Deift-Johansson
[BDJ], asymptotics similar to the ones of Theorem 1.2
(with the error term of the order of $N^{-2/3}$)
have been obtained for the double scaling limit
of a Riemann-Hilbert problem on the circle.  We provide some comments
on the relation of our approach and the one of [BDJ] in
Appendix F.

A different type of the double scaling limit in discrete
string equation (1.23), which leads to the appearance of
the Painlev\'e I equation, was discovered in
[BK], [DS], and [GM] in connection with the matrix
model of 2D quantum gravity. This limit, on the level of solutions
of the string equation, was analysed in [FIK] in the
framework of the Riemann-Hilbert isomonodromy approach.
Our method is very close to the scheme of [FIK]. An important
difference is, however,  that we construct an approximation
of the orthogonal polynomials of the order of $N^{-1}$ 
 and in a finite neighborhood
of the critical point. The original approach of [FIK] 
gives only estimates of the order of $N^{-1/3}$
and in a neighborhood of the size of $N^{-1/3}$.

Finally, it is interesting to notice that the Hastings-McLeod
solution to Painlev\'e II also appears in the Tracy-Widom distribution
function at the edge of the spectrum (see [TW2]).

{\it Double scaling limit for correlation functions.} 
From Theorem 1.2 we will derive the following
results concerning the double scaling limit for correlation
functions of the quartic matrix model.

{\bf Theorem 1.3.} {\it Let $\vec\Phi(z;y)=
\pmatrix \Phi^1(z;y) \\ \Phi^2(z;y) \endpmatrix $ be the solution for $n=0$
to system (1.81) and relations  (1.85-87). 
Then the following double scaling limit holds:       
$$\eqalign{
\lim_{N\to\infty} \frac{1}{\(cN^{1/3}\)^{m-1}}
&K_{Nm}\(\frac{u_1}{cN^{1/3}},\dots,\frac{u_m}{cN^{1/3}};
t_c+c_0yN^{-2/3}\)\cr
&=\det\(Q_c(u_i,u_j;y)\)_{i,j=1,\dots,m},}
\eqno (1.90)
$$
where $c=\z'(0)>0$, and
$$
Q_c(u,v;y)=\frac{\Phi^1(u;y)\Phi^2(v;y)
-\Phi^1(v;y)\Phi^2(u;y)}{\pi(u-v)}\,.
\eqno (1.91)
$$
Furthermore, if $z$ is in the bulk of the spectrum, i.e.,
$0<|z|<z_0$, then
$$\eqalign{
\lim_{N\to\infty} \frac{1}{(p(z)N)^{m-1}}
&K_{Nm}\(z+\frac{u_1}{p(z)N},\dots,z+\frac{u_m}{p(z)N};
t_c+c_0yN^{-2/3}\)\cr
&=\det\(Q_b(u_i,u_j)\)_{i,j=1,\dots,m},}
\eqno (1.92)
$$ 
where
$$
Q_b(u,v)=\frac{\sin\pi(u-v)}{\pi(u-v)}\,,
\eqno (1.93)
$$
the sine kernel.
At the edge of the spectrum,
$$\eqalign{
\lim_{N\to\infty} \frac{1}{\(cN^{2/3}\)^{m-1}}
&K_{Nm}\(z_0+\frac{u_1}{cN^{2/3}},\dots,z_0+\frac{u_m}{cN^{2/3}};
t_c+c_0yN^{-2/3}\)\cr
&=\det\(Q_e(u_i,u_j)\)_{i,j=1,\dots,m},}
\eqno (1.94)
$$ 
where $c=w'(z_0)>0$ and
$$
Q_e(u,v)=
\frac{\Ai(u)\Ai'(v)-\Ai(v)\Ai'(u)}{u-v}\,,
\eqno (1.95)
$$ 
the Airy kernel.}

{\it Remark.} The sine-kernel in the bulk of the spectrum is
established for a general (fixed) $V(M)$ by Pastur and Shcherbina [PS]
in a very different approach (see also a nonrigorous derivation
of the sine kernel
in  the physical work of Br\'ezin and Zee [BZ]).
In the Riemann-Hilbert approach the sine-kernel in the bulk of the
spectrum is obtained for a general (fixed) $V(M)$
by Deift, Kriecherbauer, McLaughlin, Venakides,
and Zhou in [DKMVZ1] 
(see also earlier paper [BI] for the quartic noncritical case).
The Airy kernel at the edge is  established for the Gaussian
model by  Bowick and Br\'ezin [BB], Forrester [For],
Moore [Moo], and 
Tracy and Widom [TW2], and for the quartic
model by Bleher and Its [BI]. 

{\it Plan for the rest of the paper.}  In Section 2 we will give a
formal (perturbative) derivation 
of  asymptotics (1.45) for $R_n$. It will be
justified later in subsequent sections. In Section 3 we will discuss
a normal form for the system of differential equations
on the psi-function at the critical point. In 
Section 4 
we will develop a three step construction of the approximate
solution in a neighborhood of the critical point.
At the critical point four turning points coalesce and this
makes the asymptotic analysis rather nonstandard.
We will be looking for the solution in the form 
$V(z)\Phi(N^{1/3}\z(z))$, where $\Phi(z)$ is a psi-function
for the Hastings-McLeod solution to Painlev\'e II, and first we 
will construct
$\z(z)$ in the zeroth order approximation,
then we will construct $V(z)$ in the zeroth order approximation, 
and after that we will correct
$\z(z)$ to include terms of the first order.
The basic role in this construction will be played
(cf. [Ble2], [Kap3])
by the equation of the equality of periods [see (4.64)].

In Section 5 we will construct the WKB approximate solution
in $\Om^c$ and we will prove that it matches
the critical point approximate solution  up to terms of the
order of $N^{-1}$. In Section 6 we will construct a
turning points approximate solution in the region $\Om^1$, 
and we will show that it matches the
WKB approximate solution. In Section 7 we will construct
the WKB solution of cosine type in the intermediate 
region $\Om_1$ and we will prove that it matches
the CP and TP solutions. In addition,
we will prove that it matches the WKB solution
on the top and on the bottom of $\Om_1$.

The analysis of Sections 5-7 provides us with
an {\it explicit} matrix-valued function $\Psi^{0}_{n}(z)$
which solves asymptotically, as $N \to \infty$, the basic 
Riemann-Hilbert problem (1.35-38). In Section 8
we will prove that the quotient  
$\Psi_{n}(z)[\Psi^{0}_{n}(z)]^{-1}$, 
is equal to $I + O(N^{-1}(1+|z|)^{-1})$ and this will 
conclude the proof of Theorem 1.2. We emphasize that
we only use differential  equation (1.27) to motivate
our choice of the function  $\Psi^{0}_{n}(z)$. The uniform
estimate for the difference $\Psi_{n}(z)[\Psi^{0}_{n}(z)]^{-1}
 - I$ is proved by means independent of the
WKB theory of differential equations. Indeed, the proof of the estimate
is based on the analysis of the Riemann-Hilbert problem which is solved
by the quotient  $\Psi_{n}(z)[\Psi^{0}_{n}(z)]^{-1}$.
The ideas and techniques used in Section 8 are close to 
the Deift-Zhou nonlinear steepest descent method [DZ1],
although there is an essential difference as well.
We give more details on this matter in Appendix F.
Finally, in Section 9 we will prove Theorem 1.3.

For what follows it will be convenient to make the substitution
$$
\Psi_n(z)=\widetilde\Psi_n(z)
\pmatrix
(2\pi)^{-1/2} & 0 \\
0 & (2\pi)^{1/2}
\endpmatrix.
\eqno (1.96)
$$
This does not change the Lax pair equations (1.34) and
this simplifies equation (1.35) to
$$
\wt \Psi_{n+}(z) =\wt\Psi_{n-}(z)
\pmatrix
1 & -i \\
0 & 1 
\endpmatrix.
\eqno (1.97)
$$
For the sake of brevity we will redenote $\wt\Psi_n(z)$ back
to $\Psi_n(z)$ and we will account substitution
(1.96) at the very end.

\beginsection 2. Formal Painlev\'e II Asymptotics Near the 
Critical Point \par

String equation (1.24) is supplemented by the initial conditions
$$
R_0=0,\quad R_1=\frac{\int_{-\infty}^\infty z^2 e^{-NV(z)}dz}
{\int_{-\infty}^\infty e^{-NV(z)}dz}\,
\eqno (2.1)
$$
[see (1.12)]. The critical point 
$
\lacr=\frac{t^2}{4g}
$
is a bifurcation point for $R_n$. Namely,
for $\la<\lacr$ the numbers $R_n$ are attracted to two different
branches for odd and even $n$, so that
$$
\lim_{n,N\to\infty\:\; n/N\to\la} R_n
=\left\{
\eqalign{
&R(\la)\iff n=2k+1,\cr
&L(\la)\iff n=2k,\cr}\right.
\eqno (2.2)
$$
where 
$$
R,L={-t\pm\sqrt{t^2-4\la g}\over 2g},\qquad\la<\lacr.
\eqno (2.3)
$$
see [BI]. Contrariwise, for $\la>\lacr$, the numbers $R_n$ are 
attracted to one branch, 
$$
\lim_{n,N\to\infty\:\; n/N\to\la} R_n=R(\la),
\eqno (2.4)
$$
where
$$
R={-t+\sqrt{t^2+3g\la}\over 6g},\qquad \la>\lacr
\eqno (2.5)
$$
At the critical point,
$$
R=L=-{t\over 2g},\qquad \la=\lacr\,.
$$

\centinsert{\pscaption{\psboxto (6in;3.6in){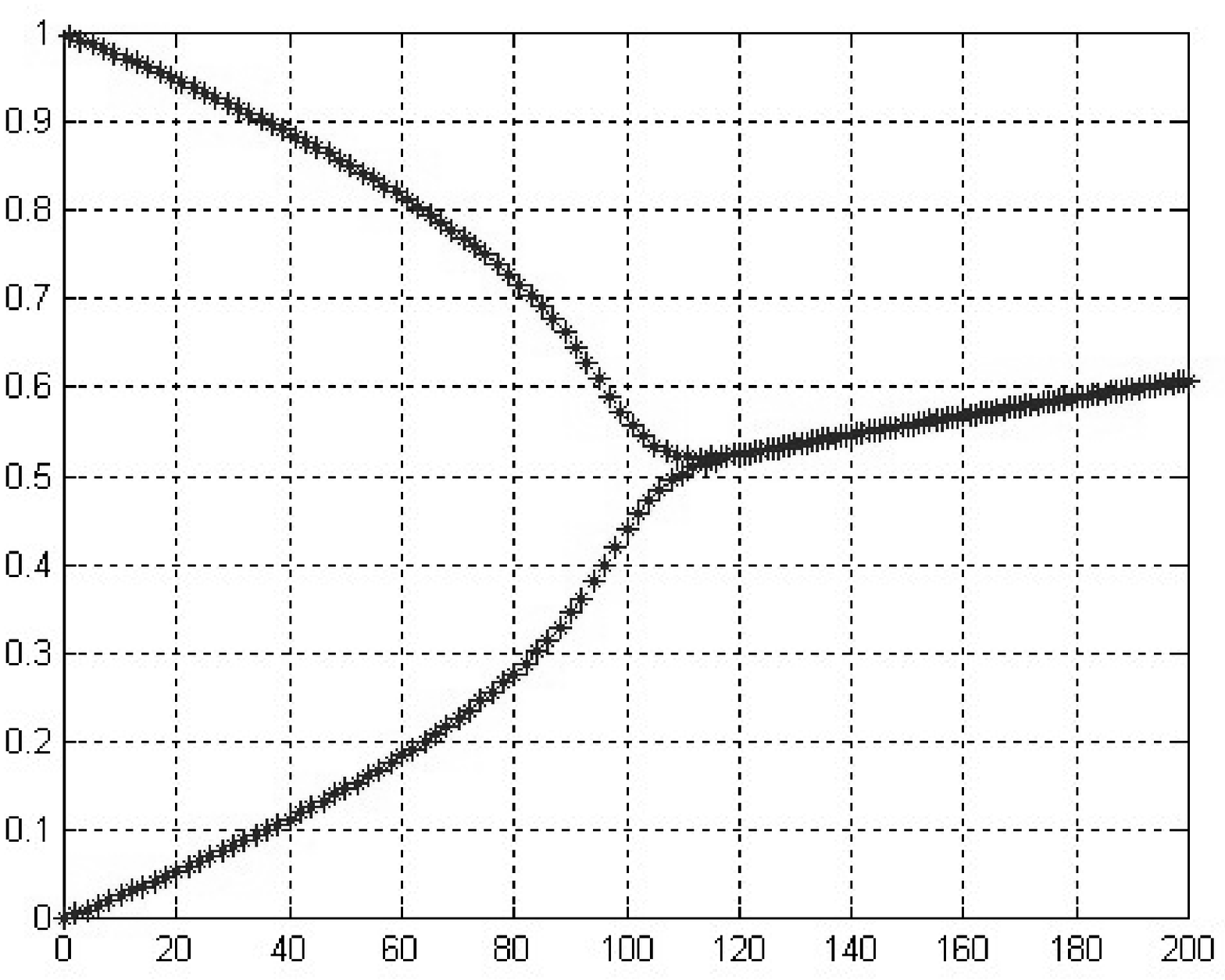}}
{\srm Fig 3:  Computer integration of the string equation: $R_n$ versus
$n$. The solution shown corresponds to $t=-1$, $g=1$, and $N=400$. }}

\vskip 3mm

\rm
\noindent
Fig. 3 presents results of numerical integration of string equation
(1.24) with the parameters
 $t=-1,\; g=1$, and $N=400$.
In this case $\la_c=1/4$, so that $n_c=100$.
 The authors thank Bobby Ramsey
for his help in carrying out the numerical integration. It is worth
noticing that the numerical integration has been done by minimizing
a global variational functional on trajectories
$\{R_n,\;n=0,1,\dots,n_0\}$
 which proved to be an efficient method of integration. 

The problem of the double scaling limit is to
find the asymptotic behavior of $R_n$ near $\lacr$.
First we will calculate the double scaling formally
(by perturbation theory) and later we
will prove it rigorously. 
Let us assume that
$$
R_n=\left\{
\eqalign{
&R(y)\equiv -{t\over 2g}+N^{-\b}u(y)+N^{-\g}v(y),\iff n=2k+1,\cr
&L(y)\equiv-{t\over 2g}-N^{-\b}u(y)+N^{-\g}v(y),\iff n=2k,
\cr}
\right.
\eqno (2.6)
$$
where $y$ is determined by the equation
$$
{n\over N}=\la_c+N^{-\a}y,
$$
with some exponents  $\a,\b,\g>0$
to be determined. Our assumption here is that $u(y)$ and $v(y)$ are
smooth functions of $y\in\R$. Now we substitute ansatz (2.6) into (1.24),
first for odd $n$ and then for even,
$$\eqalign{
&\lacr+N^{-\a}y=R(y)
\[ t+g(2L(y)+R(y))+g\,N^{2\a-2}\De L(y)\],\cr
&\lacr+N^{-\a}y=L(y)
\[ t+g(2R(y)+L(y))+g\, N^{2\a-2}\De R(y)\],\cr}
\eqno (2.7)
$$
where $\De$ is defined as
$$
\De f(y)={f\(y-N^{\a-1}\)-2f(y)+f\(y+N^{\a-1}\)\over N^{2\a-2}}\,.
$$
For our calculations we can replace $\De$ by the operator of
the second derivative. 
Subtracting the second equation in (2.7) from the first one,
we obtain that
$$
(R-L)\,(t+gR+gL)+g\,N^{2\a-2}(RL''-LR'')=0.
$$
Substituting (2.6) and neglecting higher order terms in $N^{-1}$
  we obtain that
$$
4guv+N^{2\a-2+\g}\,tu''=0.
\eqno (2.8)
$$
To get a nontrivial scaling we set
$$
2\a-2+\g=0\,;
\eqno (2.9)
$$
then (2.8) reduces to the equation
$$
v=-{tu''\over 4gu}\,.
\eqno (2.10)
$$
The first equation in (2.7) gives that
$$
N^{-\a}y=-N^{-\g}2tv
-N^{-2\b}gu^2
\eqno (2.12)
$$
(modulo smaller terms). To get a nontrivial scaling we put
$$
\a=\g=2\b
$$
Combining these relations with (2.9), we obtain that
$$
\a=\g={2\over 3},\quad \b={1\over 3}\,,
\eqno (2.13)
$$ 
so that (2.6) is written as
$$
R_n=
 -{t\over 2g}+(-1)^{n+1}N^{-1/3}u(y)+N^{-2/3}v(y),
\quad
y=N^{2/3}\({n\over N}-\la_c\),
\eqno (2.14)
$$
and (2.12) reduces to the equation
$$
y=-2tv-gu^2.
\eqno (2.15)
$$
Substituting $v$ from (2.10), we obtain that  
$$
y={t^2\over 2g}\,{u''(y)\over u(y)}-gu^2(y),
\eqno (2.16)
$$
which is the Painlev\'e II equation. From (2.10),
$$
v(y)=-{t\over 4g}\,{u''(y)\over u(y)}
={y+gu^2(y)\over (-2t)}\,.
\eqno (2.17)
$$
To bring (2.16,7) to a standard form of the Painlev\'e II equation
we make a rescaling of $u,v$ and $y$. To that end we rewrite (2.14)
as
$$ 
R_n=
 -{t\over 2g}+c_1(-1)^{n+1}N^{-1/3}u(y)+c_2N^{-2/3}v(y),
\quad
y=c_0^{-1}N^{2/3}\({n\over N}-{t^2\over 4g}\),
\eqno (2.18)
$$
where  
$$
c_0=\({t^2\over 2g}\)^{1/3},\qquad 
c_1=\({2|t|\over g^2}\)^{1/3},\qquad
c_2={1\over 2}\({1\over 2|t|g}\)^{1/3}\,.
\eqno (2.19)
$$
Then (2.16) reduces to
$$
u''=uy + 2u^3,
\eqno (2.20)
$$
which is a standard form of the Painlev\'e II equation
(in fact, a particular case of the Painlev\'e II).
As we have already indicated in introduction, 
ansatz (2.18) had been suggested in physical papers
by Douglas, Seiberg, Shenker [DSS], Crnkovi\'c, Moore [CM],
 and Periwal, Shevitz [PeS].
 
The equations (2.3) and (2.5) give the boundary 
conditions
$$
u(y)\left\{
\eqalign{
&\sim \sqrt{-y/2},\iff y\to-\infty,\cr
& \to 0,\iff y\to\infty.\cr}
\right.
\eqno (2.21)
$$
This selects a special solution to the Painlev\'e equation,
the Hastings-McLeod solution [HM] (cf. Section 1). It has the following
asymptotics:
$$\eqalign{
&u(y)=\({-y\over 2}\)^{1/2}\(1+{1\over 4y^3}+\dots\),\quad 
v(y)=-{1\over 4y^2}+O(y^{-4}),\quad y\to-\infty\,,\cr
&u(y)=\Ai(y)\(1+O\(e^{-(2/3)y^{3/2}}\)\),\quad
v(y)=y+O\(e^{-(4/3)y^{3/2}}\),\quad y\to\infty,\cr}
\eqno (2.22)
$$
where $\Ai(y)$ is the Airy function.
Of course,
the numbers $R_n$ in (2.18) do not satisfy the 
Freud equation (1.24) exactly,
and we have for them the Freud equation with an error term,
$$
{n\over N}=R_n\(t+gR_{n-1}+gR_n+gR_{n+1}\)+O\(N^{-4/3}\).
\eqno (2.23)
$$
More precisely, substitution of (2.18) into (1.24) gives that
$$\eqalign{
R_n&\(t+gR_{n-1}+gR_n+gR_{n+1}\)-{n\over
N}=N^{-2/3}c_0\[v(y)-y-2u^2(y)\]\cr 
&+N^{-1}(-1)^n\[u''(y)-u(y)v(y)\]+O\(N^{-4/3}\),}
\eqno (2.24)
$$
which implies (2.23).

To get an asymptotic expansion for $R_n$ we use (2.18) with
$$
u(y)=\sum_{j=0}^\infty N^{-(2/3)j}u_j(y),\qquad 
v(y)=\sum_{j=0}^\infty N^{-(2/3)j}v_j(y).
\eqno (2.25)
$$
Then from the terms of the order of $N^{-2/3}$ and $N^{-1}$
we obtain Painlev\'e equations (2.20) on $u_0,\;v_0$.
After that from the terms of the order of $N^{-4/3}$ and
$N^{-5/3}$ we obtain a linear system on $u_1,\;v_1$; then
from the terms of the order of $N^{-2}$ and
$N^{-7/3}$ we obtain a linear system on $u_2,\;v_2$,
etc. To illustrate this process assume that $g=1$
and $t=-2$ (general case can be reduced to this one).
Then
$$\eqalign{
R_n&\(t+gR_{n-1}+gR_n+gR_{n+1}\)-{n\over
N}=N^{-2/3}2^{1/3}(v_0-2u_0^2-y)+N^{-1}(u_0''-u_0v_0)\cr
&+N^{-4/3}2^{1/3}\(-4u_0u_1+v_1+\frac{2^{1/3}}{16}(2v''_0-16u_0u_0''
+3v_0^2)\)\cr
&-N^{-5/3}\(v_0u_1+u_0v_1+\frac{2^{1/3}}{4}(u_0v_0''
-v_0u_0'')\)+\dots}
\eqno (2.26)
$$
Equating coefficients on the right to zero we obtain the equations
$$
v_0=y+2u_0^2,\qquad u_0''=u_0v_0,
\eqno (2.27)
$$
and
$$
\eqalign{
4u_0u_1-v_1&=\frac{2^{1/3}}{16}(2v''_0-16u_0u_0''
+3v_0^2),\cr
v_0u_1+u_0v_1&=-\frac{2^{1/3}}{4}(u_0v_0''
-v_0u_0''),}
\eqno (2.28)
$$
etc., from which we subsequently determine the functions $u_j(y)$ and
$v_j(y)$.

\beginsection 3. The $\Psi$-Functions for Painlev\'e II \par 

Our next step is to derive a model (normal form) equation for the
matrix differential equation
$$
\Psi'_n(z)=NA_n(z)\Psi_n(z)
$$
at the critical point $z=0$. To that end we substitute (2.18)
into the matrix elements of $A_n(z)$,  
rescale $z$ as
$$
z=CN^{-1/3}s,\qquad C=\({32\over |t|g}\)^{1/6},
\eqno (3.1)
$$ 
and keep the leading terms. This gives the equation
$$
\Psi'(s)=A(s)\Psi(s),
\eqno (3.2)
$$
with
$$
A(s)=
\pmatrix
(-1)^n4u(y)s & 4s^2+(-1)^n2w(y)+v(y) \\
-4s^2+(-1)^n2w(y)-v(y)) & -(-1)^n4u(y)s \\
\endpmatrix,
\eqno (3.3)
$$
where
$$
v(y)=y+2u^2(y),\qquad w(y)=u'(y),
$$
and $u(y)$ is the Hastings-McLeod solution of the Painlev\'e II
equation (2.20). For another form of the model equation see
the physical paper by Akemann, Damgaard, Magnea, and Nishigaki
[ADMN].

{\it Turning Points of the Model Equation.} The turning points
of (3.2)
are the zeros of $\det A(s)$ on the complex plane. Observe that
$$
\det A(s)=16s^4+8ys^2+v^2(y)-4w^2(y),
\eqno (3.4)
$$
hence the turning points are solutions of the biquadratic equation
$$
s^4+{y\over 2}s^2+{v^2(y)-4w^2(y)\over 16}=0.
\eqno (3.5)
$$
Denote for this equation,
$$\eqalign{
&p=p(y)\equiv {y\over 2},\quad 
q=q(y)\equiv {v^2(y)-4w^2(y)\over 16},\cr
&D=D(y)\equiv p^2(y)-4q(y)=
 {y^2-v^2(y)\over 4}+w^2(y).}
\eqno (3.6)
$$

{\bf Proposition 3.1} {\it (i) The discriminant $D(y)$ is positive for
all $y\in\R$, so that equation (3.5) has 4 roots, 
$\pm s_{1,2}=\pm s_{1,2}(y),$ such that 
$$
-\infty <s_1^2(y)< s_2^2(y)<\infty.
$$ 
(ii) For all $y\in\R$,
$
s_1^2(y)<0.
$ 
There exists some $y_0>0$ such that
$$
s_2^2(y)>0,\quad y<y_0;\qquad s_2^2(y)<0,\quad y>y_0.
$$
(iii) The following asymptotics hold:
$$\eqalign{
&s_1^2\sim -{1\over 16y^2};\qquad
s_2^2\sim -{y\over 2},\qquad
y\to-\infty,\cr 
&s_{1,2}^2=-{y\over 4}\mp{1\over 4(2\pi)^{1/2}y^{1/2}}
e^{-(2/3)y^{3/2}}+\dots,\qquad 
y\to\infty.\cr} 
\eqno (3.7)
$$}

{\it Proof.} We have: 
$$
D'(y)=-u^2(y)-2u'(y)\( yu(y)+2u^3(y)-u''(y)\)=-u^2(y),
$$
and $D(\infty) =0$, hence
$$
D(y)=\int_y^\infty u^2(x)\,dx>0,
\eqno (3.8)
$$
which proves (i). To prove (ii) and (iii)
observe that by (3.8),
$$\eqalign{
&q'(y)={y\over 8}-{D'(y)\over 4}={y+2u^2(y)\over 8}={v(y)\over 8};\cr
&q(0)=-{D(0)\over 4}<0.}
\eqno (3.9) 
$$
Since $q(-\infty)=0$, we obtain that
$$
q(y)={1\over 8}\int_{-\infty}^y v(y)\,dy={1\over 32 y}+O\(|y|^{-3}\),
\quad y\to -\infty.
$$
This gives the first line in (3.7). 
From (3.8) and (2.22),
$$
D(y)={1\over 8\pi y}e^{-(4/3)y^{3/2}}\(1+O\(y^{-3/2}\)\),\quad
y\to\infty.
$$
This gives the second line in (3.7). Thus, (iii) is proved.

To prove (ii) we will show that $q(y)$ has a unique
zero $y_0$ on the real axis. Then (ii) follows. 
Since 
$$
v(y)=y+2u^2(y)>0,\quad y>0,
$$
we obtain from (3.9)
that $q(y)$ is increasing for $y\ge 0$ and it has one zero
$y_0>0$ on the positive half-axis (observe that $q(\infty)=\infty$).
We claim that $q(y)$ has no zeros on the negative half-axis. Indeed,
$$
v''(y)=4(u'(y))^2+4u''(y)u(y)=4(u'(y))^2+4u^2(y)v(y),
$$
hence if $v(y)> 0$ then $v''(y)> 0$. From (2.22) we have that
$v(y)$ is negative for negative $y$ sufficiently
large in absolute value. On the other hand, $v(0)=2u^2(0)\ge 0$,
hence there is $y_1\le 0$ such that $v(y_1)=0$. We claim that $v(y)$
can have only one zero. Assume that there are two. Then there exists
$y_2<0$ such that $v(y_2)=0$ and $v(t)>0$ for $y_2-\delta<t<y_2$,
for some $\delta>0$. But then $v'(y_2)\le 0$ and $v''(t)>0$
for $y_2-\delta<t<y_2$. Take now any
$y<y_2$. By the Taylor expansion,
$$
v(y)=v(y_2)+v'(y_2)(y-y_2)+{1\over 2}\int_y^{y_2}(t-y)v''(t)\,dt,
$$
hence $v(y)> 0$ as long as $v(t)\ge 0$ on $[y,y_2]$, 
which implies that $v(y)>0
$ for all $y\le y_2$ which is not true. The contradiction proves the
unicity of the zero of $v(y)$. This in turn proves that $q(y)$ has no zero
for negative $y$'s. Indeed, $q(-\infty)=q(y_0)=0$ hence if $q(y_3)=0$
for some $y_3\le 0$ then $v(y)=8q'(y)$ would have at least two zeros
which is not true. This finishes the proof of Proposition 3.1.

{\it Standard Form of the Model Equation.}
The matrix (3.3) differs from the standard 
Flaschka-Newell form of the $\Psi$-equation
for Painlev\'e II [FN]. To bring it to the standard form we make the gauge
transformation 
$$
\Psi(s)=U\widetilde \Psi(s), \qquad U=
\pmatrix
1 & -i \\
-i & 1
\endpmatrix
\pmatrix
1 & 0 \\
0 & (-1)^n i
\endpmatrix=
\pmatrix
1 & (-1)^n \\
-i & (-1)^n i
\endpmatrix.
\eqno (3.10)
$$
Then 
$$
\widetilde \Psi'(s)=\widetilde A(s)\widetilde\Psi(s),
\eqno (3.11)
$$
where
$$
\widetilde A(s)=U^{-1}A(s)U=
\pmatrix
-4is^2-iv & 4us+2iw \\
4us-2iw & 4is^2+iv
\endpmatrix,\quad v=y+2u^2.
\eqno (3.12)
$$

Solutions to equation (3.11) are entire functions of the
complex variable $s$. Their behaivor at $s=\infty$ is govern
by six (the infinity is an irregular singular point of (3.11)
of Poincare index 3) Stokes matrices. A key fact for
our analysis is that the Stokes matrices of (3.11), with
$u$, $v$, and $w$ determined via the Hastings-McLeod
Painlev\'e function, are known. More precisely (see [IN], [DZ2]),
there exist six Stokes' solutions $\widetilde\Psi_j(s)$ to equation
(3.11) such that
$$
\lim_{|s|\to\infty}\widetilde\Psi_j(s)e^{i\((4/ 3)s^3+ys\)\sg_3}=I
\iff
\left| \arg s-{(j-1)\pi\over 3}\right|\le {\pi\over 3}-\ep\,.
\eqno (3.13)
$$
The Stokes solutions are uniquely determined by (3.13),
and they are related as follows:
$$\eqalign{
&\widetilde\Psi_1(s)=\widetilde\Psi_6(s)
\pmatrix
1 & -1 \\
0 & 1 
\endpmatrix,\qquad
\widetilde\Psi_2(s)=\widetilde\Psi_1(s)
\pmatrix
1 & 0 \\
1 & 1 
\endpmatrix,\qquad
\widetilde\Psi_3(s)=\widetilde\Psi_2(s),\cr
&\widetilde\Psi_4(s)=\widetilde\Psi_3(s)
\pmatrix
1 & 0 \\
-1 & 1 
\endpmatrix,\qquad
\widetilde\Psi_5(s)=\widetilde\Psi_4(s)
\pmatrix
1 & 1 \\
0 & 1 
\endpmatrix,\qquad
\widetilde\Psi_6(s)=\widetilde\Psi_5(s).\cr}
\eqno (3.14)
$$

{\it Remark .} The existence of the Stokes solutions
$\widetilde\Psi_j(s)$ is a fact of the general
theory of systems of linear ODEs with rational
coefficients (see e.g. [Sib]), and it has nothing to do with
the particular choice of the parameters
$u$, $y$, and $w$ in (3.11). For any
triple $u$, $y$, and $w$ the solutions
$\widetilde\Psi_j(s)$ exist, they are holomorphic with respect
to $u$, $y$, and $w$, and the asymptotics (3.13)
is uniform with respect to $u$, $y$, and $w$ varying
in a compact. Also, in general case, the relations
(3.14) are replaced by the general equation,
$$
\widetilde\Psi_{j+1}(s) = \widetilde\Psi_j(s)S_{j},
$$
where the Stokes matrices $S_{j}$ are some
{\it transcendental} functions of parameters
$u$, $y$, and $w$. A remarkable fact ([FN], [JMU];
see also [IN] and indeed classical works of
R. Garnier [G]) is that the Stokes matrices
$S_{j}$ form a complete set of the first integrals
for the Painlev\'e equation (2.20). The particular choice
of Stokes matrices indicated in (3.14) corresponds
to a selection of the Hastings-McLeod solution
of the Painlev\'e II equation (2.20).

Equations (3.14) lead in turn to three special vector 
solutions to (3.2). 
 
{\bf Proposition 3.2.} {\it Assume that  $A(s)$ is defined as in
(3.3). Then for every real $y$, there exist vector
solutions $\vec\Phi(s)=\vec\Phi(s;y)$ and
 $\vec \Phi_j(s)=\vec \Phi_j(s;y),\; j=1,2,$ of the
equation $\vec \Phi'(s)=A(s)\vec \Phi(s)$ on the complex plane such
that: 

(i) $\vec \Phi(s)$ is real, i.e., 
$$
\vec\Phi(\overline{s})=\overline{\vec\Phi(s)},
\eqno (3.15)
$$
and it satisfies the parity equation,
$$
\vec\Phi(-s)=(-1)^n\sg_3
\vec\Phi(s).
\eqno (3.16)
$$
As $|s|\to\infty$,
$$
\vec \Phi(s)\sim \(\sum_{k=0}^\infty {\vec\G^0_k\over s^k}\)
e^{-i\((4/ 3)s^3+ys\)},
\iff \ep<\arg s<\pi-\ep,
\eqno (3.17)
$$
($\forall\,\ep>0$), with $\vec \G^0_0=\pmatrix i^n \\ i^{n-1}
\endpmatrix$, and 
$$\eqalign{
\vec \Phi(s)&\sim \(I+\sum_{k=1}^\infty \frac{\G_{2k}}{s^{2k}}\)
\pmatrix
\cos\(\frac{4s^3}{3}+ys-\frac{\pi n}{2}\) \\
-\sin\(\frac{4s^3}{3}+ys-\frac{\pi n}{2}\)
\endpmatrix,\cr
&\iff |\arg s|<\frac{\pi}{3}-\ep\quad\text{\rm or}\quad
|\arg s-\pi|<\frac{\pi}{3}-\ep,}
\eqno (3.17')
$$
where $\G_{2k}$ are some matrix valued coefficients.

(ii) $\vec\Phi_1(s)$ has the asymptotics as $|s|\to\infty$,
$$
\vec\Phi_1(s)
\sim \(\sum_{k=0}^\infty {\vec\G^1_k\over s^k}\)
e^{i\((4/ 3)s^3+ys\)},
\iff -{\pi\over 3}+\ep<\arg s<{4\pi\over 3}-\ep, 
\eqno (3.18)
$$ 
with $\vec \G^1_0=\pmatrix (-i)^{n+1} \\ (-i)^n
\endpmatrix$. 

(iii) $\vec\Phi_2(s)=\overline{\vec\Phi_1(\overline{s})}$ and
$$
\vec\Phi_2(-s)=\pmatrix (-1)^{n+1} & 0 \\ 0 & (-1)^n
\endpmatrix \vec\Phi_1(s).
\eqno (3.19)
$$

\noindent In addition,
$$
\vec \Phi_1(s)-\vec\Phi_2(s)=-i\vec \Phi(s).
\eqno (3.20)
$$}

In what follows we will use the matrix valued functions
$$
\Phi^{u}(s)= \(\vec \Phi(s),\vec\Phi_{1}(s)\)
,\qquad
\Phi^{d}(s)= \(\vec \Phi(s),\vec\Phi_{2}(s)\).
\eqno (3.21)
$$
As follows from Proposition 3.2 they satisfy the relations,
$$\eqalign{
&\Phi^u(s)=\Phi^d(s)S, \qquad S=
\pmatrix
1 & -i \\
0 & 1
\endpmatrix,\cr
&\Phi^{u,d}(-s)=(-1)^n\sg_3 \Phi^{d,u}(s)\sg_3,\qquad 
\overline{\Phi^u(\overline s)}=\Phi^d(s).}
\eqno (3.22)
$$
At infinity for all $\ep>0$ the function $\Phi^u(s)$ has the
asymptotics  
$$\eqalign{
\Phi^u(s)&=
\pmatrix
1 & -i \\
-i & 1
\endpmatrix
\(\sum_{j=0}^\infty {m_j\over s^j}\)
e^{-\({4\over 3}is^3+iys+\g\)\sg_3},\cr
&\text{as}\quad
s\to\infty,\quad \ep\le\arg s\le \pi-\ep,\cr}
\eqno (3.23)
$$
where 
$$
\g=-i{n\pi\over 2}\,,\qquad m_0=I,
$$
and the rest of the matrix coefficients $m_j$ can be found
recursively by substituting series (3.23) into equation (3.2).
The first two coefficients are given by the following equations:
$$\eqalign{
m_1&=-{iD\over 2}\sg_3-{(-1)^nu\over 2}\sg_1,\quad
m_2={u^2-D^2\over 8}\,I +(-1)^n{w+uD\over 4}\,\sg_2;\cr
D&=w^2-u^4-yu^2,}
\eqno (3.24)
$$
where $\sg_j$ are the Pauli matrices,
$$
\sg_1=
\pmatrix
0 & 1 \\
1 & 0
\endpmatrix,\quad
\sg_2=\pmatrix
0 & -i \\
i & 0
\endpmatrix,\quad
\sg_3=
\pmatrix
1 & 0 \\
0 & -1
\endpmatrix.
\eqno (3.25)
$$
Proof of Proposition 3.2  is given in Appendix A
below.

\beginsection 4. Semiclassical Approximation Near the Critical Point
\par 

In this section, which plays the central role in the whole paper,
we will construct a semiclassical approximation to the equation
$$
\Psi'(z)=NA_n^0(z)\Psi(z),
\eqno (4.1)
$$
in a fixed neighborhood of the critical point $z=0$.
It is worth to underline from the very beginning that
we will not prove any results concerning properties
and asymptotics of solutions to equation (4.1). It
will be used only as a tool 
in the construction of an approximate solution to the
RH problem. 

 Denote
$$
\t_n^0\equiv t+gR_n^0+gR_{n+1}^0.
\eqno (4.2)
$$
From (1.66) we get that
$$\eqalign{
\t_n^0&=N^{-2/3}c_5+N^{-1}c_6+O\(N^{-4/3}\),
\quad c_5=\({g^2\over 2|t|}\)^{1/3}\[v(y)
+2(-1)^nw(y)\],\cr
c_6&=\({g\over 2|t|}\)\[2(-1)^nu(y)v(y)+v'(y)\]\,;\cr
\t_{n-1}^0&=N^{-2/3}c_7+N^{-1}c_8+O\(N^{-4/3}\),
\quad c_7=\({g^2\over 2|t|}\)^{1/3}\[v(y)
-2(-1)^nw(y)\],\cr
c_8&=\({g\over 2|t|}\)\[2(-1)^nu(y)v(y)-v'(y)\].}
\eqno (4.3)
$$
This gives the matrix elements of $A_n^0(z)$ as
$$\eqalign{
a_{11}^0(z)
&=-{gz^3\over 2}-\[c_1g(-1)^{n+1}N^{-1/3}u(y)
+c_2gN^{-2/3}v(y)\]z
\,,\cr
a_{12}^0(z)&=(R_n^0)^{1/2}
\[gz^2+N^{-2/3}c_5+N^{-1}c_6+O\(N^{-4/3}\)\]\,,\cr
a_{21}^0(z)&=-(R_n^0)^{1/2}
\[gz^2+N^{-2/3}c_7+N^{-1}c_8+O\(N^{-4/3}\)\]\,,\cr}
\eqno (4.4)
$$
where the constants $c_1,\;c_2$ are defined in (1.45).
In the semiclassical approximation the function $\det A_n^0(z)$
will be important.  We will use the function
$$
d(z)\equiv
-\frac{g^2z^4}{4}(z^2-z_0^2)+g\(\frac{n}{N}-\la_c\) z^2-
\(c_3N^{-4/3}+c_4N^{-5/3}\),
\eqno (4.5)
$$
as a suitable approximation to $\det A_n^0(z)$ [cf. (1.58)].
From (1.67) we have that
$$
\det A_n^0(z)=d(z)+\a_n^0 z^2+\b_n^0;
\qquad \a_n^0=O(N^{-4/3}),\quad \b_n^0=O(N^{-2}).
\eqno (4.5')
$$

{\it Three Step Critical Point Solution.}
In a neighborhood of $z=0$ we are looking for a {\it critical point
approximate solution} to equation (4.1) in the following form:
$$
\Psi_{\CP}(z)=V(z)\Phi\(N^{1/3}\z(z)\),
\eqno (4.6)
$$
where $V(z)$ is a gauge matrix-valued function, $\Phi(z)$ is a
matrix-valued solution to the model equation 
$$
\Phi'(z)=A(z)\Phi(z),
\eqno (4.7)
$$
with $A(z)$ defined as in (3.3), and $\z(z)$ is an analytic change of
variable with $\z'(0)\not=0$. We will choose $\Phi(z)$ differently for
$\Im z\ge 0$ and 
$\Im z\le 0$ to secure the necessary multiplicative jump. Here we
will carry out a general 
analysis of (4.1) and we will assume that $\Phi(z)$ is any solution to
(4.7). 
Substituting (4.6) into (4.1) we obtain the equation
$$
V(z)\[\z'(z)N^{-2/3}A\(N^{1/3}\z(z)\)\]V^{-1}(z)
=A_n^0(z)-N^{-1}V'(z)V^{-1}(z).
\eqno (4.8)
$$
From this equation we determine iteratively $\z(z)$ and $V(z)$
in three steps: (1) the zeroth order approximation for $\z(z)$,
(2) the zeroth order approximation for $V(z)$, and (3) the first order
approximation for $\z(z)$. 

{\it Zeroth Order Approximation for $\z(z)$.} Taking the determinant 
of the both sides in (4.8) we obtain that
$$
[\z'(z)]^2N^{-4/3}\det
A\(N^{1/3}\z(z)\)=\det\[A_n^0(z)-N^{-1}V'(z)V^{-1}(z)\]. 
\eqno (4.9)
$$
In the zeroth order approximation we will neglect terms of the
order of $N^{-1}$. So we  drop the term
$N^{-1}V'(z)V^{-1}(z)$ on the right,
$$
[\z'(z)]^2N^{-4/3}\det
A\(N^{1/3}\z(z)\)=\det A_n^0(z),
\eqno (4.10)
$$
which is an equation on $\z(z)$ alone. By (3.4),
$$
N^{-4/3}\det A\(N^{1/3}\z\)
=f(\z)\equiv
16\z^4(z)+8N^{-2/3}y\z^2(z)+N^{-4/3}\[v^2(y)-4w^2(y)\].
\eqno (4.11)
$$
We replace $\det A_n^0(z)$ by $d(z)$ [see (4.5)], reducing
(4.10) to the equation
$$
(\z')^2f(\z)=d(z).
\eqno (4.12)
$$
Now we drop the $O(N^{-1})$ terms in $f$ and $d$ 
and we introduce the functions
$$
f^0(\z)=16\z^4+8N^{-2/3}y\z^2.
\eqno (4.13)
$$
and
$$
d^0(z)=\frac{g^2z^4}{4}\(z_0^2-z^2\)+g\({n\over N}-\lacr\)z^2.
\eqno (4.14)
$$
Equation (4.12) reduces then to
$$
(\z')^2 f^0(\z)=d^0(z),
\eqno (4.15)
$$
or  
$$
\z'\sqrt{f^0(\z)}=\sqrt{d^0(z)}\,.
\eqno (4.16)
$$
To fix a branch for the square roots, observe that
both $f^0(z)$ and $d^0(z)$ are positive at $z=z_0/2$
for large $N$. We will assume that the both square roots
are also positive at $z_0/2$. Equation (4.16) is separable
and it is easy to solve it. The problem is to find an
analytic solution. Consider this problem more carefully.
 Let us make the change of variables
$$
z=CN^{-1/3}s,\qquad \z=N^{-1/3}\sg,
\eqno (4.17)
$$
where $C$ is the same as in (3.1). We will call $(s,\sg)$ local
coordinates (at the critical point) and $(z,\z)$ global ones.
In the local coordinates equation (4.16) reduces to
$$
\sg'\sqrt{\f^0(\sg)}=\sqrt{\de^0(s)},
\eqno (4.18)
$$
where
$$\eqalign{
\f^0(\sg)&\equiv N^{4/3}f^0(N^{-1/3}\sg)=16\sg^4+8y\sg^2,\cr
\de^0(s)&\equiv C^2N^{4/3}d^0(CN^{-1/3}s)
=16s^4+8c_0^{-1}N^{2/3}\({n\over N}-\lacr\)s^2
-N^{-2/3}c_9s^6,\cr}
\eqno (4.19)
$$
where $c_0$ is defined in (1.43) and
$$
c_9={2^{14/3}g^{2/3}\over |t|^{4/3}}.
\eqno (4.20)
$$
Using the definition of $y$ in (1.43) we reduce (4.19) to
$$\eqalign{
\f^0(\sg)&=16\sg^4+8y\sg^2,\cr
\de^0(s)&=16s^4+8ys^2
-N^{-2/3}c_9s^6.\cr}
\eqno (4.21)
$$
Observe that $\f^0$ and $\de^0$ differ only by a term of the order of
$N^{-2/3}$. Therefore, 
we are looking for an analytic solution to (4.18) in the form
$$
\sg(s)=s+N^{-2/3}\sg_1(s)+N^{-4/3}\sg_2(s)+\dots 
\eqno (4.22)
$$
However, there is an obstruction to the analytic solution. Namely, the
function $\f^0(\sg)$ has a double zero at $\sg=0$ and two simple zeros at  
$\sg=\pm \sg_0,\;\sg_0=(-y/2)^{1/2}$.
Similarly, the function $\de^0(s)$ has a double zero at $s=0$ and two simple
zeros at some points 
$s=\pm s_0$ that are close to $\pm \sg_0$, $s_0=\sg_0+O(N^{-2/3})$.
Consider a fixed (i.e., 
independent of $N$) closed contour $\G$ on the complex plane that goes around
all these zeros. Then both $\sqrt{\f^0(\sg)}$ and $\sqrt{\de^0(s)}$ are analytic
on $\G$, and if equation (4.18) holds and $\sg(s)$ is analytic, then
necessarily,
$$
\oint_{\G}\sqrt{\f^0(\sg)}\,d\sg=\oint_{\G}\sqrt{\de^0(s)}\,ds.
\eqno (4.23)
$$
This equation is also sufficient for the existence of an analytic solution
$\sg(s)$. Namely, integrating the both sides of (4.18) we obtain the equation
$$
\int_{\sg_0}^{\sg}\sqrt{\f^0(\tau)}\,d\tau+C_1
=\int_{s_0}^s\sqrt{\de^0(\tau)}\, d\tau+C_2.
\eqno (4.24)
$$
The integrals on the both sides have the same $s^{3/2}$-singularities
at $\sg=\sg_0$ and $s=s_0$, respectively, and to have an analytic
solution $\sg(s)$ at $s=s_0$ we take $C_1=C_2$. Similar considerations
hold at $s=-s_0$ and the necessary and sufficient
condition to have analyticity of $\sg(s)$ at both $s=s_0$ and $s=-s_0$
is that
$$
\int_{\sg_0}^{-\sg_0}\sqrt{\f^0(\tau)}\,d\tau
=\int_{s_0}^{-s_0}\sqrt{\de^0(\tau)}\, d\tau,
\eqno (4.25)
$$
which is equivalent to (4.23). We will call (4.23) the equation of
periods.

From (4.21) it is not difficult to evaluate the periods by perturbation
theory in $1/N^{2/3}$ and 
they are certainly different.
 Therefore we need a parameter to adjust the periods. 
To that end we change the relation between $n$ and $y$ from (1.43) to
$$
y=c_0^{-1}N^{2/3}\left(\frac{n}{N}-\la_c\right)+\a N^{-2/3},
\eqno (4.26)
$$
where $\a$ is a parameter. Then (4.21) changes as follows:
$$\eqalign{  
\f^0(\sg)&=16\sg^4+8y\sg^2,\cr
\de^0(s)&=16 s^4+8\(y-\a N^{-2/3}\)s^2-c_9N^{-2/3}s^6,\cr}
\eqno (4.27)
$$
and we find the value of the parameter $\a=\a(y)$ from
equation (4.23). Let us analyze this procedure in more detail.

When $y=0$ we take $\a=0$. Then both both $\f^0$ and $\de^0$
have a quadruple zero at the origin and equation (4.23) holds.
For $y\not=0$, let us scale
the functions $\f^0(\sg)$ and $\de^0(s)$ to the ones
$$\eqalign{
&\hat \f(\sg)\equiv -{1\over 16 y^2}\,\f^0(\sqrt {-2y}\,\sg)
=-\sg^4+\sg^2,\cr
&\hat \de(s)\equiv -{1\over 16 y^2}\,\de^0(\sqrt {-2y}\,s)
=-s^4+(1-\hat\a N^{-2/3})s^2-{N^{-2/3}c_9y\over 2}s^6,
\quad \a=y\hat\a,}
\eqno (4.28)
$$
where $\sqrt {-2y}>0$ for $y<0$ and $\Im \sqrt {-2y}>0$ for $y>0$. 
Then (4.23) reduces to
$$
\int_0^1\sqrt{\hat \f(\tau)}\,d\tau
=\int_0^{\hat s}\sqrt{\hat \de(\tau)} \, d\tau.
\eqno (4.29)
$$
where $\hat s$ is the close to 1 zero of $\hat\de(s)$,
$$
\hat \de(\hat s)=0,\quad \hat s=1+O(N^{-2/3}).
\eqno (4.30)
$$
To analyze (4.29) we use the implicit function theorem. To that end,
introduce, for small $|x_1|,|x_2|$, the function
$$
I(x_1,x_2)=\int_0^{\hat s(x_1,x_2)}\sqrt{-s^4+(1-x_1)s^2-x_2s^6}\,ds,
\eqno (4.31)
$$
where $\hat s(x_1,x_2)$ is the close to 1 zero of the function under 
the radical. We claim that $I(x_1,x_2)$ is an analytic function
at $x_1=x_2=0$ and
$$
\left.{\partial I(x_1,x_2)\over\partial x_1}
\right|_{x_1=x_2=0}\not=0\,,\quad
\left.{\partial I(x_1,x_2)\over\partial x_2}
\right|_{x_1=x_2=0}\not=0\,.
\eqno (4.32)
$$
To prove the analyticity apply the Cauchy theorem and rewrite
$I(x_1,x_2)$ as
$$
I(x_1,x_2)=\frac 12\oint_{\G_0}\sqrt{-s^4+(1-x_1)s^2-x_2s^6}\,ds,
\eqno (4.31')
$$
where $\Gamma_0$ is a circle on the complex plane of radius 1 centered
at $s=1$. In this form one can differentiate $I(x_1,x_2)$ with respect
to $x_1,x_2$ and prove the analyticity at $x_1=x_2=0$.
To prove (4.32) differentiate (4.31${}'$), set $x_1=x_2=0$, and return
back to the real integral:
$$
\left.
{\partial I(x_1,x_2)\over\partial x_1}
\right|_{x_1=x_2=0}=-\int_0^{1}\frac {s^2}{2\sqrt{-s^4+s^2}}\,ds\not=0,
$$
and similar for $\di{\partial I(x_1,x_2)\over\partial x_2}$.
Equation (4.29) is equivalent to
$$
I(x_1,x_2)=I(0,0),\quad x_1=\hat\a N^{-2/3},\quad 
x_2={N^{-2/3}c_9y\over 2}\,.
$$
The implicit function theorem ensures the existence
of an analytic solution $x_1=x_1(x_2)$ with 
$dx_1/dx_2(0)\not=0$. This in turn gives an analytic
solution $\a=\a(y)\sim cy^2$, $c\not=0$, to (4.23).

Let us summarize our calculations.
In the zeroth order approximation the change of variable
function $\z_0(z)$ is determined from the initial
value problem
$$
\z_0'\sqrt{f^0(\z_0)}=\sqrt{d^0(z)},\quad \z_0(0)=0,
\eqno (4.33)
$$
where
$$\eqalign{
f^0(z)&=16 z^4+8N^{-2/3}yz^2,\quad
\frac{n}{N}=\la_c+c_0N^{-2/3}\(y-\a N^{-2/3}\),\cr
d^0(z)
&=\frac{g^2z^4}{4}(z_0^2-z^2)+g\({n\over
N}-\la_c\)z^2,\cr}
\eqno (4.34)
$$
and $\a$ is determined from (4.23). 
We consider the solution to (4.33) in the region
$$
(-\Om_1)\cup\Om^0\cup\Om_1=
\{ z\,:\;-z_0+d_1\le\Re z\le z_0-d_1,\; |\Im z|\le d_2\},
$$
see Fig.2 above. We also notice that, because of the uniqueness
of the solution of initial value problem (4.33), the
following symmetry relation talks place:
$$
\z_{0}(-z) = -\z_{0}(z),
\eqno(4.35)
$$

When $z$ is separated from 0, i.e. $|z|\ge \om_0>0$ where
$\om_0 < d_{1}$ does not depend on $N$, (4.34) gives that
$$\eqalign{
&\sqrt{f^0(z)}=4z^2+N^{-2/3}yz+O(N^{-4/3}),\cr
&\sqrt{d^0(z)}=\sqrt{d(z)}+O(N^{-4/3}),}
$$
[cf. (4.5,9)], 
hence equation (4.33) implies that 
$$
4\z_0'(z)\z_0^2(z)+N^{-2/3}y\z_0'(z)=\sqrt{d(z)}+O(N^{-4/3}),
\quad |z|\ge\om_0.
$$
In what follows we use the function
$$
\mu_{0}(z)\equiv \sqrt {-d(z)},
\eqno (4.36)
$$
rather than $\sqrt{d(z)}$, so we rewrite the last equation as 
$$
4i\z_0'(z)\z_0^2(z)+N^{-2/3}iy\z_0'(z)=\mu_{0}(z)+O(N^{-4/3}),
\quad z\in\((-\Om_1)\cup\Om^0\cup\Om_1\)\setminus\{|z|<\om_0\}.
\eqno (4.37)
$$
Observe that
the function $-d(z)$ is positive for large $N$ if $z\ge z_0+d_1$
[cf. (4.5,9)] and we take the branch of the square root
for $\mu_{0}(z)$ which is positive for $z\ge z_0+d_1$. In (4.37)
we continue $\mu_{0}(z)$ analytically
going around $z_0$ from above. We note that, for sufficiently large
$N$, $\mu_{0}(z)$ is holomorphic in the domain
$$
\{|z| >\omega_{0}\}\setminus \Bigl((-\infty, -z^{N}]\cup
[z^{N}, +\infty)\Bigr),
$$
where $z^{N}$ denote the zero of $d(z)$ which approaches $z_{0}$
as $N \to \infty$. In particular, $\mu_{0}(z)$ is holomorphic in
the domain
$\((-\Om_1)\cup\Om^0\cup\Om_1\)\setminus\{|z|<\om_0\}$. 

Equation (4.37) can be used to describe, within an error term of 
the order of $N^{-4/3}$, the function
$\z_{0}(z)$ by an elementary explicit formula.
Indeed, as long as $z$ is
separated from 0, the function $\mu_{0}(z)$ admits the asymptotic
representation [cf. (4.5,36)],  
$$\eqalign{
\mu_{0}(z)& = {igz^2 \over{2}}\sqrt{z^{2}_{0} - z^2}
+{ic_{0}y \over{\sqrt{z^{2}_{0} - z^2}}}N^{-2/3} + O(N^{-4/3}),
\cr
z&\in \La(\om_0)\equiv\((-\Om_1)\cup\Om^0\cup\Om_1\)\setminus\{|z|<\om_0\}.}
\eqno (4.38)
$$ 
where $\sqrt{z^{2}_{0} - z^2}$ denote the branch of the
root which is analytic in $\C\setminus \((-\infty, z_{0}]\cup
[z_{0}, \infty)\)$ and positive for $-z_{0} < z < z_{0}$.
This estimate allows us to rewrite  (4.37) as 
$$
4\z_0'(z)\z_0^2(z)+N^{-2/3}y\z_0'(z)
={gz^2 \over{2}}\sqrt{z^{2}_{0} - z^2}
+{c_{0}y \over{\sqrt{z^{2}_{0} - z^2}}}N^{-2/3} + O(N^{-4/3}),
\quad z\in \La(\om_0).
\eqno (4.39)
$$
By integrating the last equation and taking into account the
symmetry $z \to -z$ we arrive to the relation
$$
{4\over 3}\z_0^3(z)+N^{-2/3}y\z_0(z)
= D_{\infty}(z) + N^{-2/3}yD_{1}(z) + O(N^{-4/3}),
\quad z\in \La(\om_0),
\eqno (4.40)
$$
where we have introduced the notations
$$
D_{\infty}(z) := \int_{0}^{z}{gu^2 \over{2}}\sqrt{z^{2}_{0} - u^2}\, du
\eqno (4.41)
$$
and
$$
D_{1}(z) := c_{0}\int_{0}^{z} {du\over{\sqrt{z^{2}_{0} - u^2}}}.
\eqno (4.42)
$$
Notice that both the functions $D_{\infty}(z)$ and $D_{1}(z)$
are analytic and odd in $\C\setminus \((-\infty, z_{0}]\cup
[z_{0}, \infty)\)$. (Of course, $D_{\infty}(z)$ and $D_{1}(z)$
can be expressed in terms of elementary functions, but we will
not need these expressions; the integral representations are
already elementary enough and quite convinient for any further
analysis.) Equation (4.40), in its turn, implies the estimate
$$
\z_{0}(z) = \z_{\infty}(z) +  N^{-2/3}y\z_{1}(z) + O(N^{-4/3}),
\quad z\in \La(\om_0),
\eqno (4.43)
$$
where the functions $\z_{\infty}(z)$ and  $\z_{1}(z)$ are
defined by the equations
$$
\z_{\infty}(z) = \[{3\over 4}D_{\infty}(z)\]^{1/3}
\eqno (4.44)
$$
and
$$
\z_{1}(z) = {{D_{1}(z) - \z_{\infty}(z)}\over{4\z^{2}_{\infty}(z)}}\,,
\eqno (4.45)
$$
respectively. By a straightforward calculation one can see that
both $\z_{\infty}(z)$ and  $\z_{1}(z)$ are
analytic at $z=0$. In fact, the first terms of
the relevant Taylor series are
$$
\z_{\infty}(z) = C^{-1}z  - {1\over{10Cz_{0}^{2}}}z^{3} + \ldots,
\quad D_{1}(z) = C^{-1}z  - {1\over{6Cz_{0}^{2}}}z^{3} + \ldots,
\eqno (4.46)
$$
and
$$
\z_{1}(z) = {1\over{60c_{0}^{2}}}z + \ldots,
\eqno (4.47)
$$ 
(note the absence of the singularity of $\z_{1}(z)$ at $z=0$). 
Therefore, the functions $\z_{\infty}(z)$ and  $\z_{1}(z)$ 
are analytic in the full rectangle $(-\Om_1)\cup\Om^0\cup\Om_1$.
This, in virtue of maximum principle, allows us to 
extend the asymptotic formulae (4.43) to the full rectangle,
$$
\z_{0}(z) 
=\z_{\infty}(z) +  N^{-2/3}y\z_{1}(z) + O(N^{-4/3}),
\quad
z\in (-\Om_1)\cup\Om^0\cup\Om_1.
\eqno (4.48)
$$

Initial value problem (4.33) which we have used to define
the function $\z_{0}(z)$ was obtained by neglecting terms
of the order of $N^{-1}$ in basic equation (4.9). Therefore, 
in view of  uniform  asymptotics (4.46), it makes perfect 
sense to redefine 
 $\z_{0}(z)$ as  
$$
\z_{0}(z) \equiv \z_{\infty}(z) +  N^{-2/3}y\z_{1}(z),
\quad y = c_{0}^{-1}N^{2/3}\({n\over N} - \lambda_{c}\),
\eqno(4.49)
$$
(note that we dropped the parameter $\alpha$ in the definition
of $y$). With this
new definition we preserve all the three basic properties of
the function $\z_{0}(z)$ which we will use in the next
step and later in matching the critical point and WKB
asymptotics (see section 5 and Appendix C).  These properties are:

\item {(i)} analyticity and $\z'_0(z)\not=0$ in the rectangle
$(-\Om_1)\cup\Om^0\cup\Om_1$,
\item {(ii)} estimate (4.37),
\item {(iii)} estimate (4.22) for the function $\sigma(s)
\equiv N^{1/3}\z_{0}(CN^{-1/3}s)$,
\item {(iv)} symmetry relation (4.35).
 
\noindent
Our next step will be a construction of the gauge matrix $V(z)$ in the
zeroth order approximation. 

{\it Gauge Matrix in the Zeroth Order Approximation.}
In the zeroth order approximation equation (4.8)
reduces to 
$$
V(z)\[\z_0'(z)N^{-2/3}A\(N^{1/3}\z_0(z)\)\]V^{-1}(z)
=A_n^0(z),
\eqno (4.50)
$$
which can be viewed as a generalized eigenvector problem,
because by Step 1 the matrices 
 $$
\z_0'(z)N^{-2/3}A\(N^{1/3}\z_0(z)\)
$$
and $A_n^0(z)$ have the same (up to terms of the order 
of $N^{-4/3}$) eigenvalues.
We will be looking for an approximate solution $V(z)$ to (4.50),
with a 
possible error term in the equation of the order of $O(N^{-1})$, and 
we will replace the matrix elements of $A^0_n(z)$ by
their suitable approximations [cf. (4.4)]. We put
$$\eqalign{
a_{11}^0(z)
&=-a_{22}^0(z)=-{gz^3\over 2}-\[c_1g(-1)^{n+1}N^{-1/3}u(y)
+c_2gN^{-2/3}v(y)\]z
\,,\cr
a_{12}^0(z)&=(R_n^0)^{1/2}
\(gz^2+N^{-2/3}c_5\)\,,\cr
a_{21}^0(z)&=-(R_n^0)^{1/2}
\(gz^2+N^{-2/3}c_7\)\,.\cr}
\eqno (4.51)
$$

 {\bf Lemma 4.1} {\it Let $B=\( b_{ij} \)$ and $D= \( d_{ij}\)$ be two
$2\times 2$ matrices such that
$$
\tr B=\tr D=0,\qquad \det B=\det D
$$
Then the equation $VB=DV$ has the following two explicit solutions:
$$
V_1=
\pmatrix
d_{12} & 0 \\
b_{11}-d_{11} & b_{12}
\endpmatrix,
\qquad
V_2=\pmatrix
b_{21} & d_{11}-b_{11} \\
0 & d_{21}
\endpmatrix
\eqno (4.52)
$$}

Proof of Lemma 4.1 is given in Appendix B below. We apply Lemma
4.1 to solve equation (4.50), with 
$$
B=\z_0'(z)N^{-2/3}A\(N^{1/3}\z_0(z)\),\qquad D=A_n^0(z)
$$
The problem is that in (4.50) we need an
analytic matrix valued function $V(z)$ which is invertible 
in some fixed neighborhood of the origin.
Neither $V_1$ nor $V_2$ in (4.52) are
invertible. Nevertheless, we will find a linear combination of
$V_1$ and $V_2$ (plus some negligibly small terms)
which is analytic and invertible.

Let us rewrite (4.50) in the local coordinates 
$(s,\sg)$ defined in (4.17):
$$\eqalign{
&V_0(s)\[\sg'(s)A(\sg(s))\]V^{-1}_0(s)=CN^{2/3}A_n^0(CN^{-1/3}s),\cr
&V_0(s)=V(CN^{-2/3}s).}
\eqno (4.53)
$$
Define
$$\eqalign{
&B_0(s)=\sg'(s)A\(\sg(s)\)=
\pmatrix
b_{11} & b_{12} \\
b_{21} & b_{22}
\endpmatrix \cr
&b_{11}=-b_{22}=(-1)^n4u\sg'(s)\sg(s),\cr
&b_{12}= \sg'(s)(4\sg^2(s)+(-1)^n2w+v),  \cr
&b_{21}=\sg'(s)(-4\sg^2(s)+(-1)^n2w-v),}
\eqno (4.54)
$$
and
$$\eqalign{
&D_0(s)=CN^{2/3}A_n^0(CN^{-1/3}s)=
\pmatrix
d_{11} & d_{12} \\
d_{21} & d_{22}
\endpmatrix \cr
&d_{11}=-d_{22}= 
(-1)^n4us-N^{-1/3}c_0^{-1}(vs+4s^3)\cr
& d_{12}=\({R_n^02g\over |t|}\)^{1/2}(4s^2+(-1)^n2w+v),\cr
& d_{21}=\({R_n^02g\over |t|}\)^{1/2}(-4s^2+(-1)^n2w-v)
\cr}
\eqno (4.55)
$$
To solve (within the error $O(N^{-2/3})$) the equation 
$$
V_0(s)B_0(s)V_0^{-1}(s)=D_0(s)
\eqno (4.56)
$$
we can use any linear combination of $V_1$ and $V_2$ in (4.52). To
determine an appropriate linear combination consider the matrix
element $b_{11}-d_{11}$ which appears in $V_1$ and with minus sign in
$V_2$. By (4.54,55), 
$$
b_{11}-d_{11}=(-1)^n4u\sg'(s)\sg(s)-(-1)^n4us
+N^{-1/3}c_0^{-1}s(4s^2+v).
\eqno (4.57)
$$
By (4.22), $\sg(s)=s+N^{-2/3}\sg_1(s)+\dots$, hence
$$
b_{11}-d_{11}=N^{-1/3}c_0^{-1}s(4s^2+v)+O(N^{-2/3}).
\eqno (4.58)
$$
Thus,  the function
$b_{11}(s)-d_{11}(s)$ has three zeros: $s=0$ 
 and two other zeros which are determined, in the zeroth order
approximation in $N^{-1/3}$, by the quadratic equation
$$
4s^2+v=0.
\eqno (4.59)
$$
(The function $b_{11}(s)-d_{11}(s)$ may have more zeros at the distance
of the order of $N^{1/3}$ from the origin but we are not interested
in those now.) Let us take
$$
V_0(s)={1\over \sqrt{\det W_0(s)}}W_0(s),\qquad 
W_0(s)=
\pmatrix
d_{12}(s)-b_{21}(s) & b_{11}(s)-d_{11}(s) \\
b_{11}(s)-d_{11}(s) & b_{12}(s)-d_{21}(s)
\endpmatrix,
\eqno (4.60)
$$
which is a linear combination of $V_1$ and $V_2$ in (4.52).
We want $V_0(s)$ 
to be an analytic invertible matrix-valued function. To that end
we will slightly correct $W_0(s)$. 
Consider the matrix elements $w_{ij}(s)$ of $W_0(s)$. From (4.54,55)
and (1.66),
$$\eqalign{
w_{11}(s)&=d_{12}(s)-b_{21}(s)\cr
&=8s^2+2v+N^{-1/3}c_0^{-1}(-1)^{n+1}u(4s^2+(-1)^n2w+v)
+O\(N^{-2/3}\);\cr
w_{12}(s)&=w_{21}(s)=b_{11}(s)-d_{11}(s)
=N^{-1/3}c_0^{-1}s(4s^2+v)+O(N^{-2/3});\cr
w_{22}(s)&=b_{12}(s)-d_{21}(s)\cr
&=8s^2+2v+N^{-1/3}c_0^{-1}(-1)^{n+1}u(4s^2-(-1)^n2w+v)
+O\(N^{-2/3}\).\cr}
\eqno (4.61)
$$
Observe that in the leading order of
approximation in $N^{-1/3}$ the zeros of $w_{ij}(s)$ 
are determined by the same quadratic equation (4.59). Let us define now
$$
W_0(s)=
\pmatrix
d_{12}(s)-b_{21}(s)-N^{-1/3}\a_{11} & 
b_{11}(s)-d_{11}(s)-N^{-2/3}\a_{12}s \\
b_{11}(s)-d_{11}(s)-N^{-2/3}\a_{21}s & 
b_{12}(s)-d_{21}(s)-N^{-1/3}\a_{22}
\endpmatrix
\eqno (4.62)
$$
where the numbers $\a_{11},\a_{12}=\a_{21}$, and $\a_{22}$ do not
depend on $s$ and are chosen in such a way that all the elements
in the matrix on the right vanishes at the zeros of $4s^2+v$.
Then the matrix valued function
$$
\widetilde W_0(s)={W_0(s)\over 8s^2+2v},
$$
is analytic in $s$ and for finite $s$,
$$
\widetilde W_0(s)=
\pmatrix
1 & 0\\
0 & 1
\endpmatrix
+N^{-1/3}
\pmatrix 
(2c_0)^{-1}(-1)^{n+1}u & (2c_0)^{-1}s \\
(2c_0)^{-1}s & (2c_0)^{-1}(-1)^{n+1}u
\endpmatrix
+O\(N^{-2/3}\).
\eqno (4.63)
$$
This implies that 
$$
V_0(s)={1\over \sqrt{\det W_0(s)}}W_0(s)
={1\over \sqrt{\det \widetilde W_0(s)}}\widetilde W_0(s)
$$
is analytic in $s$ as well and for finite $s$,
$$
V_0(s)=
\pmatrix
1 & 0\\
0 & 1
\endpmatrix
+N^{-1/3}
\pmatrix 
0 & (2c_0)^{-1}s \\
(2c_0)^{-1}s & 0
\endpmatrix
+O\(N^{-2/3}\).
\eqno (4.64)
$$
If we go back to the global coordinates $z$ and $\z_0$ we obtain that
$$
V(z)=V_0(C^{-1}N^{1/3}z)={1\over \sqrt{\det W(z)}}W(z),
\qquad W(z)=C^{-1}N^{-2/3}W_0(C^{-1}N^{1/3}z).
\eqno (4.65)
$$
To ensure that $V(z)$ is analytic in the region
$(-\Om_1)\cup\Om^0\cup\Om_1$ we will prove the following lemma. 

{\bf Lemma 4.2.} {\it Define
$$
\wt W(z)=C^{-1}N^{-2/3}\widetilde W_0(C^{-1}N^{1/3}z)
=\frac{W(z)}{8(C^{-1}N^{1/3}z)^2+2v}\,.
$$
Then there exists $d_2^0>0$ such that if $0<d_2\le 
d_2^0$ then for sufficiently large $N$ for all
$z\in(-\Om_1)\cup\Om^0\cup\Om_1$, $\det \wt W(z)\not=0.$}

Proof of Lemma 4.2 is given in Appendix E below.
Observe that for $|z|\le\ep_0$,
$$\eqalign{
W(z)&=
\pmatrix
-\wt a_{21}(z)+a_{12}^0(z) &
\wt a_{11}(z)-a_{11}^0(z) \\
\wt a_{11}(z)-a_{11}^0(z) &
\wt a_{12}(z)-a_{21}^0(z) 
\endpmatrix+ O(N^{-1}),\cr
\wt a_{ij}(z)&=\z_0'(z) N^{-2/3}a_{ij}\(N^{1/3}\z_0(z)\).}
\eqno (4.66)
$$
An important feature here is that the correcting terms $N^{-1/3}\a_{11}$,
$N^{-2/3}\a_{12}s$, etc., in (4.62) become $O(N^{-1})$ after multiplication by
$N^{-2/3}$ and substitution $s=C^{-1}N^{1/3}z$. From (3.3),
$$\eqalign{
&\wt a_{11}(z)=-\wt a_{22}(z)=N^{-1/3}(-1)^n4u\z_0'(z)\z_0(z),\cr
&\wt a_{12}(z)=\z_0'(z)\[ 4\z_0^2(z)+N^{-2/3}\((-1)^n2w+v\)\],\cr
&\wt a_{21}(z)=\z_0'(z)\[ -4\z_0^2(z)+N^{-2/3}\((-1)^n2w-v\)\].\cr}
\eqno (4.67)
$$

{\it First Order Approximation for $\z(z)$.} 
We will solve equation (4.9) with an error term of the order
of $N^{-2}$. Using notation (4.11) we write (4.9) as
$$
(\z')^2f(\z)=\det[A_n^0(z)-N^{-1}V'(z)V^{-1}(z)].
\eqno (4.68)
$$ 
As a suitable approximation to the determinant on the right
we will consider the function
$$
a(z)\equiv d(z)-N^{-1}\[a_{11}^0(z)q_{22}(z)+
a_{22}^0(z)q_{11}(z)-a_{12}^0(z)q_{21}(z)-
a_{21}^0(z)q_{12}(z)\],
\eqno (4.69)
$$
where $q_{ij}(z)$ are the matrix elements of the matrix
$$
Q(z)\equiv V'(z)V^{-1}(z),
\eqno (4.70)
$$
where $V(z)$ is defined in (4.65).
We replace (4.68) by the equation
$$
\z'\sqrt{f(\z)}=\sqrt{a(z)}.
\eqno (4.71)
$$ 
In the local coordinates
$(s,\sg)$ it is written as
$$
\sg'\sqrt{f_0(\sg)}=\sqrt{a_0(s)}.
\eqno (4.72)
$$ 
where
$$
f_0(\sg)\equiv N^{4/3}f\(N^{-1/3}\sg\)=
\det A(\sg)=16\sg^4+8y\sg^2+\[v^2(y)-4w^2(y)\]
\eqno (4.73) 
$$
and
$$
a_0(s)=C^2N^{4/3}a\(CN^{-1/3}s\).
\eqno (4.74)
$$
From (4.69),
$$
a_0(s)=d_0(s)-
\[d_{11}(s)q_{22}^0(s)+
d_{22}(s)q_{11}^0(s)-d_{12}(s)q_{21}^0(s)-
d_{21}(s)q_{12}^0(s)\],
\eqno (4.75)
$$
where
$$\eqalign{
d_0(s)&\equiv  C^2N^{4/3}d\(CN^{-1/3}s\)
=16s^4+8c_0^{-1}N^{2/3}\({n\over N}-\lacr\)s^2\cr
&+\[v^2(y)-4w^2(y)\]-N^{-1/3}c_0^{-1}(-1)^n2w(y) + O\(N^{-2/3}\)} 
\eqno (4.76)
$$
[use (4.5)], $d_{ij}(s)$ are the matrix elements of the matrix
$D_0(s)$ defined in (4.55), and $q_{ij}^0(s)$ are the matrix
elements of the matrix
$$
Q_0(s)\equiv CN^{2/3}Q\(CN^{-1/3}s\)=
V_0'(s)V^{-1}(s)
=N^{-1/3}\pmatrix
0 & (2c_0)^{-1} \\
(2c_0)^{-1} & 0
\endpmatrix+O\(N^{-2/3}\)
$$
[use (4.50)]. From (4.75) and (4.55),
$$\eqalign{
a_0(s)&
=d_0(s)+N^{-1/3}(2c_0)^{-1}\[d_{12}(s)+d_{21}(s)\]+O(N^{-2/3})\cr
&=d_0(s)+N^{-1/3}c_0^{-1}(-1)^n2w(y)
+O\(N^{-2/3}\),}
$$
hence by (4.76),
$$
a_0(s)=
16s^4+8c_0^{-1}N^{2/3}\({n\over N}-\lacr\)s^2
+\[v^2(y)-4w^2(y)\]+N^{-2/3}r_N(s),
\qquad r_N(s)=O(1). 
\eqno (4.77)
$$
Here the notation $r_N(s)=O(1)$ has the following meaning:
for any $R>0$ there exist $N_0=N_0(R)$ and $C=C(R)$ such that
for all $N\ge N_0$, the function $r_N(s)$ is analytic in the disk
$|s|\le R$ and $|r_N(s)|\le C(R)$ for all $s$ in the disk.

To get an analytic solution to (4.72) we have to secure the 
equality of periods of $f_0(s)$ and $a_0(s)$. The function $f_0(s)$
has four zeros $\pm s_j,\; j=1,2,$ (see Proposition 3.1). The function
$a_0(s)$ is an $O(N^{-2/3})$-perturbation of $f_0(s)$ and
$a_0(s)$ also has four zeros $\pm t_j,\; j=1,2,$ such that
$|t_j-s_j|\to 0$ as $N\to\infty$.
We have to secure the equality of the periods, 
$$
\int_0^{s_j}\sqrt{f_0(s)}\,ds=\int_0^{t_j} \sqrt{a_0(s)}\,ds,\quad
j=1,2. 
\eqno (4.78)
$$
To that end we need two parameters. As before we can use the
parameter $\a$ in (4.26) which slightly changes the
relation between $n$ and $y$. Where can we take the second 
parameter? The idea is to change $a_0(s)$ by a constant
$\b N^{-2/3}$ putting
$$
a_0(s)=16 s^4+8\(y-\a N^{-2/3}\) s^2+\[v^2(y)-4w^2(y)
\]+N^{-2/3}r_N(s)+\b N^{-2/3}.
\eqno (4.79)
$$
Observe that by (4.74),
$$
a(z)=C^{-2}N^{-4/3}a_0(C^{-1}N^{1/3}z),
$$
therefore,
when we will go back to the global coordinates, the function
$a(z)$ will change by $\b N^{-2}$ which is of the order of
the error term.  We find
the values of the parameters $\a$ and $\b$ from 
condition (4.78). Let us analyze this condition more carefully.

Assume that $y\not= y_0$, so that $s_2\not=0$.
Introduce the auxiliary functions
$$
I_j(x_1,x_2,x_3)=\int_0^{t_j}
\sqrt{ 16s^4+8ys^2+\[v^2(y)-4w^2(y)\]+x_1s^2+x_2+x_3r_N(s)}\,ds,
\eqno (4.80)
$$
where $t_j$ is the corresponding zero of the function under the
radical, $j=1,2$. Then (4.78) reads
$$
I_j\(-8\a N^{-2/3},\b N^{-2/3},N^{-2/3}\)=I_j(0,0,0),\quad j=1,2.
\eqno (4.81)
$$
Observe that the Jacobian
$$
J\equiv \left .\det
\pmatrix 
\di {\d I_1\over \d x_1} & \di {\d I_1\over \d x_2} \\   
\di {\d I_2\over \d x_1} & \di {\d I_2\over \d x_2} 
\endpmatrix\right|_{x_1=x_2=x_3=0}
$$
is not zero. Otherwise, this would mean
that there exists a number $c$ such that
the periods of the second kind abelian (elliptic in fact)
integral,
$$
I(s)=\int_{0}^{s}{{u^{2} + c }\over {\sqrt{ 16u^4+8yu^2+
\[v^2(y)-4w^2(y)\]}}}du,
$$ 
on the genus 1 algebraic curve $w = 16s^4+8ys^2+ \[v^2(y)-4w^2(y)\]$ 
are all zero. This in turn would mean that the sum,
$$
{s\over 4} + I(s),
$$
makes a meromorphic function
on the curve $(w, s)$ having a simple pole at exactly one point,
namely at one of the two points lying over $s =\infty$. This is imposible
since non of these points is a Weierstrass point of the elliptic curve
$(w,s)$ (see e.g. [FK]). (In fact using Riemann's bilinear relations
for the periods of abelian integrals, one can show that $J = {{\pi i}\over
128}$.)
Therefore, the implicit function theorem gives
 the solvability of equations (4.78) for $\a$ and $\b$.

If $y=y_0$, so that $s_{2} = 0$ and the curve $(w,s)$ degenerates, the
parameter
$\beta$ should be taken equal to $-r_{N}(0)$ to ensure that $t_{2}=0$ as well.
The parameter $\alpha$ is determined from equation (4.78), $j=1$,
which is the only period equation left. Its solvability follows from
the obvious inequality (we recall that $y_{0}>0$),
$$
\int_{0}^{s_{1}}{{s^{2} }\over {\sqrt{ 16s^4+8y_{0}s^2}}}du \neq 0,
\quad s_{1} = i\sqrt{{y_{0}}\over 2}.
$$
(One can also argue that since the Jacobian $J= {{\pi i}\over 128}$
and hence does not depends on $y$,
the solvability for $y\neq y_{0}$ implies as well 
the solvability for $y=y_{0}$.)

Let us summarize our calculations. The function $\z(z)$ is the unique
analytic solution  of the initial value problem
$$\eqalign{
\z'\sqrt{f(\z)}&=\sqrt{a(z)},
\qquad \z(0)=0;\cr
f(\z)&=
16\z^4(z)+8N^{-2/3}y\z^2(z)+N^{-4/3}\[v^2(y)-4w^2(y)\],\cr
a(z)&= d(z)-N^{-1}\[a_{11}^0(z)q_{22}(z)+
a_{22}^0(z)q_{11}(z)-a_{12}^0(z)q_{21}(z)\right.\cr
&\left.-
a_{21}^0(z)q_{12}(z)\]-\b N^{-2},\cr
y&=c_0^{-1}N^{2/3}\({n\over N}-\lacr\)+\a N^{-2/3},\cr}
\eqno (4.82)
$$
where $d(z)$ is defined in (4.5), the functions $a_{jk}^{0}(z)$
are defined in (4.51), the functions $q_{jk}(z)$ are the matrix entries
of the matrix $Q(z)$ defined in (4.70) 
and the constants $\a$ and $\b$ are uniquely determined by
equation of periods (4.78). The branches of square roots
in (4.82) are determined for large $N$
by the condition that they
are positive on the interval $[(z_0/4),(3z_0/4)]$. 
The domain for $\z(z)$ is the rectangle $(-\Om_{1})\cup\Om^0\cup\Om_1$.

By exactly the same arguments as in the case of initial 
problem (4.33) and taking into account
that the term in the brackets in (4.82) is
of the order of $N^{-2/3}$ [cf. equation (C.10) in
Appendix C], we derive from  (4.82) the estimate
$$
\z(z) = \z_{\infty}(z) +  N^{-2/3}y\z_{1}(z) + O(N^{-4/3}),
\quad z \in (-\Om_{1})\cup\Om^0\cup\Om_1,
\eqno (4.83)
$$
that is [cf. (4.48)]
$$
\z(z) = \z_{0}(z) + O\(N^{-4/3}\), \quad z \in  (-\Om_{1})\cup\Om^0\cup\Om_1.
\eqno (4.84)
$$
In addition, the following symmetry relations take
place: 
$$
\z(-z) = - \z(z),
\eqno (4.85)
$$
and
$$
V(-z) = \sigma_{3} V(z) \sigma_{3}.
\eqno (4.86)
$$
Equation (4.86) follows from (4.35)
and the explicit 
formulae (4.62), (4.65) defining $V(z)$. The uniqueness of a solution to 
initial value problem (4.82) implies equation (4.85).

{\it Remark .} Unlike the zeroth order change of variable
function $\z_{0}(z)$, the first-order function $\z(z)$ 
can not be replaced by the first two terms in the
right hand side of (4.83). As we will see in Section 5
(Proposition 5.2 and Theorem 5.4), in order to 
match the critical point solution
and the WKB solution within an $O(N^{-1})$ error, 
the approximation given by (4.83) is
not enough. It can be, however, used for a more explicit
but  less accurate approximation to the solution $\Psi_{n}(z)$ of
Riemann-Hilbert problem (1.35-38). We discuss this
matter in more detail in Appendix F below.

\beginsection 5. Matching CP and WKB Solutions \par

{\it WKB Solution.} The WKB solution is an approximate solution
to (4.1) in $\Om^c$ and it is defined as (cf. [BI])
$$
\Psi_{\WKB}(z)=\tilde C_0T(z)e^{-\[N\int_\infty^z\mu(u)\,du\sg_3
+\int_\infty^z\diag T^{-1}(u)T'(u)\,du+C_1\sg_3\]}
\eqno (5.1)
$$
where $\tilde C_0\not=0,\,C_1$ are some constants (parameters
of the solution),
$$
\mu(z)=\[-d(z)\]^{1/2},\qquad
T(z)=
\pmatrix
1 & {a_{12}^0(z)\over \mu-a_{11}^0(z)} \\
-{a_{21}^0(z)\over \mu-a_{11}^0(z)} & 1
\endpmatrix,
\eqno (5.2)
$$ 
and $\diag A$ means the diagonal part of the matrix $A$. 
Recall that $d(z)$ is a suitable approximation of $\det A_n^0(z)$
 [see (4.5)] and it is a polynomial of the sixth degree in $z$.
Observe that both the function $\mu_0(z)$ in ($4.34'$) and $\mu(z)$
in (5.2) are defined as $\[-d(z)\]^{1/2}$. The difference is in
their domains: for $\mu_0(z)$ the domain is 
$((-\Om_1)\cup\Om^0\cup\Om_1)\setminus
\{|z|\le\om_0\}$ while for $\mu(z)$ it is $\Om^c$. The function $\mu_0(z)$
is an analytic continuation of $\mu(z)$ along the contour which goes by
the positive half-axis from $\infty$ to $z_0+d_1$ and then around $z_0$
from above. For $z$ lying in $\[((-\Om_1)\cup\Om^0\cup\Om_1)\setminus
\{|z|\le\om_0\}\]\cap\Om^c$,  
$$
\mu(z) = \left\{
\eqalign{&\mu_{0}(z), \quad \Im z >0,\cr
& -\mu_{0}(z), \quad \Im z <0.}\right.
\eqno (5.2')
$$
The integral
$\int_\infty^z\mu(u)\,du$ in (5.1)
diverges at infinity and its regularization 
is defined as follows. We obtain from (4.5) that
$$\eqalign{
&\mu(z)=\mu_3 z^3+\mu_1 z+\mu_{-1} z^{-1}+\widetilde \mu(z),\quad
\widetilde\mu(z)=\sum_{j=0}^\infty \mu_{-3-2j} z^{-3-2j}\,,\cr
&\mu_3={g\over 2}\,,\quad \mu_1={t\over 2}\,,\quad
\mu_{-1}=-{n\over N}\,,}
$$
and we define
$$
\int_{\infty}^z \mu(u)\,du\equiv {\mu_3\over 4}z^4+{\mu_1\over 2}z^2
+\mu_{-1}\ln z+\int_\infty^z\widetilde\mu(u)\,du\,.
\eqno (5.3)
$$
Because of the logarithmic term,
this defines $\int_{\infty}^z \mu(u)\,du$ as a multivalued
function on
the complex plane but the function $e^{-N\int_{\infty}^z
\mu(u)\,du}$
is one-valued if $n$ is integer. 
  
The integral $\int_\infty^z\diag T^{-1}(u)T'(u)\,du$ in (5.1)
converges at infinity, because by (4.4,5), as $z\to\infty$,
$$
{a_{12}^0(z)\over \mu(z)-a_{11}^0(z)}=\frac{(R_n^0)^{1/2}}{z}+O(z^{-3}),\qquad
{a_{21}^0(z)\over \mu(z)-a_{11}^0(z)}=-\frac{(R_n^0)^{1/2}}{z}+O(z^{-3}).
$$
Formula (5.1) can be brought into the following form:
$$
\Psi_{\WKB}(z)=C_0T_0(z)e^{-\[N\xi(z)+\tau(z)+C_1\]\sg_3},
\eqno (5.4)
$$
where $C_0=\tilde C_0/\sqrt 2$ and
$$\eqalign{
\xi(z)&=\int_\infty^z\mu(u)\,du,\qquad 
\tau(z)=\int_\infty^z 
\frac{a^0_{12}(u){a^0_{21}}'(u)- {a_{12}^0}'(u)a_{21}^0(u)}
{4\mu(u)\[\mu(u)-a^0_{11}(u)\]}\,du,\cr
T_0(z)&=
\({\mu(z)-a_{11}^0(z)\over \mu(z)}\)^{1/2}T(z),\qquad
\det T_0(z)\equiv 2\,.}
\eqno (5.5)
$$
The square root branch in (5.5) is determined by the condition
that for large $z>0$,
$$
\({\mu(z)-a_{11}^0(z)\over \mu(z)}\)^{1/2}>0.
$$
The integral for $\tau(z)$ in (5.5) converges at
infinity, so no regularization is needed. 
The reduction of (5.1) to (5.4) is based on the formula
$$
\diag T^{-1}(z)T'(z)=\tau'(z)\sg_3
-{1\over 2}\[\ln\({\mu(z)-a^0_{11}(z)\over \mu(z)}\)\]'I
\eqno (5.5')
$$
which follows from (5.2) by a straightforward calculation 
(cf. Appendix A in [BI]).
The numbers $C_0\not=0,\,C_1$ in (5.4) are free constants which
will be chosen later.

{\bf Proposition 5.1.} {\it There exists $N_0=N_0(d_1,d_2)$ such that
for all $N\ge N_0$, (5.4,5) define $\Psi_{\WKB}(z)$ as an analytic
function in $\Om^c$. As $z\to\infty$,
$$
\Psi_{\WKB}(z)=\sqrt 2\,C_0\(I+z^{-1}(R_n^0)^{1/2}
\pmatrix 
0 & 1 \\
1 & 0
\endpmatrix +O(|z|^{-2})\)
e^{-\[N(V(z)/2)-n\ln z
+C_1\]\sg_3}.
\eqno (5.6)
$$
In addition, the symmetry relation,
$$
\Psi_{\WKB}(-z) = (-1)^{n} \sigma_{3}\Psi_{\WKB}(z)\sigma_{3},
\eqno (5.7)
$$
holds.}

{\it Proof.} From (5.3), (5.4)
$$
\Psi_{\WKB}(z)=C_0T_0(z)e^{-\[N(V(z)/2)-n\ln z
+N\wt\xi(z)+\tau(z)+C_1\]\sg_3},
\eqno (5.4')
$$
where 
$$
\wt\xi(z)=\int_\infty^z\wt\mu(u)\,du.
$$
Let us prove that for large $N$,
$$
\mu(z)-a_{11}^0(z)\not=0,\quad  z\in\Om^c.
\eqno (5.8)
$$
Indeed, by (4.4,5),
$$\eqalign{
\mu^2(z)&-\(a_{11}^0(z)\)^2=-d(z)-\(a_{11}^0(z)\)^2
={g^2z^4(z^2-z_0^2)\over 4}-{g^2z^6\over 4}\cr
&+O\(N^{-1/3}(1+|z|)^4\)
=-{g^2z^4z_0^2\over 4}+O\(N^{-1/3}(1+|z|)^4\)\not=0,\qquad z\in\Om^c,}
$$
hence (5.8) holds. In addition, as $z\to \infty$,
$$
{\mu(z)-a_{11}^0(z)\over \mu(z)} = 2 + O\(|z|^{-2}\).
\eqno (5.9)
$$
Therefore, the function $\({\mu(z)-a_{11}^0(z)\over
\mu(z)}\)^{1/2}$ is analytic in $\Om^c$. Since $\tilde{\mu}(z)$
is analytic in  $\Om^c$ and $\tilde{\mu}(z) 
= O(z^{-3})$ as $z\rightarrow \infty$ it follows that 
$\tilde{\xi}(z)$ is analytic in   $\Om^c$. Similar arguments
prove the analyticity of $\tau(z)$ and hence the analyticity 
of $\Psi_{\WKB}(z)$. Asymptotics (5.6) follows from (5.4$'$,9). 
 The equations
$$
\mu(-z) = -\mu(z), \quad a^{0}_{11}(-z) = -a^{0}_{11},
\quad a^{0}_{12}(-z) = a^{0}_{12}(z),
\quad
a^{0}_{21}(-z) = a^{0}_{21}(z),
$$
and
$$
\xi(-z) = \pm {n\over N}\pi i + \xi(z), \quad
\tau(-z) = \tau(z)
$$
imply symmetry (5.7). Proposition 5.1 is
proved.

{\it Critical Point Solution.} 
We define  the critical point solution in 
the region $\Om_c\equiv (-\Om_1)\cup\Om^0\cup\Om_1$ as
$$
\Psi_{\CP}(z)=
\left\{
\eqalign{
&\tilde{C}V(z)\Phi^u\(N^{1/3}\z(z)\),\qquad \Im z\ge 0,\cr 
&\tilde{C}V(z)\Phi^d\(N^{1/3}\z(z)\),\qquad \Im z\le 0,\cr}
\right. 
\eqno (5.10)
$$
where $\tilde{C}$ is a constant, a parameter of the solution. The functions
$\z(z)$ and 
$V(z)$ are defined by (4.68) and (4.51), respectively, and the model
solutions $\Phi^{u,d}(z)$ are defined and described in Section 3 [see
(3.21)]. It is important to notice
that equations (4.85), (4.86), and (3.22) yield 
$$
\Psi_{\CP}(-z) = (-1)^{n} \sigma_{3}\Psi_{\CP}(z)\sigma_{3},
\eqno (5.11)
$$
i.e. the same symmetry relation as for the function 
$\Psi_{\WKB}(z)$.

{\it Matching CP and WKB Solutions.} Let $\G_c^{\pm}$ be 
the horizontal sides of the rectangle $\Om_c$,
$$
\G_c^{\pm}=\{ z\in\Om_c\,:\; \Im z =\pm d_2\}.
$$
Our goal is to show that 
we can choose constants $C_0\not=0$ and $C_1$ in (5.1) and $
\tilde C\not=0$
in (5.10) such that the
CP solution $\Psi_{\CP}(z)$ coincides, up to terms of the order
of $N^{-1}$, with the WKB solution $\Psi_{\WKB}(z)$ on $\G_c^{\pm}$. 
Because of symmetries
(5.7,11),  it is enough to consider $\G_c^+$.
Replacing in (5.10) the model function $\Phi^u(z)$ by its
asymptotics (3.23), we obtain that
$$\eqalign{
\Psi_{\CP}(z)&=\tilde{C}V(z)
\pmatrix
1 & -i \\
-i & 1
\endpmatrix
Y_0(z)
e^{-i\(N(4/3)\z^3(z)+N^{1/3}y\z(z) - i\gamma\)\sg_3},\quad z\in \G_c^+,}
\eqno (5.12)
$$
where 
$$
Y_0(z)=I + {N^{-1/3}m_{1}\over{\z(z)}} + {N^{-2/3}m_{2}\over{\z^{2}(z)}}
+O\(N^{-1}\) .
\eqno (5.13)
$$
To transform $\Psi_{\CP}(z)$ to the WKB solution we use the following
proposition. Denote, as before, by $z_{0}^N$ the zero of 
the potential $U^{0}(z)$ (see(1.68)) that approaches
$z_0$ as $N\to\infty$. 
 
{\bf Proposition 5.2} {\it For $z\in \G^+_c$,
$$\eqalign{
i\(N(4/3)\z^3(z)+N^{1/3}y\z(z)+ N^{-1/3}{{D}\over 2}\z^{-1}(z)\)
&=N\int_{\infty}^z\mu(u)\,du+\tau(z) +C^0\cr
&+ {{i\pi n}\over 2}
+N^{-2/3}\De(z) +O(N^{-1}),}
\eqno (5.14) 
$$
where $D = D(y)$ is given in (3.6), 
$$
C^0=N\int_{z_0^N}^\infty \mu^c(u)\,du-\frac{1}{4}\ln R_n^0,
\eqno (5.15)
$$
$\mu^c(z)=\sqrt {U^0(z)}$ [see (1.68)], and
$$\eqalign{
\Delta(z)& \equiv {c^0\over
{ia^{0}_{11}(z)-a^{0}_{21}(z) - i\mu(z)}}
- 2i(-1)^{n}w\z'(z){a^{0}_{11}(z)\over b(z)},\cr
c^0&=\(\frac{2|t|}{g}\)^{1/6}(-1)^nw,\qquad
b(z) \equiv -2\mu^{2}(z) -i\mu(z)\(a^{0}_{12}(z) - a^{0}_{21}(z)\).} 
\eqno (5.15')
$$
In addition,
$$
V(z)
\pmatrix
1 & -i \\
-i & 1
\endpmatrix
=T_0(z)V_0(z)
\eqno (5.16)
$$
where  
$$\eqalign{
V_{0}(z) &
= I + {N^{-1/3}n_{1}\over{\z(z)}} + {N^{-2/3}n_{2}(z)\over{\z^{2}(z)}}
+O\(N^{-1}\);\cr
n_1&=(-1)^n{u\over 2}\sg_{1},\quad
n_2(z)={u^2\over 8}I
+ (-1)^{n+1}{w\over 4}\sg_{2}
+ \Delta(z)\z^2(z)\sg_{3}.}
\eqno (5.16')
$$}

Proof of Proposition 5.2 is given in Appendix C below. Applying
this proposition to (5.12) we obtain that
$$
\Psi_{\CP}(z)=\tilde{C}T_0(z)Y_1(z)e^{-[N\xi(z)+\tau(z)+C^0]\sg_3},
\eqno (5.17)
$$
where
$$
Y_1(z)=V_0(z)Y_0(z)e^{\( N^{-1/3}{{iD}\over{2\z(z)}} 
-N^{-2/3}\De(z) + O(N^{-1})\)\sg_3}.
\eqno (5.18)
$$

{\bf Proposition 5.3.} $Y_1(z)=I+O(N^{-1})$ for $z\in \G^+_c$.

{\it Proof.} From (5.18,$16'$,13) we obtain that
$$\eqalign{
Y_1(z)&=\[I + {N^{-1/3}n_{1}\over{\z(z)}} +
{N^{-2/3}n_{2}\over{\z^{2}(z)}}\]
\[I + {N^{-1/3}m_{1}\over{\z(z)}} + {N^{-2/3}m_{2}\over{\z^{2}(z)}}
 \]\cr
&\times\[ I + N^{-1/3}{{iD}\over{2\z(z)}}\sg_3 
- N^{-2/3}\({D^{2}\over {8\z^{2}(z)}}I + \De(z)\sg_3\)\]
+O\(N^{-1}\)\cr
&=I+\frac{N^{-1/3}Y_1^{(1)}}{\z(z)}+\frac{N^{-2/3}Y_1^{(2)}(z)}{\z^2(z)}
+O\(N^{-1}\),
}
$$
where
$$\eqalign{
Y_1^{(1)}&=n_1+m_1+\frac{iD}{2}\sg_3,\cr
Y_1^{(2)}(z)&=n_2(z)+m_2(z)+n_1m_1-\frac{D^2}{8}I
-\Delta(z)\z^2(z)\sg_3+\frac{(n_1+m_1)iD}{2}\sg_3.}
$$
By formulae (3.24) and (5.$16'$),
$$\eqalign{
n_1&={(-1)^nu\over 2}\sg_{1},\quad
m_1=-{iD\over 2}\sg_3-{(-1)^nu\over 2}\sg_1,
\quad D=w^2-u^4-yu^2,\cr
n_2(z)&={u^2\over 8}I
- (-1)^n{w\over 4}\sg_{2}
+ \Delta(z)\z^2(z)\sg_{3},\quad
m_2={u^2-D^2\over 8}\,I +(-1)^n{w+uD\over 4}\,\sg_2,\cr
n_1m_1&=-\frac{(-1)^nuD}{4}\sg_2-\frac{u^2}{4}I,\quad
\frac{(n_1+m_1)iD}{2}\sg_3=\frac{D^2}{4}I.}
$$
This gives $Y_1^{(1)}=0$, $Y_1^{(2)}(z)=0$. 
Proposition 5.3 is proved. 

Now we can formulate the main result of this section about the
match of $\Psi^u_{\CP}(z)$ with $\Psi_{\WKB}(z)$ on $\G^+_c$.

{\bf Theorem 5.4.}  {\it If we take 
$$
C_0=\tilde C;\qquad C_1=
 N\int_{z_{0}^{N}}^\infty\sqrt{U^{0}(u)}du-\frac{1}{4}\ln R_n^0,
\eqno (5.19)
$$
where the potential function $U^{0}(z)$ is defined in (1.68), then
$$
\Psi_{\CP}(z)=\(I+O(N^{-1})\)\Psi_{\WKB}(z)
\eqno (5.20)
$$
uniformly with respect to $z\in \G^+_c$.}

{\it Proof.} From (5.17) and Proposition 5.3,
$$
\Psi_{\CP}(z)=\(I+O(N^{-1})\)\tilde{C}T_0(z)Y_1(z)
e^{-[N\xi(z)+\tau(z)+C^0]\sg_3}.
\eqno (5.21)
$$
Therefore, if we take the constants $C_0$ and $C_1$ in formula (5.4)
 as in (5.19), then equation (5.20) follows.
Theorem 5.4 is proved.

{\it Remark.} Since the parameters $d_{1}$, $d_{2}$ of the
rectangle $\Omega$ can vary, estimate (5.20) is valied 
uniformly in
an $\ep$-neighborhood of $\Gamma^{+}_{c}$.

\beginsection 6. Matching TP and WKB Solutions \par

{\it Turning Point Solution.} Let $\Om_0=\Om_1\cup\Om^1$ (see Fig.2) and
let $\Om_0^{u,d}$ be the upper and lower halves
of $\Om_0$,
$$
\Om_0^{u,d}=\{ z\in\Om_0\,:\; \pm \Im z\ge 0\}.
$$
The turning point solution $\Psi_{\TP}(z)$ is defined in $\Om_0$ 
as
$$
\Psi_{\TP}(z)=
\left\{
\eqalign{
&\tilde{C}_{1}W(z)Y_u(w(z)),\qquad z\in\Om_0^u,\cr
&\tilde{C}_{1}W(z)Y_d(w(z)),\qquad z\in\Om_0^d.\cr}
\right.
\eqno (6.1)
$$
The constant $\tilde C_1\not=0$ is a parameter of the solution.
The function of change of variable
$w(z)$ and the gauge matrix $W(z)$  have been defined by 
formulae (1.78,79) above.
The matrix valued functions
 $Y_{u,d}(z)$ are model solutions defined in terms of
the Airy function,
$$
Y_{u,d}(z)=
\pmatrix
N^{1/6} & 0 \\
0 & N^{-1/6}
\endpmatrix
\pmatrix
y_0(N^{2/3}z) & y_{1,2}(N^{2/3}z) \\
y'_0(N^{2/3}z) & y'_{1,2}(N^{2/3}z) 
\endpmatrix,
\eqno (6.2)
$$
where
$$\eqalign{
y_0(z)&=\Ai(z),\cr
y_1(z)&=e^{-\pi i/6}\Ai\( e^{-2\pi i/3}z\),\cr
y_2(z)&=e^{\pi i/6}\Ai\( e^{2\pi i/3}z\).\cr}
\eqno (6.3)
$$
Let us remind that $\Ai(z)$ is a solution
to the Airy equation $y''=zy$, which has the following asymptotics as
$z\to\infty$:
$$
\Ai(z)= {1\over 2\sqrt\pi z^{1/4}}
\exp\(-{2z^{3/2}\over 3}+O(|z|^{-3/2})\),\quad 
-\pi+\ep\le \arg z\le\pi-\ep.
\eqno (6.4)
$$
The functions $y_j(z)$ 
satisfy the relation
$$
y_1(z)-y_2(z)=-iy_0(z).
\eqno (6.5)
$$
Let us explain (6.1).

The turning point solution can be obtained by solving the
Schr\"odinger equation [see (1.50)],
$$
-\eta''+N^2U^0\eta=0,
\eqno (6.6)
$$
near the turning point. Namely, we are looking 
for solutions in the form
$$
\eta(z)={\tilde{C}_{1}N^{1/6}\over \sqrt {w'(z)}} y_j\(N^{2/3}w(z)\),\quad
j=0,1,2,
\eqno (6.7)
$$
(cf. [Ble1]).
Equation (6.6) reduces then to the following
equation on $w(z)$:
$$
\(w'(z)\)^2w(z)=U^0(z)+{1\over 2N^2}\{ w,z\}
\eqno (6.8)
$$
where $\{ w,z\}$ is the Schwarzian,
$$
\{w,z\}={w'''(z)\over w'(z)}-{3\over
2}\({w''(z)\over w'(z)}\)^2.
\eqno (6.9)
$$   
Dropping the Schwarzian term we get
$$
\(w'(z)\)^2w(z)=U^0(z),
\eqno (6.10)
$$
or taking the square root,
$$
\left(\frac{2}{3}w^{3/2}(z)\right)'=\sqrt{U^0(z)}
\eqno (6.11)
$$
To secure the analyticity of $w(z)$ at $z=z_0^N$ we take a solution
as
$$
w(z)=\({3\over 2}\int_{z_0^N}^z \sqrt {U^0(u)}\,du\)^{2/3}. 
\eqno (6.12)
$$
The function $U^0(z)$ is positive for $z>z_0^N$ and we take the
positive branch for $\sqrt {U^0(u)}$ for $u>z_0^N$ and also
the positive branch for the power 2/3. 
Formula (6.12) defines $w(z)$
as an analytic function in $\Om_0$ for large $N$. Observe that
because the dropped Schwarzian  term in (6.8) is of the order of $N^{-2}$,
(6.7,12) solves (6.6) with an error $O(N^{-2})$.

To obtain the gauge matrix $W(z)$, recall
that equation (6.6) was derived from the system
$$
\vec\Psi'=NA_n^0\vec\Psi,\qquad \vec\Psi=\pmatrix
\psi_1 \\ \psi_2
\endpmatrix
$$
by solving $\psi_2$ in terms of $\psi_1$,
$$
\psi_2=\frac{1}{a_{12}^0}\,\(N^{-1}\psi_1'-a_{11}\psi_1\),
$$
and by 
substitution $\psi_1=\sqrt{a_{12}^0}\,\eta$ [cf. (1.49)].
 For $\psi_1$ and $\psi_2$
we obtain from (6.7) the following 
formulae:
$$\eqalign{
\psi_1(z)&=\tilde{C}_{1}N^{1/6}\({a_{12}^0(z)\over w'(z)}\)^{1/2}
y_j\(N^{2/3}w(z)\),\quad j=0,1,2;\cr
\psi_2(z)&=\tilde{C}_{1}\({a_{12}^0(z)\over w'(z)}\)^{1/2}\[N^{-1/6}
{w'(z)\over a_{12}^0(z)}y'_j\(N^{2/3}w(z)\)
-N^{1/6}{a_{11}^0(z)\over a_{12}^0(z)}y_j\(N^{2/3}w(z)\)\],\cr}
\eqno (6.13)
$$
(we omit the term of the order of $N^{-5/6}$ in $\psi_2$),
or in the vector form,
$$
\vec\Psi(z)
=\tilde{C}_{1}W(z)\pmatrix
N^{1/6} & 0 \\
0 & N^{-1/6}
\endpmatrix
\pmatrix
y_j\(N^{2/3}w(z)\) \\
y'_j\(N^{2/3}w(z)\)
\endpmatrix,
\eqno (6.14)
$$
where
$$
W(z)=\({a_{12}^0(z)\over w'(z)}\)^{1/2}
\pmatrix
1 & 0 \\
-{a_{11}^0(z)\over a_{12}^0(z)} & {w'(z)\over a_{12}^0(z)}
\endpmatrix.
\eqno (6.15)
$$
This gives the gauge matrix $W(z)$.
 
It is important to notice that the turning point solution $\Psi_{\TP}(z)$
has the ``right'' jump on the real axis:
$$
\Psi_{\TP}^+(z)=\Psi^-_{\TP}(z)
\pmatrix
1 & -i \\
0 & 1
\endpmatrix,\qquad z\in \Om_0\cap\{\Im z=0\},
\eqno (6.16)
$$
which follows from equation (6.5).
We want to check that if we choose appropriately 
the constants $C_0$ and $C_1$ in (5.1), then the turning point solution
$\Psi_{\TP}(z)$ matches the WKB-solution $\Psi_{\WKB}(z)$
on the boundary of $\Om_0$, excluding a neighborhood of the 
interval $[0,z_0]$.
Let $\G_0^r$ be the union of the three sides of the square
$\Om_0$, excluding the one crossing $[0,z_0]$,
$$
\G_0^r=\{ z\in\Om_0\,:\; |\Im z_0|=d_2\;\;\text{\rm or}\;\;
 \Re z=z_0+d_1\}.
\eqno (6.17)
$$

{\bf Lemma 6.1.} {\it If we take the WKB solution with the constants
$$
C_0={\tilde{C}_{1}\over 2\pi^{1/2}};\qquad 
C_1=N\int_{z^N_0}^\infty \sqrt{U^0(u)}\,du-\frac{1}{4}\ln R_n^0,
\eqno (6.18)
$$
then 
$$
\Psi_{\TP}(z)=\(I+O(N^{-1})\)\Psi_{\WKB}(z),\qquad z\in\G_0^r,
\eqno (6.19)
$$
uniformly with respect to $z\in \G_0^r$.}

A proof of Lemma 6.1 will be given below in Appendix D.
It will be based on an alternative form of the WKB solution
in $\Om^c$.

{\it Alternative Form of WKB solution.} The standard WKB
form for a solution of the Schr\"odinger equation (6.6) is
$$
\eta(z)={C_0\over \(\kappa'(z)\)^{1/2}}\,e^{\pm N\kappa(z)}.
\eqno (6.20)
$$
Equation (6.6) reduces then to the following equation on $\kappa(z)$:
$$
\(\kappa'(z)\)^2=U^0(z)+{1\over 2N^2}\{\kappa,z\},
\eqno (6.21)
$$
where $\{\kappa,z\}$ is the Schwarzian derivative.
Dropping the term with the Schwarzian derivative we obtain 
$$
\kappa'(z)=\( U^0(z)\)^{1/2},
\eqno (6.22)
$$
which gives 
$$
\eta(z)={C_0\over \(U^0(z)\)^{1/4}}\,
e^{\pm\(N\int_{z^N_0}^z \(U^0(u)\)^{1/2}du+\hat C_1\)},
\eqno (6.23)
$$
and the WKB solution in the form
$$
\hat\Psi_{\WKB}(z)=C_0T^c(z)E(N\xi^c(z)),
\eqno (6.24)
$$
where the gauge matrix $T^c(z)$ is defined in (1.70),
$$
\xi^c(z)=\int_{z^N_0}^z \mu^c(u)\,du+\hat C_1,
\qquad \mu^c(z)=\(U^0(z)\)^{1/2},
\eqno (6.25)
$$
and the model solution $E(z)$ is
$$
E(z)=
\pmatrix 
e^{-z} & e^z \\
-e^{-z} & e^z
\endpmatrix.
\eqno (6.26)
$$ 
Lemma 6.1 is an obvious corollary of the
following result which will be proved in Appendix D below.

{\bf Lemma 6.2.} {\it If we take $C_0,C_1$ as in Lemma 6.1
and $\hat C_1=0$ then for $z\in\Om^c$,
$$
\Psi_{\WKB}(z)=\(1+O(N^{-1}|z|^{-1})\)\hat \Psi_{\WKB}(z),
\qquad z\in \Om^c,
\eqno (6.27)
$$
and for $z\in \G_0^r$,
$$
\Psi_{\TP}(z)=\(I+O(N^{-1})\)\hat\Psi_{\WKB}(z),\qquad z\in\G_0^r.
\eqno (6.28)
$$}

It is worth noticing that the derivation of (6.28)
will be based on WKB type asymptotics for the Airy
function. In the vector form the latter is formulated
as follows. Let
$$
\vec {\Ai}(z)=
\pmatrix
\Ai(z) \\
\Ai'(z)
\endpmatrix,\qquad \vec E(z)=
\pmatrix 
e^{-z} \\
-e^{-z}
\endpmatrix,\qquad \vec C(z)=\pmatrix 
\cos z \\
-\sin z
\endpmatrix.
\eqno (6.29)
$$ 

{\bf Proposition 6.3.} {\it For any $\ep>0$, as $z\to\infty$,
$$
\vec{\Ai}(z)=\(1+O(|z|^{-3/2})\)\frac{1}{2\sqrt\pi}B(z)
\vec E\(\frac{2z^{3/2}}{3}\),
\qquad -\pi+\ep\le\arg z\le \pi-\ep,
\eqno (6.30)
$$
where
$$
B(z)=
\pmatrix
z^{-1/4} & 0 \\
0 & z^{1/4}
\endpmatrix,
\eqno (6.31)
$$
and
$$
\vec{\Ai}(-z)=\(1+O(|z|^{-3/2})\)\frac{1}{\sqrt\pi}B(z)
\vec C\(\frac{2z^{3/2}}{3}\,-\frac{\pi}{4}\),
\qquad -\frac{2\pi}{3}+\ep\le\arg z\le \frac{2\pi}{3}-\ep.
\eqno (6.32)
$$}

The notation $O(|z|^{-3/2})$ in (6.31,32) means a $2\times 2$
matrix valued function $r(z)$ such
that $|r(z)|\le C_0|z|^{-3/2}$ for $|z|\ge C_1$,
where $C_0,C_1>0$ are some constants. The branches
for fractional powers are fixed by the condition
that they are positive on the positive half-axis.
We will not prove Proposition 6.3 because it is
just a reformulation of well-known asymptotics
for the Airy function.

\beginsection 7. Matching CP and TP Solutions \par

In this section we will define a WKB solution $\Psi_{\WKB}(z)$
in the region $\Om_1$ and we will show that it matches both $\Psi_{\CP}(z)$
and $\Psi_{\TP}(z)$. We begin with introduction of
an auxiliary CP solution. Let $\Om_c^r$ be
the part of $\Om_c=\Om^0\cup\Om_1$ between the diagonals of the first and 
 fourth quadrants,
$$
\Om_c^r=\{ z\in\Om_c\,:\; |\Im z|\le \Re z\}.
$$
We define the auxiliary CP solution in $\Om_c^r$ as
$$
\Psi_{\CP}^a(z)= \tilde C V(z)\(\vec \Phi_2(N^{1/3}\z(z)),
\vec\Phi_1(N^{1/3}\z(z))\)
\eqno (7.1)
$$ 
where $\vec\Phi_1(z)$ and $\vec\Phi_2(z)$ are the Painlev\'e II
psi-functions defined in Proposition 3.2.  Using (5.10) and (3.21) we
can relate $\Psi_{\CP}^a(z)$ to $\Psi_{\CP}(z)$ as
$$
\Psi_{\CP}(z)=\Psi_{\CP}^a(z)S_{1,2},\qquad \pm \Im z\ge 0,
\eqno (7.2)
$$
where
$$
S_1=
\pmatrix
-i & 0 \\
i & 1
\endpmatrix ,
\qquad
S_2=
\pmatrix
-i & 1 \\
i & 0
\endpmatrix .
\eqno (7.3)
$$
Observe that $\Psi_{\CP}^a(z)$ is defined in terms of
 $\vec\Phi_1(z)$
and $\vec\Phi_2(z)$ for which we know from (3.18) their asymptotics
in the sector 
$$-(\pi/3)+\ep\le \arg z \le (\pi/3)-\ep.
$$
This allows us to find asymptotics
of $\Psi_{\CP}^a(z)$ in $\Om_1$, assuming that $d_2$ is small enough.
 Namely, $\z'(0)\ge \ep>0$ and if $d_2$ is small enough then
$z\in\Om_1$ implies that 
$$-(\pi/3)+\ep\le \arg \z(z) \le (\pi/3)-\ep.$$
Using the asymptotics of $\Phi_{1,2}(z)$ we obtain the
following result.

{\bf Proposition 7.1}. {\it If we take $C_0$ and $C_1$ as in
(5.19) then uniformly with respect to $z\in \Om_1$,
$$
\Psi_{\CP}^a(z)=\(I+O(N^{-1})\)\Psi^0_{\WKB}(z)S_0,\qquad
S_0=
\pmatrix
i & 0 \\
0 & 1
\endpmatrix,
\eqno (7.4)
$$
where the function $\Psi^0_{\WKB}(z)$ is an analytic
continuation of the function $\Psi_{\WKB}(z)$ in (5.4) 
from the upper half-plane to $\Om_1$.} 

Proof of Proposition 7.1 repeats the one of Theorem 5.4
and we omit it (take into account that formula (5.14)
is extended to $\Om_1$, see Appendix C).
 Formula (7.4) is easy to check without
any calculations if $\Im z\ge\ep>0$, $z\in\Om_1$. In this case
$\vec\Phi_1(N^{1/3}\z(z))$ is exponentially small,
while $\vec\Phi_2(N^{1/3}\z(z))$ is exponentially big
(cf. Proposition 3.2),
hence in (7.1) the function
$$
\vec\Phi_2(N^{1/3}\z(z))=\vec\Phi_1(N^{1/3}\z(z))
+i\vec\Phi(N^{1/3}\z(z))
$$
can be replaced by $i\vec\Phi(N^{1/3}\z(z))$ with an
exponentially small error. This gives $\Psi_{\CP}(z)S_0$
which can be further replaced by $\Psi_{\WKB}(z)S_0$, with an $O(N^{-1})$
error,  due to Lemma 6.1. Thus, we obtain (7.4).
  
We will call the function on the right in (7.4) the auxiliary
WKB solution,
$$
\Psi_{\WKB}^a(z)=\Psi^0_{\WKB}(z)S_0,
\eqno (7.5)
$$
so that
$$
\Psi_{\CP}^a(z)=(I+O(N^{-1}))\Psi_{\WKB}^a(z),\qquad z\in \Om_1.
$$ 
With the help of $\Psi^a_{\WKB}(z)$ we introduce the WKB solution 
in $\Om_1$ by pattern (7.2),
$$
\Psi_{\WKB}(z)=\Psi_{\WKB}^a(z)S_{1,2},\qquad \pm \Im z\ge 0.
\eqno (7.6)
$$
It shares the following nice properties:

{\bf Proposition 7.2.} {\it Uniformly with respect to $z\in\Om_1$,
$$
\Psi_{\CP}(z)=(1+O(N^{-1}))\Psi_{\WKB}(z).
\eqno (7.7)
$$
On the real axis,
$$
\Psi_{\WKB}^+(z)=\Psi_{\WKB}^-(z)
\pmatrix
1 & -i \\
0 & 1 
\endpmatrix,\qquad d_1\le z\le z_0-d_1,
\eqno (7.8)
$$
Finally, on the horizontal sides of $\Om_1$,
$$
\Psi_{\WKB}^+(z)=(1+O(N^{-1}))\Psi_{\WKB}^-(z), \qquad
z\in\G^+_1\cup\G^-_1,
$$
where
$$
\G_1^{\pm}=\{z\in\Om_1\,:\; \Im z=\pm d_2\}.
\eqno (7.9)
$$}

{\it Proof.} Equation (7.7) follows from Proposition 7.1; 
(7.8) from (7.6); and (7.9) from (7.7) and Theorem 5.4. 
 
We introduce next the auxiliary TP solution by the same pattern
as the CP one,
$$
\Psi_{\TP}^a(z)=\tilde C_1W(z)Y_a(w(z)),\qquad z\in\Om_0^l,
$$
where
$$
Y_a(z)=
\pmatrix
N^{1/6} & 0 \\
0 & N^{-1/6}
\endpmatrix
\pmatrix
y_2(N^{2/3}z) & y_1(N^{2/3}z) \\
y'_2(N^{2/3}z) & y'_1(N^{2/3}z) 
\endpmatrix.
\eqno (7.10)
$$ 
The function $\Psi_{\TP}^a(z)$ is defined in the domain
$$
\Om_0^l=\{ z=\Om_0\,:\; |\Im z|\le z_0-\Re z\},
$$
between two $45^{\circ}$-lines through $z_0$ to the left. 
Using (6.1,2,5) we can relate $\Psi_{\TP}^a(z)$
to $  \Psi_{\TP}(z)$ as
$$
\Psi_{\TP}(z)=\Psi_{\TP}^a(z)S_{1,2},\qquad \pm \Im z\ge 0.
\eqno (7.11)
$$
By (6.3,4) we know the asymptotics
of $y_{1,2}(z)$ in the sector $|\arg z-\pi|\le 2\pi/3-\ep$
and from this asymptotics we obtain, like in Lemma 6.1,
 that $\Psi_{\TP}^a(z)$ matches
$\Psi_{\WKB}^a(z)$ in $\Om_1$,
$$
\Psi_{\TP}^a(z)=(I+O(N^{-1}))\Psi_{\WKB}^a(z),\qquad z\in \Om_1.
$$ 
This, together with (7.11), proves the following addition to
Proposition 7.2:

{\bf Addition to Proposition 7.2.} {\it Uniformly with respect to $z\in\Om_1$,
$$
\Psi_{\TP}(z)=(1+O(N^{-1}))\Psi_{\WKB}(z).
\eqno (7.12)
$$}

\beginsection 8. Proof of the Main Theorem \par

Define the matrix-valued function $\Psi^0_n(z)$ on the
complex plane by the formulae:
$$
\Psi^0_n(z)=\left\{
\matrix
& \Psi_{\WKB}(z), & z\in\Om^c\cup\Om_1;\\
& (-1)^n\sg_3\Psi_{\WKB}(-z)\sg_3, & z\in(-\Om_1);\\
& \Psi_{\TP}(z), & z\in\Om^{1};\\
& (-1)^n\sg_3\Psi_{\TP}(-z)\sg_3, & z\in\(-\Om^{1}\);\\
& \Psi_{\CP}(z), & z\in\Om^{0}.
\endmatrix
\right.
\eqno (8.1)
$$
Let
$$
X_n(z)=\Psi_n(z)[\Psi_n^0(z)]^{-1}.
\eqno (8.2)
$$
From (1.37) and (5.6) we obtain that as $z\to\infty$,
$X_n(z)$  admits the asymptotic expansion
$$
X_n(z)\sim\sum_{j=0}^\infty \frac{\Theta_j}{z^j},
\eqno (8.3)
$$
with
$$\eqalign{
&\Theta_0={1\over{\sqrt{2}}}C_0^{-1}\G_0e^{(C_1-\la_n)\sg_3},\cr
&\Theta_1={1\over{\sqrt{2}}}C_0^{-1}\[\G_1 e^{(C_1-\la_n)\sg_3}
-\G_0e^{(C_1-\la_n)\sg_3}(R_n^0)^{1/2}\sg_1\].}
\eqno (8.4)
$$
In particular,
$$
\lim_{z\to\infty} X_n(z)=\Theta_0.
\eqno (8.5)
$$
From (8.1) we obtain that $\Psi_n^0(z)$ satisfies the following 
parity equation:
$$
\Psi_n^0(-z)=(-1)^n\sg_3\Psi_n^0(z)\sg_3,
\eqno (8.6)
$$
the same as $\Psi_n(z)$ [cf. (1.42)]. Hence
$$
X_n(-z)=\sg_3 X_n(z)\sg_3.
\eqno (8.7)
$$
The function $X_n(z)$ has multiplicative jumps on a number of contours,
because of the piecewise definition (8.1) and because of the jump 
of $\Psi_n(z)$ 
on the real axis. Let us discuss this question more carefully.
Let $\g_0$ be the union of boundaries of the regions
$\Om$, $\Om_1$, and $-\Om_1$,
$$ 
\g_0=\partial\Om\cup\partial \Om_1\cup(-\partial\Om_1)
$$
Then by Proposition 7.2,
$$
[X_{n-}(z)]^{-1}X_{n+}(z)=[\Psi_{n-}^0(z)]^{-1}
\Psi_{n+}^0(z)=I+O(N^{-1}),\qquad z\in\g_0.
$$
The function $X_n(z)$ has no jump on the segment
$[-z_0-d_1,z_0+d_1]$ because here the jumps of
 $\Psi_n^0(z)$  and $\Psi_n(z)$ cancel out each other.
Consider also the set
$$
\g_1=\{z\le -z_0-d_1\}\cup\{z\ge z_0+d_1\}.
$$
and let $\g=\g_0\cup\g_1$ (see Fig.4).

\centinsert{\pscaption{\psboxto (6in;1.5in){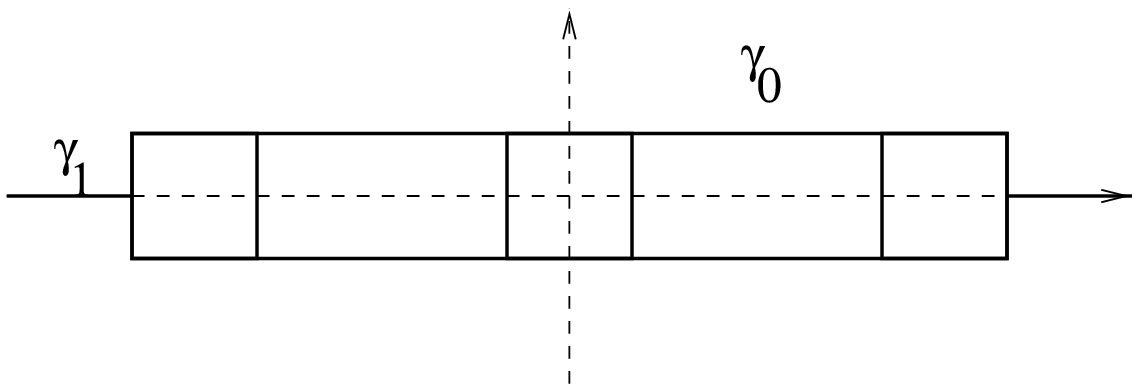}}
{\srm Fig 4:  The contour $\g=\g_0\cup\g_1$.   }}

\vskip 3mm

\noindent
On $\g_1$,
$$   
[X_{n-}(z)]^{-1}X_{n+}(z)=[\Psi_{n}^0(z)]^{-1}
\pmatrix
1 & -i \\
0 & 1
\endpmatrix
\Psi_{n}^0(z).
$$
Observe that if $z\ge z_0+d_1$ then 
$$
\Psi_n^0(z)=\Psi_{\WKB}(z)=\hat T(z)e^{-N\xi^c(z)\sg_3}
$$
where the matrix $\hat T(z)$ is uniformly bounded, hence
$$
[X_{n-}(z)]^{-1}X_{n+}(z)=[\hat T(z)]^{-1}
\pmatrix
1 & -ie^{-2N\xi^c(z)} \\
0 & 1
\endpmatrix
\hat T(z)=I+O\(e^{-aN|z|}\).
$$   
with some $a>0$.
Thus, we obtain the following proposition.

{\bf Proposition 8.1.} {\it The function $X_n(z)$ is an analytic
matrix-valued function on the complex plane with multiplicative
jumps on the contours $\g_0$ and $\g_1$
such that
$$
[X_{n-}(z)]^{-1}X_{n+}(z)=\left\{
\eqalign{
& I+O(N^{-1}),\quad z\in \g_0,\cr
& I+O\(e^{-aN|z|}\),\quad z\in\g_1.}\right.
\eqno (8.8)
$$
At infinity $X_n(z)$  admits asymptotic
expansion  (8.3).}

Proposition 8.1 implies that the fucntion $X_{n}(z)$
solves the Rimann-Hilbert problem on the contour $\gamma=
\gamma_{0}\cup \gamma_{1}$,
$$
X_{n}(\infty) \equiv \lim_{z\to \infty}X_{n}(z)
= \Theta_{0}
\eqno (8.9)
$$
$$
X_{n+}(z)=X_{n-}(z)G(z), \quad z \in \gamma,
\eqno (8.10)
$$
with the jump matrix given by the equations,
$$
G(z) = [\Psi_{n-}^{0}(z)]^{-1}\Psi_{n+}^{0}(z),\quad z \in \gamma_{0},
\eqno (8.11)
$$
$$
G(z) = [\Psi_{n}^0(z)]^{-1}
\pmatrix
1 & -i \\
0 & 1
\endpmatrix
\Psi_{n}^0(z),\quad z \in \gamma_{1},
$$
and satisfying the estimates,
$$
||I-G(z)||_{L_{2}(\gamma)\cap L_{\infty}(\gamma)} = O(N^{-1}).
\eqno (8.12)
$$

The Riemann-Hilbert problem shares a remarkable property
of wellposedness (see e.g. [BDT], [CG], [LiS], and [Zh]). Namely,
estimate (8.12)  imply the following estimate of $X_n(z)$
on the full complex plane (see Appendix D in [BI]).

{\bf Proposition 8.2.} {\it For all $z\in\C$,
$$
X_n(z)=\Theta_0\(I+O\(\frac{1}{N(1+|z|)}\)\).
\eqno (8.13)
$$}

Comparing (8.13) with (8.3) we obtain that 
$$
\Theta^{-1}_{0}\Theta_1=O(N^{-1}),
$$
that is, according to (8.4),
$$
e^{-(C_1-\la_n)\sg_3}\G_{0}^{-1}\G_1 e^{(C_1-\la_n)\sg_3}
-(R_n^0)^{1/2}\sg_1=O(N^{-1}).
\eqno (8.14)
$$
This is a $2\times 2$ matrix equation. If we write it for the matrix
elements we obtain from (1.38) that two equations are trivial, $0=0$,
and the other two are
$$\eqalign{
&e^{-2C_1+2\la_n}=(R_n^0)^{1/2}+O\(N^{-1}\),\cr
&R_ne^{2C_1-2\la_n}=(R_n^0)^{1/2}
+O\(N^{-1}\).}
\eqno (8.15)
$$
Observe that by (4.2),
$$
R_n^0=-\frac{t}{2g}+O\(N^{-1/3}\),
$$
so that $R_n^0>0$ is uniformly bounded and separated from 0
as $N\to\infty$, hence from the
first equation in (8.15) we obtain that
$$
e^{\la_n-C_1}=(R_n^0)^{1/4}+O(N^{-1}).
\eqno (8.16)
$$
Substituting of this into the second equation in (8.15) gives that
$$
R_n=R_n^0+O\(N^{-1}\).
\eqno (8.17)
$$
From (8.16) we also obtain  that
$$
e^{\la_n}=e^{C_1}(R_n^0)^{1/4}\(1+O(N^{-1})\).
$$
Since $h_n=e^{2\la_n}$ (see Proposition 1.1), we obtain that
$$
h_n=e^{2C_1}(R_n^0)^{1/2}\(1+O(N^{-1})\).
\eqno (8.18)
$$
The constant $C_1$ is given in (5.19). Substituting its value into (8.18)
we obtain that
$$
h_n=e^{2N\int_{z^N_0}^\infty\mu^c(u)\,du}\(1+O(N^{-1})\).
\eqno (8.19)
$$
This proves the part of Theorem 1.2 concerning the asymptotics
of $R_n$ and $h_n$.

Let us take
$$
C_0=\frac{1}{2^{1/2}(R_n^0)^{1/4}}.
\eqno (8.20)
$$
Then from (8.16) and (8.4) we obtain that
$$
\Theta_0=I+O(N^{-1}),
$$
hence
$$
\Psi_n(z)=\(I+O\({1\over{N(1+|z|)}}\)\)\Psi_n^0(z).
\eqno (8.21)
$$
Due to (8.1) and (1.96) this proves asymptotic relations (1.64), (1.76),
(1.80), and (1.89). Theorem 1.2 is proved.

\beginsection 9. Universality \par

{\it Critical Point.} The double scaling limit at the critical
point $z=0$ is determined by the kernel
$$
Q_c(u,v)=\lim_{N\to\infty}\frac{1}{cN^{1/3}}Q_N\(\frac{u}{cN^{1/3}},
\frac{v}{cN^{1/3}}\),
\eqno (9.1)
$$
where 
$$
Q_N(z,w)=R_N^{1/2}\frac{\psi_N(z)\psi_{N-1}(w)-\psi_{N-1}(z)\psi_N(w)}
{z-w}
$$
and $c>0$ is a normalizing constant.
To evaluate (9.1) we apply CP solution (1.88) which gives that 
modulo $O(N^{-1/3})$ terms,
$$\eqalign{
\psi_N(z)&=
\frac{1}{\pi^{1/2}}\, (R_N^0)^{1/4}\,
\[V_{11}(z)\Phi^1(N^{1/3}\z(z))
+V_{12}(z)\Phi^2(N^{1/3}\z(z))\],\cr  
\psi_{N-1}(z)&=
\frac{1}{\pi^{1/2}}\, (R_N^0)^{-1/4}\,
\[V_{21}(z)\Phi^1(N^{1/3}\z(z))
+V_{22}(z)\Phi^2(N^{1/3}\z(z))\].}  
\eqno (9.2)
$$
To evaluate (9.1) we can replace $V_{ij}(z)$ by $V_{ij}\equiv V_{ij}(0)$
and $\z(z)$ by $\z'(0)z$. We take
$$
c=\z'(0).
$$
Then, modulo $O(N^{-1/3})$ terms,
$$\eqalign{
\frac{1}{cN^{1/3}}&Q_N\(\frac{u}{cN^{1/3}},
\frac{v}{cN^{1/3}}\)\cr
&=C (u-v)^{-1}\[
\(V_{11}\Phi^1(u)+V_{12}\Phi^2(u)\)
\(V_{21}\Phi^1(v)+V_{22}\Phi^2(v)\)\right.\cr
&\left.-\(V_{11}\Phi^1(v)+V_{12}\Phi^2(v)\)
\(V_{21}\Phi^1(u)+V_{22}\Phi^2(u)\)\]\cr
&=C' \frac{\Phi^1(u)\Phi^2(v)-\Phi^1(v)\Phi^2(u)}{u-v}\,,}
\eqno (9.3)
$$
where $C=1/\pi $ and 
$$
C'=C(V_{11}V_{22}-V_{12}V_{22})=C\det V=C.
$$
Observe that as they are defined in Section 3, the functions
$\Phi^j(z)$ depend on $n\mod 4$. Nevertheless, the
combination 
$$
\Phi^1(u)\Phi^2(v)-\Phi^1(v)\Phi^2(u)
\eqno (9.4)
$$
does not depend on $n$. Indeed, by formula (A.5) in Appendix A
below,
$$
\vec\Phi(z)=U_0\vec B_0(z),\qquad U_0=
\pmatrix
i^n & (-i)^n \\
i^{n-1} & (-i)^{n-1}
\endpmatrix,
$$
where $\vec B_0(z)=\pmatrix B_0^1(z) \\ B_0^2(z) \endpmatrix$ does not
depend on $n$. 
A direct computation gives that
$$
\Phi^1(u)\Phi^2(v)-\Phi^1(v)\Phi^2(u)=(\det U_0)\[B_0^1(u)B_0^2(v)
-B_0^1(v)B_0^2(u)\].
$$
Since $\det U_0=2i$, this proves that combination (9.4) does
not depend on $n$.
Thus,
$$
Q_c(u,v)=\frac{\Phi^1(u)\Phi^2(v)-\Phi^1(v)\Phi^2(u)}{\pi(u-v)}\,.
\eqno (9.5)
$$

{\it Edge of the  spectrum.} The double scaling limit at the edge
is determined by the kernel
$$
Q_e(u,v)=\lim_{N\to\infty}\frac{1}{cN^{2/3}}Q_N\(z_0+\frac{u}{cN^{2/3}},
z_0+\frac{v}{cN^{2/3}}\),
\eqno (9.6)
$$
where $Q_n(z,w)$ is the same as in (9.2) 
and $c>0$ is a normalizing constant.
To evaluate (9.6) we use the TP solution
$$\eqalign{
\psi_n(z)&=
\frac{ (R_n^0)^{1/4}}{2\pi^{1/2}}\,
\[W_{11}(z)N^{1/6}\Ai(N^{2/3}\z(z))
+W_{12}(z)N^{-1/6}\Ai'(N^{1/3}\z(z))\],\cr  
\psi_{n-1}(z)&=
\frac{ (R_n^0)^{1/4}}{2\pi^{1/2}}\,
\[W_{21}(z)N^{1/6}\Ai(N^{1/3}\z(z))
+W_{22}(z)N^{-1/6}\Ai'(N^{1/3}\z(z))\].}  
\eqno (9.7)
$$
If we take
$
c=\z'(z_0)
$
and repeat the above calculation at the critical point,
then we will obtain that
$$
Q_e(u,v)=\frac{\Ai(u)\Ai'(v)-\Ai(v)\Ai'(u)}{u-v}\,,
\eqno (9.8)
$$ 
the Airy kernel.

{\it Bulk of the spectrum}. The double scaling limit in the bulk
is determined by the kernel
$$
Q(u,v)=\lim_{N\to\infty}\frac{1}{cN}Q_N\(z+\frac{u}{cN},
z+\frac{v}{cN}\),
\eqno (9.9)
$$
where  $c=p(z)$, the value of the density function at $z$.
To evaluate (9.9) we use the WKB solution and in the
same way as above we obtain that
$$
Q_b(u,v)=\frac{\sin\pi(u-v)}{\pi(u-v)},
$$
the sine kernel.

{\it Acknowledgements.} Pavel Bleher 
was supported in part by the NSF grant  DMS-9970625.
The final part of this project was done when he
was visiting  Service de Physique Th\'eorique,
CEA/Saclay, and the support during his stay there
is gratefully acknowledged. 
Alexander Its was supported in part by the NSF grants  DMS-9801608
and  DMS-0099812.

\beginsection Appendix A. Proof of Proposition 3.2 \par

{\it Proof.}
Let $\widetilde\Psi_j(z)$, $j = 1, ..., 6$
 be the Stokes' solutions to equation
(3.11) introduced in Section 3. We recall that
the solutions $\widetilde\Psi_j(z)$ are uniquely
determined by the asymptotic condition,
$$
\lim_{|z|\to\infty}\widetilde\Psi_j(z)e^{i\((4/ 3)z^3+yz\)\sg_3}=I
\iff
\left| \arg z-{(j-1)\pi\over 3}\right|\le {\pi\over 3}-\ep\,.
\eqno (A.1)
$$
and that they are related as follows:
$$\eqalign{
&\widetilde\Psi_1(z)=\widetilde\Psi_6(z)
\pmatrix
1 & -1 \\
0 & 1 
\endpmatrix,\qquad
\widetilde\Psi_2(z)=\widetilde\Psi_1(z)
\pmatrix
1 & 0 \\
1 & 1 
\endpmatrix,\qquad
\widetilde\Psi_3(z)=\widetilde\Psi_2(z),\cr
&\widetilde\Psi_4(z)=\widetilde\Psi_3(z)
\pmatrix
1 & 0 \\
-1 & 1 
\endpmatrix,\qquad
\widetilde\Psi_5(z)=\widetilde\Psi_4(z)
\pmatrix
1 & 1 \\
0 & 1 
\endpmatrix,\qquad
\widetilde\Psi_6(z)=\widetilde\Psi_5(z),\cr}
\eqno (A.2)
$$
(see (3.13) and (3.14)).
Let us rewrite these equations in terms of vector solutions. Let
$$
\widetilde\Psi_2(z)=\(\vec B_0(z),\vec B_1(z)\).
$$
 Then by (A.2),
$$\eqalign{
&\widetilde\Psi_1(z)=\widetilde\Psi_4(z)=\(\vec B_2(z),\vec B_1(z)\),\qquad 
\vec B_2(z)=\vec B_0(z)-\vec B_1(z),\cr
&\widetilde\Psi_2(z)=\widetilde\Psi_3(z)=\(\vec B_0(z),\vec B_1(z)\),\cr
&\widetilde\Psi_5(z)=\widetilde\Psi_6(z)=\(\vec B_2(z),\vec B_0(z)\),\cr}
\eqno (A.3)
$$
and
$$\eqalign{
\lim_{|z|\to\infty}\vec B_0(z)e^{i\((4/ 3)z^3+yz\)}
&=\pmatrix
1\\ 0
\endpmatrix,
\iff \ep<\arg z<\pi-\ep,\cr
\lim_{|z|\to\infty}\vec B_0(z)e^{-i\((4/ 3)z^3+yz\)}
&=\pmatrix
0\\ 1
\endpmatrix,
\iff \pi+\ep<\arg z<2\pi-\ep,\cr
\lim_{|z|\to\infty}\vec B_1(z)e^{-i\((4/ 3)z^3+yz\)}
&=\pmatrix
0\\ 1
\endpmatrix,
\iff -{\pi\over 3}+\ep<\arg z<{4\pi\over 3}-\ep,\cr
\lim_{|z|\to\infty}\vec B_2(z)e^{i\((4/ 3)z^3+yz\)}
&=\pmatrix
1\\ 0
\endpmatrix,
\iff {2\pi\over 3}+\ep<\arg z<{7\pi\over 3}-\ep.\cr}
\eqno (A.4)
$$
Define
$$\eqalign{
&\vec\Phi(z)=i^nU\vec B_0(z)=
\pmatrix
i^n & (-i)^n \\
i^{n-1} & (-i)^{n-1}
\endpmatrix \vec B_0(z),\cr 
&\vec\Phi_1(z)=i^{n-1}U\vec B_1(z)=
\pmatrix
i^{n-1} & (-i)^{n+1} \\
i^{n-2} & (-i)^n
\endpmatrix \vec B_1(z),\cr 
&\vec\Phi_2(z)=i^{n+1}U\vec B_2(z)=
\pmatrix
i^{n+1} & (-i)^{n-1} \\
i^n & (-i)^{n-2}
\endpmatrix \vec B_2(z),\cr} 
\eqno (A.5)
$$
Then
$$
\vec\Phi_j'(z)=A(z)\vec\Phi_j(z)
\eqno (A.6)
$$
and
$$\eqalign{
\lim_{|z|\to\infty}\vec \Phi(z)e^{i\((4/ 3)z^3+yz\)}
&=\pmatrix
i^n\\ i^{n-1}
\endpmatrix,
\iff \ep<\arg z<\pi-\ep,\cr
\lim_{|z|\to\infty}\vec \Phi(z)e^{-i\((4/ 3)z^3+yz\)}
&=\pmatrix
(-i)^n \\ (-i)^{n-1}
\endpmatrix,
\iff \pi+\ep<\arg z<2\pi-\ep,\cr
\lim_{|z|\to\infty}\vec \Phi_1(z)e^{-i\((4/ 3)z^3+yz\)}
&=\pmatrix
(-i)^{n+1} \\ (-i)^n
\endpmatrix,
\iff -{\pi\over 3}+\ep<\arg z<{4\pi\over 3}-\ep,\cr
\lim_{|z|\to\infty}\vec \Phi_2(z)e^{i\((4/ 3)z^3+yz\)}
&=\pmatrix
i^{n+1}\\ i^n
\endpmatrix,
\iff {2\pi\over 3}+\ep<\arg z<{7\pi\over 3}-\ep.\cr}
\eqno (A.7)
$$
Observe that the function $\overline{\vec\Phi\(\overline{z}\)}$
satisfies the equation (A.6), because the matrix $A(z)$ is real, and,
in addition, $\overline{\vec\Phi\(\overline{z}\)}$
 has the same asymptotics (A.7) as $\vec\Phi(z)$.
Hence (3.15) holds. Similarly, $\overline{\vec\Phi_1\(\overline{z}\)}$
satisfies the equation (A.6), and
it has the same asymptotics as $\vec\Phi_2(z)$.
Hence (iii) holds.

The function $V\vec\Phi(-z)$, where $V=\pmatrix (-1)^n & 0 \\ 0
& (-1)^{n-1} \endpmatrix $ satisfies (A.6) and it
 has the same
asymptotics as $\vec\Phi(z)$. Hence (3.16) holds. Similar argument
proves (3.19). Finally, (3.20)
follows from (A.3). Proposition 3.2 is proved.

\beginsection Appendix B. Proof of Lemma 4.1 \par

Direct computation:
$$
V_1B=
\pmatrix
d_{12} & 0 \\
b_{11}-d_{11} & b_{12}
\endpmatrix
\pmatrix 
b_{11} & b_{12} \\
b_{21} & -b_{11}
\endpmatrix=
\pmatrix
b_{11}d_{12} & b_{12}d_{12} \\
b_{11}^2-b_{11}d_{11}+b_{12}b_{21} &
-b_{12}d_{11} 
\endpmatrix
$$
and
$$
DV_1=
\pmatrix 
d_{11} & d_{12} \\
d_{21} & -d_{11}
\endpmatrix
\pmatrix
d_{12} & 0 \\
b_{11}-d_{11} & b_{12}
\endpmatrix=
\pmatrix
b_{11}d_{12} & b_{12}d_{12} \\
d_{11}^2-b_{11}d_{11}+d_{12}d_{21} &
-b_{12}d_{11} 
\endpmatrix.
$$
Since 
$$
b_{11}^2+b_{12}b_{21}=-\det B=-\det D=d_{11}^2+d_{12}d_{21},
$$
we obtain that $V_1B=DV_1$ which was stated. Similarly,
$V_2B=DV_2$. Lemma 4.1 is proved.

\beginsection Appendix C. Proof of Proposition 5.2 \par

Let  $z \in \Gamma_{c}^{+}$. By (4.37) and (5.$2'$),
$$
\mu(z) = 4i\z_{0}'(z) \z_{0}^{2}(z) + N^{-2/3}iy\z_{0}'(z) + O\(
N^{-4/3}\), 
\eqno (C.1)
$$
hence from  (4.66,67) we obtain the following asymptotics
of the matrix $W(z)$ as $N\to\infty$:
$$\eqalign{
W(z) &= 
\pmatrix
a^{0}_{12}(z) - i\mu(z) & -a^{0}_{11}(z) \\
-a^{0}_{11}(z) & -a^{0}_{21}(z) - i\mu(z)
\endpmatrix
+ N^{-1/3}(-1)^{n}4u\z_{0}'(z)\z_{0}(z)\sg_{1}\cr
&+ N^{-2/3}\[-(-1)^{n}2w\z_{0}'(z)\sg_{3}
+(v-y)\z_{0}'(z)I\] + O\(N^{-1}\).}
\eqno (C.1')
$$
Substituting this formula into (4.65), we obtain the
equation,
$$
V(z)\pmatrix
1 & -i \\
-i & 1
\endpmatrix
= {1\over{\sqrt{b(z)}}}
\pmatrix
a^{0}_{12}(z) - i\mu(z) +ia^{0}_{11}(z) 
& -ia^{0}_{12}(z) - \mu(z)-a^{0}_{11}(z) \\
-a^{0}_{11}(z)+ia^{0}_{21}(z) - \mu(z) 
& ia^{0}_{11}(z)-a^{0}_{21}(z) - i\mu(z)
\endpmatrix
U_{0}(z),
\eqno (C.2)
$$
where
$$
b(z) = -2\mu^{2}(z) -i\mu(z)[a^{0}_{12}(z) - a^{0}_{21}(z)],
\eqno (C.2') 
$$
and the matrix-valued function $U_{0}(z)$ admits the asymptotic
representation,
$$\eqalign{
U_{0}(z) &= I + N^{-1/3}(-1)^{n}{u\over 2\z_{0}(z)}\sg_{1}\cr
&+ N^{-2/3}\[{u^{2}\over 8\z_0^2(z)}I+ (-1)^{n+1}{w\over 4\z_0^2(z)}\sg_2
-2i(-1)^{n}{a^{0}_{11}(z)w\z_{0}'(z)\over b(z)}\sg_{3}\]
+ O\(N^{-1}\).}
\eqno (C.3)
$$
When deriving (C.3), we use (C.1), the formula
$$
b(z) = -2i\mu(z)[a_{12}^{0}(z) -i\mu(z)] + O\(N^{-2/3}\),
$$
which follows from (C.$2'$), and the formula
$$
\frac{a_{12}^0(z)+a_{21}^0(z)}{2}=N^{-2/3}c^0+O(N^{-1}),
\qquad c^0=\(2g|t|\)^{1/6}(-1)^nw,
\eqno (C.4)
$$
which follows from (4.4).
Taking into account that by ($4.5'$),
$$
\mu^{2}(z) =(a^{0}_{11})^2(z)
+a^{0}_{12}(z)a^{0}_{21}(z)+O(N^{-4/3}),
$$
 a simple algebra shows that
$$
{1\over{\sqrt{b(z)}}}
\pmatrix
a^{0}_{12}(z) - i\mu(z) +ia^{0}_{11}(z)
 & -ia^{0}_{12}(z) - \mu(z)-a^{0}_{11}(z) \\
-a^{0}_{11}(z)+ia^{0}_{21}(z) - \mu(z)
 & ia^{0}_{11}(z)-a^{0}_{21}(z) - i\mu(z)
\endpmatrix
= T_{0}(z) {\delta(z)^{\sg_3}}+O(N^{-4/3}),
\eqno (C.4')
$$
where 
$$
\delta(z) = \({{a^{0}_{12}(z) - i\mu(z) +ia^{0}_{11}(z)}\over
{ia^{0}_{11}(z)-a^{0}_{21}(z) - i\mu(z)}}\)^{1/2}
\equiv \( 1 + {{a^{0}_{12}(z) + a^{0}_{21}(z)}\over
{ia^{0}_{11}(z)-a^{0}_{21}(z) - i\mu(z)}}\)^{1/2}
$$
and $T_{0}(z)$ is the WKB gauge matrix [see  (5.5)]. 
From (C.4),
$$
\delta(z)= 1 + N^{-2/3}{c^0\over
{ia^{0}_{11}(z)-a^{0}_{21}(z) - i\mu(z)}} + O\(N^{-1}\).
$$
Using this equation, equations
(C.3) and (C.4'), and also equation (4.84),
we can rewrite (C.2)
as 
$$
V(z)\pmatrix
1 & -i \\
-i & 1
\endpmatrix
= T_{0}(z)V_{0}(z),
\eqno (C.5)
$$
where
$$\eqalign{
V_{0}(z) &= I + N^{-1/3}(-1)^{n}{u\over 2\z(z)}\sg_{1}
+ N^{-2/3}\[{u^{2}\over 8\z^2(z)}I
+ (-1)^{n+1}{w\over 4\z^2(z)}\sg_{2}+\Delta(z)\sg_{3}\]\cr
& + O\(N^{-1}\),}
\eqno (C.6)
$$
and
$$
\Delta(z) ={c^0\over
{ia^{0}_{11}(z)-a^{0}_{21}(z) - i\mu(z)}}
- 2i(-1)^{n}w\z_{0}'(z){a^{0}_{11}(z)\over b(z)}.
\eqno (C.6')
$$
This proves equation (5.16). Let us prove (5.14).
 
By (4.68,69),
$$
a(z) = \det \[A_{n}^{0}(z) - N^{-1}V'(z)V^{-1}(z)\] + O\(N^{-2}\).
$$
Using (C.5) and the diagonalizing
property of the matrix $T_{0}(z)$,
$$
T_{0}^{-1}A_{n}^{0}T_{0} = -\mu\sg_{3},
$$
we obtain that
$$\eqalign{
\det \[A_{n}^{0} - N^{-1}V'V^{-1}\]&=
\det \[A_{n}^{0} - N^{-1}T_{0}'T_{0}^{-1}
-N^{-1}T_{0}V_{0}'V_{0}^{-1}T_{0}^{-1}\]\cr
&= \det \[T_{0}^{-1}A_{n}^{0}T_{0} - N^{-1}T_{0}^{-1}T_{0}'
-N^{-1}V_{0}'V_{0}^{-1}\]\cr
&=\det \[-\mu \sg_{3} - N^{-1}T_{0}^{-1}T_{0}'
-N^{-1}V_{0}'V_{0}^{-1}\]\cr
&=-\mu^{2}\det \[I + {1\over{\mu}}N^{-1}\sg_{3}T_{0}^{-1}T_{0}'
+{1\over{\mu}}N^{-1}\sg_{3}V_{0}'V_{0}^{-1}\]\cr
&= -\mu^{2} -\mu N^{-1}\Tr \[\sg_{3}T_{0}^{-1}T_{0}'\]
- \mu N^{-1}\Tr \[\sg_{3}V_{0}'V_{0}^{-1}\] + O\(N^{-2}\).}
\eqno (C.7)
$$
From (C.6) we conlude at once that
$$
\Tr \[\sg_{3}V_{0}'V_{0}^{-1}\] =2\Delta'(z)N^{-2/3} +  O\(N^{-1}\).
\eqno (C.8)
$$
At the same time,  equation (5.5$'$) and the identity
(following from the equations $\det T_{0} \equiv 2$ and $ (T_{0})_{11} =
 (T_{0})_{22})$,
$$
\diag T_{0}^{-1}T_{0}' =
{1\over 2}\(\Tr \[\sg_{3}T_{0}^{-1}T_{0}'\]\)\sg_{3},
$$
imply
$$
\Tr \[\sg_{3}T_{0}^{-1}T_{0}'\] = 2\tau'(z).
\eqno (C.9)
$$
From (C.7-9), and taking into account our
convention about the branches of the square roots
in (4.82), we derive the asymptotic formula,
$$\eqalign{
\sqrt{a(z)}& = \[ \det \(A_{n}^{0}(z) - N^{-1}V'(z)V^{-1}(z)\) + O\(
N^{-2}\) \]^{1/2}\cr 
&= -i\mu(z) -iN^{-1}\tau'(z) -iN^{-5/3}\Delta'(z) + O\(N^{-2}\),}
\eqno(C.10)
$$
$N\rightarrow \infty, \quad z\in \Gamma_{c}^{+}$.

From  (4.11)
it follows that when $z\in \Gamma_{c}^{+}$,
$$\eqalign{
\sqrt {f(\z)}&=\left\{
16\z^4+8N^{-2/3}y\z^2+N^{-4/3}\[v^2(y)-4w^2(y)\]\right\}^{1/2}\cr 
&=  4\z^2\left\{1+N^{-2/3}\frac{y}{4\z^2}+N^{-4/3}
\frac{v^2(y)-4w^2(y)-y^2}{32\z^4}+O\(N^{-2}\)\right\}\cr
&= 4\z^2 + N^{-2/3}y - N^{-4/3}{D\over 2\z^2} + O(N^{-2}),}
\eqno (C.11)
$$
where [cf. (3.6)]
$$
D\equiv D(y) ={{y^2-v^2(y)+4w^2(y)}\over 4}.
$$
Replacing the both sides of (4.82) by their asymptotics 
(C.10,11) and integrating the resulting equation, we
arrive to the formula,
$$\eqalign{
i\({4\over 3}\z^3 + N^{-2/3}y\z + N^{-4/3}{D\over 2}\z^{-1}\)
&= \int_{\infty}^{z}\mu(u)du
+N^{-1}\tau(z)\cr
&+N^{-5/3}\Delta(z) + c + O\(N^{-2}\).}
\eqno (C.12)
$$ 
Let us find the constant $c$. We will use the fact that the 
function $\z(z)$ is odd in $z$. 

Consider the contour $\g$ which goes  along the imaginary axis
from $i\infty$ to $id_2$,  then around zero on the right
to $-id_2$ and then to $-i\infty$.  Denote by $\g_1$
the part of $\g$ from $i\infty$ to $id_2$, by $\g_2$ the part of $\g$
from $id_2$ to $-id_2$, and by $\g_3$, from $-id_2$ to $-\infty$
(see Fig.5).

\centinsert{\pscaption{\psboxto (6in;1.5in){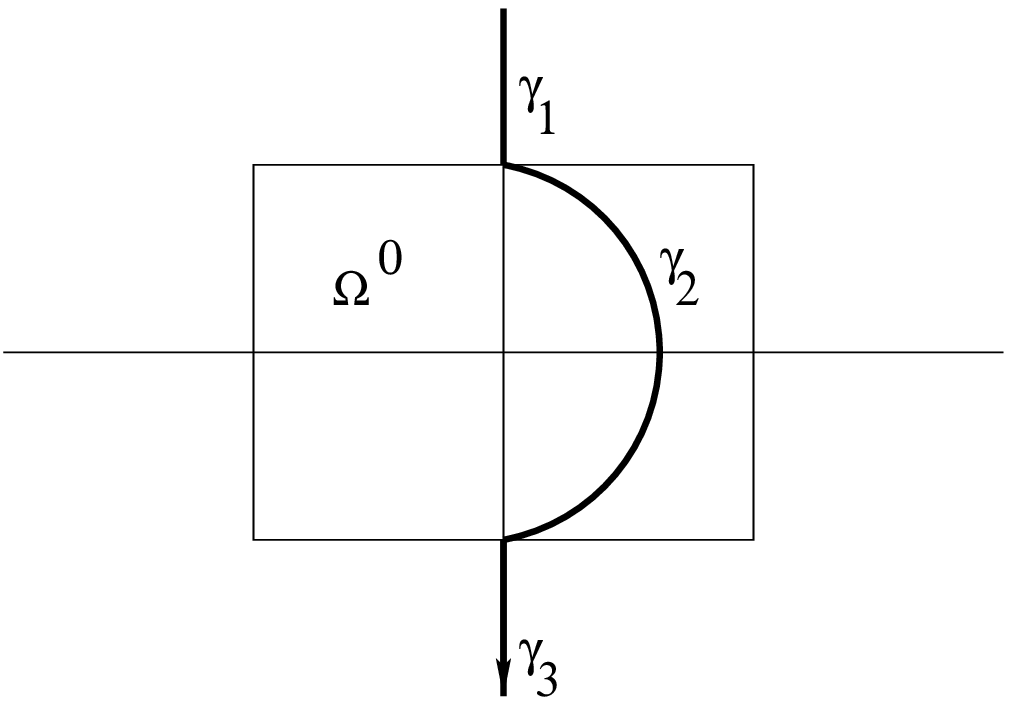}}
{\srm Fig 5:  The contour $\g$.   }}

\vskip 3mm

Let $\mu_0(z)=\sqrt{-d(z)}$ with two
cuts $(-\infty,-z^N)$ and $(z^N,\infty)$. We will consider $\mu_0(z)$
outside of a small disk $D_{\om_0}=\{|z|\le\om_0\}$ around the origin.
For $z\in\Om^c$,
$$
\mu_0(z)=\left\{
\eqalign{
&\mu(z),\quad \Im z>0,\cr
&-\mu(z),\quad \Im z<0.}\right.
\eqno (C.13)
$$
In contrast to $\mu(z)$ which is an odd function, the
function $\mu_0(z)$ is even. 
The function $\tau(z)$ is defined in (5.5). Let $\tau_0(z)$
be the function on $\C\setminus D_{\om_0}$ defined by the same
formula (5.5) with two cuts 
$(-\infty,-z^N)$ and $(z^N,\infty)$. Replacing 
in all formulae (C.1-11) the functions $\mu(z)$
and $\tau(z)$ by the functions $\mu_{0}(z)$
and $\tau_{0}(z)$ respectively, we can extend
the validity of these formulae to the points $z$ of the
annulus $\Omega_{c}\setminus D_{\omega_{0}}$. 
Hence equation (C.12) can be extended to the
equation
$$\eqalign{
i\({4\over 3}\z^3 + N^{-2/3}y\z + N^{-4/3}{D\over 2}\z^{-1}\)
&= \int_{\infty}^{z}\mu_{0}(u)du
+N^{-1}\tau_{0}(z)\cr
&+N^{-5/3}\Delta_{0}(z) + c + O\(N^{-2}\),}
\eqno (C.12')
$$
$$
z \in  \Omega_{c}\setminus D_{\omega_{0}},
$$
where
$$
\Delta_{0}(z) \equiv \Delta(z)|_{\mu(z) \to \mu_{0}(z)},
$$
and it is an analytic continuation
of  $\Delta(z)$
to $\C\setminus D_{\om_0}$ with two cuts 
$(-\infty,-z^N)$ and $(z^N,\infty)$.

We will split the function $\tau_{0}(z)$ as
$$
\tau_0(z)=\tilde\tau_0(z)+\hat\tau_0(z)+c_0+O(N^{-1}),
\eqno (C.14)
$$
where
$$\eqalign{
&\tilde\tau_0(z)={1\over 2}\ln\[{\mu_0(z)-a_{11}^0(z)\over a_{12}^0(z)} \],
\qquad
\hat\tau_0(z)=\frac{1}{2}\int_\infty^z\frac{U^1(u)}{\mu_0(u)}\,du,
\qquad c_0=\frac{1}{4}\ln R_n^0,\cr
& U^1(z)={a^0_{11}}'(z)-a^0_{11}(z)\frac{{a^0_{12}}'(z)}{a^0_{12}(z)}\,,}
\eqno (C.15)
$$
[we will justify (C.14) below, in the end of this appendix].
The branch of the logarithm in $\tilde\tau_0(z)$ is determined
by the condition  
that for sufficiently large $N$,
$$
0<\Im \tilde\tau_0(z) < {\pi \over 2},
\quad z \in \C\setminus [D_{\om_0}
\cup (-\infty,-z^N]\cup[z^N,\infty)].
$$
(Note that ${{\mu_0(z)-a_{11}^0(z)}\over a_{12}^0(z)}
= \phi(z) + O(N^{-1/3})$, where $\phi(z) ={{z+\sqrt{z^2-z^{2}_{0}}}
\over{2[R^{0}_{n}]^{1/2}}}$, and $\phi:   \C\setminus 
[(-\infty,-z_0]\cup[z_0,\infty)]
\mapsto \{\Re \phi > 0\}$.)  
Using (C.14) we rewrite (C.12$'$) as
$$\eqalign{
i\({4\over 3}\z^3 + N^{-2/3}y\z\right.
&\left.+ N^{-4/3}{D\over 2}\z^{-1}\)
= \int_{\infty}^{z}\(\mu_0(u)+N^{-1}\frac{U^1(u)}{2\mu_0(u)}\)du
+N^{-1}\tilde\tau_0(z)\cr
&+N^{-5/3}\Delta_0(z) + c_1 + O\(N^{-2}\),
\qquad c_1=c+N^{-1}c_0.}
\eqno (C.16)
$$
where $z \in \Omega_{c} \setminus D_{\omega_{0}}$. The integral 
$ \int_{\infty}^{z}\mu_0(u)\,du $ diverges at infinity
so a regularization is needed. Let us describe the regularization we
use.

Let $\g_0$ be any contour which goes from $\infty$ to $z$ in the 
region
$$
S_0\equiv \C\setminus\(D_{\om_0}\cup 
(-\infty,-z^N)\cup(z^N,\infty)\),
$$
and which starts at $\infty$ in the upper half-plane.
For our purposes it will be useful to introduce a slightly more general
regularization than (5.3), which in fact will be
equivalent to (5.3). Take any point
$a$ on $\g_0$ and define the regularized integral as  
$$
\int_{\g_0} \mu_0(u)\,du\equiv {\mu_3\over 4}a^4+{\mu_1\over 2}a^2
+\mu_{-1}\ln a+\int_\infty^a\widetilde\mu_0(u)\,du
+\int_a^z \mu_0(u)\,du\,,
\eqno (C.17)
$$
where 
$$
\wt\mu_0(z)=\mu_0(z)-\mu_3 z^3-\mu_1 z-\mu_{-1}z^{-1};
\qquad \mu_3=\frac{g}{2},\quad \mu_1=\frac{t}{2},\quad
\mu_{-1}=-\frac{n}{N}\,.
$$
The branch of $\ln a$ is taken as follows. By our assumption
$\g_0$ starts at $\infty$ in the upper half-plane. If the whole
piece of $\g_0$ from $\infty$ to $a$ lies in the upper half-plane,
then take $\ln a$
on the main branch of logarithm,
 with a cut at $(-\infty,0)$
and $\ln 1=0$. For other $a$'s extend $\ln a$ continuously along
$\g_0$.
Definition (C.17) shares the following important properties:
\item{1.} {The right-hand side of (C.17) does not depend on $a$.}
\item{2.} {The contour of integration $\g_0$
 can be deformed by the Cauchy theorem.}
\item{3.} {The integral $\int_{\infty}^z \mu_0(u)\,du$
does not depend on the contour of integration in $S_0$.} 

\noindent
Observe that the property 3 is not automatic because
$S_0$ is not simply connected. It follows obviously
from the fact that $\mu_0(z)$ is an even function.

The function $\frac{U^1(z)}{2\mu_0(z)}$ has the following asymptotics
as $z\to\infty$, $\Im z>0$:
$$
\frac{U^1(z)}{2\mu_0(z)}=-\frac{1}{2}z^{-1}+r(z),\qquad
r(z)=O(z^{-3}),
\eqno (C.18)
$$
[see (1.71)], 
and we define the regularized integral of this function as 
$$
\int_{\infty}^z \frac{U^1(u)}{2\mu_0(u)}
 \,du\equiv -\frac{\ln a}{2}+\int_\infty^a r(u)\,du
+\int_a^z \frac{U^1(u)}{2\mu_0(u)}\,du\,.
\eqno (C.19)
$$
It also shares the properties 1, 2, and 3.

Observe that
the left-hand side of (C.16)  is an odd function
in the annulus $\Om^0\setminus D_{\om_0}$. Hence the right-hand side
is odd as well. Applying this to $z=id_2$ we obtain the equation
$$\eqalign{
\int_{\g_1}&\(\mu_0(u)+N^{-1}\frac{U^1(u)}{2\mu_0(u)}\)du
+N^{-1}\tilde\tau_0(id_2)
+N^{-5/3}\Delta_0(id_2) + c_1\cr
&=
-\[\int_{\g_1\cup\g_2}\(\mu_0(u)+N^{-1}\frac{U^1(u)}{2\mu_0(u)}\)du
+N^{-1}\tilde\tau_0(-id_2)
+N^{-5/3}\Delta_0(-id_2) + c_1\]+O(N^{-2}).}
\eqno (C.20)
$$
The functions $\mu_0(z)$ and $U^1(z)$ are even, hence
$$
\int_{\g_1}\(\mu_0(u)+N^{-1}\frac{U^1(u)}{2\mu_0(u)}\)du=
\int_{\g_3}\(\mu_0(u)+N^{-1}\frac{U^1(u)}{2\mu_0(u)}\)du-\frac{\pi
in}{N}-\frac{\pi i}{2N}, 
\eqno (C.21)
$$
with the term $-\frac{\pi i n}{N}-\frac{\pi i}{2N}$ 
coming from regularization.
Indeed, due to (C.13), as $z\to\infty$, $\Im z<0$,
$$
\mu_0(z)=-\mu_3 z^3-\mu_1 z-\mu_{-1}z^{-1}+\wt \mu_0(z),
\qquad \wt\mu_0(z)=O(z^{-3}),
\eqno (C.22)
$$
and
we define the regularization of $\int_{\g_3}\mu_0(u) du$ as
$$
\int_{\g_3} \mu_0(u)\,du\equiv {\mu_3\over 4}a^4+{\mu_1\over 2}a^2
+\mu_{-1}\ln a+\int_a^\infty\wt\mu_0(u)\,du
+\int_z^a \mu_0(u)\,du\,,
\eqno (C.23)
$$
where $z=-id_2$ and $a$ is any point on $\g_3$. The
integral $\int_a^\infty$ is taken over the part of $\g_3$
from $a$ to $\infty$ (in the lower half-plane). Similarly,
as $z\to\infty$, $\Im z<0$,
$$
\frac{U^1(z)}{2\mu_0(z)}=\frac{1}{2}z^{-1}+r(z),\qquad
r(z)=O(z^{-3}),
\eqno (C.24)
$$
and we define
$$
\int_{\g_3} \frac{U^1(u)}{2\mu_0(u)}
 \,du\equiv -\frac{\ln a}{2}+\int_a^\infty r(u)\,du
+\int_z^a \frac{U^1(u)}{2\mu_0(u)}\,du\,.
\eqno (C.25)
$$
We take the main branch of logarithm for $\ln a$  in (C.23,25).
The remainder functions $\wt\mu_0(z)$ and $r(z)$ are even.
Since
$$
\ln (id_2)=\ln (-id_2)+\pi i
\eqno (C.26)
$$
and $\mu_{-1}=-\frac{n}{N}$, 
$$\eqalign{
\int_{\g_1}&\(\mu_0(u)\right.\left.+N^{-1}\frac{U^1(u)}{2\mu_0(u)}\)du=
 {\mu_3\over 4}(id_2)^4+{\mu_1\over 2}(id_2)^2
+\(\mu_{-1}-\frac{1}{2N}\)\ln (id_2)\cr
&+\int_{\infty}^{id_2}\[\wt\mu_0(u)
+N^{-1}r(u)\]\,du
= {\mu_3\over 4}(-id_2)^4+{\mu_1\over 2}(-id_2)^2
-\(\frac{n}{N}+\frac{1}{2N}\)\[\ln (-id_2)+\pi i\]\cr
&+\int_{-id_2}^\infty\[\wt\mu_0(u)+N^{-1}r(u)\]\,du
=\int_{\g_3}\(\mu_0(u)+N^{-1}\frac{U^1(u)}{2\mu_0(u)}\)du
-\frac{\pi in}{N}-\frac{\pi i}{2N},}
$$
hence (C.21) is proved.

We obtain from (C.20,21) that
$$\eqalign{
c_1=&-\frac{1}{2}\int_{\g}\(\mu_0(u)+N^{-1}\frac{U^1(u)}{2\mu_0(u)}\)du
+\frac{\pi in}{2N}+\frac{\pi i}{4N}
-N^{-1}\frac{\tilde\tau_0(id_2)+\tilde\tau_0(-id_2)}{2}\cr
&-N^{-5/3}\frac{\Delta_0(id_2)+\Delta_0(-id_2)}{2}+O(N^{-2}).}
\eqno (C.27)
$$
For  $u\in\g$,
$$
\mu_0(u)+N^{-1}\frac{U^1(u)}{2\mu_0(u)}=\[-d(u)+N^{-1}U^1(u)\]^{1/2}
+O(N^{-2}u^{-2}),
\eqno (C.28)
$$
hence
$$
\int_{\g}\(\mu_0(u)+N^{-1}\frac{U^1(u)}{2\mu_0(u)}\)du
=\int_{\g}\[-d(u)+N^{-1}U^1(u)\]^{1/2}du+O(N^{-2}).
\eqno (C.29)
$$
Deforming the contour of integration we obtain that
$$
\int_{\g}\[-d(u)+N^{-1}U^1(u)\]^{1/2}du=
-2\int_{z_0^N}^\infty\[-d(u)+N^{-1}U^1(u)\]^{1/2}du=
-2\int_{z_0^N}^\infty\mu^c(u)du.
\eqno (C.30)
$$
Thus, (C.27) reduces to
$$
c_1=\int_{z_0^N}^\infty\mu^c(u)du+\frac{\pi in}{2N}+\frac{\pi i}{4N}
+N^{-1}a+N^{-5/3}b +O(N^{-2}),
\eqno (C.31)
$$
where
$$
a=-\frac{\tilde\tau_0(id_2)+\tilde\tau_0(-id_2)}{2},
\qquad 
b=-\frac{\Delta_0(id_2)+\Delta_0(-id_2)}{2}\,.
\eqno (C.32)
$$
Let us evaluate $a$ and $b$.

From (C.15),
$$\eqalign{
\tilde\tau_0(-z)&=
{1\over 2}\ln{\mu_0(-z)-a_{11}^0(-z)\over a_{12}^0(-z)} 
={1\over 2}\ln{\mu_0(z)+a_{11}^0(z)\over a_{12}^0(z)} 
={1\over 2}\ln\frac{ a_{21}^0(z)}{\mu_0(z)-a_{11}^0(z)} \cr
&=-\tilde\tau_0(z)+\frac{1}{2}\ln\frac{a_{21}^0(z)}{a_{12}^0(z)}\,.}
\eqno (C.33)
$$
From (4.4) and (C.4),
$$
\frac{a_{21}^0(z)}{a_{12}^0(z)}=-1+\frac{a_{21}^0(z)+a_{12}^0(z)}
{a_{12}^0(z)}=-1+
N^{-2/3}\frac{2c^0}{a_{12}^0(z)}+O(N^{-1}),
\eqno (C.34)
$$
hence
$$
\ln\frac{a_{21}^0(z)}{a_{12}^0(z)}=\pi i-
N^{-2/3}\frac{2c^0}{a_{12}^0(z)}+O(N^{-1}).
\eqno (C.35)
$$
Thus, from (C.33) (and the inequality
$0<\Im \tilde\tau_0(z) < {\pi \over 2}$),
$$
\tilde\tau_0(z)+\tilde\tau_0(-z)=\frac{\pi i}{2}-
N^{-2/3}\frac{c^0}{a_{12}^0(z)}+O(N^{-1}),
\eqno (C.36)
$$ 
and
$$
N^{-1}a=-\frac{\pi i}{4N}+
N^{-5/3}\frac{c^0}{2a_{12}^0(z)}+O(N^{-2}),
\qquad z=id_2.
\eqno (C.37)
$$
Let us evaluate $\Delta_0(z)+\Delta_0(-z)$.

From (C.$6'$),
$$
\Delta_0(z)=\tilde\Delta_0(z)+\hat\Delta_0(z),
\eqno (C.38)
$$
where
$$
\tilde\Delta_0(z)={c^0\over
{ia^{0}_{11}(z)-a^{0}_{21}(z)
 - i\mu(z)}},\qquad
\hat\Delta_0(z)=- 2i(-1)^{n}w\z_{0}'(z){a^{0}_{11}(z)\over b(z)}.
\eqno (C.39)
$$
The function $\hat\Delta_0(z)$ is odd, hence
$\hat\Delta_0(z)+\hat\Delta_0(-z)=0$. For $\tilde\Delta_0(z)$
we have that modulo $O(N^{-2/3})$-terms,
$$\eqalign{
\tilde\Delta_{0}(z) +\tilde \Delta_{0}(-z)
&= c^0\({1\over
{ia^{0}_{11}-a^{0}_{21} - i\mu_{0}}} + 
{1\over
{-ia^{0}_{11}-a^{0}_{21} - i\mu_{0}}}\)\cr
&= -2c^0
\({{i\mu_{0} + a^{0}_{21}}\over
{-\mu^{2}_{0} + 2i\mu_{0}a^{0}_{21} + (a^{0}_{21})^2 +  
(a^{0}_{11})^2}}\)\cr
&= -2c^0\({{i\mu_{0} + a^{0}_{21}}\over
{-a^{0}_{12}a^{0}_{21} + 2i\mu_{0}a^{0}_{21} + (a^{0}_{21})^2}}\)\cr
&=  -2c^0\({{i\mu_{0} + a^{0}_{21}}\over
{ 2a^{0}_{21}(i\mu_{0} + a^{0}_{21})}}\)
= -\frac{c^0}{a^0_{21}(z)}=\frac{c^0}{a^0_{12}(z)}.}
\eqno (C.40)
$$
Thus,
$$
N^{-5/3}b=-N^{-5/3}\frac{c^0}{2a^0_{12}(z)}
+O(N^{-2}),\qquad z=id_2.
\eqno (C.41)
$$
Combining (C.37) and (C.41) we obtain that
$$
N^{-1}a+N^{-5/3}b=-\frac{\pi i}{4N}+O(N^{-2}),
\eqno (C.42)
$$
hence by (C.31), 
$$
c_1=\int_{z_0^N}^\infty\mu^c(u)du+\frac{\pi in}{2N} +O(N^{-2}),
$$
so that by (C.15),
$$
c=c_1-N^{-1}c_0=\int_{z_0^N}^\infty\mu^c(u)du-
\frac{1}{4N}\ln R_n^0+\frac{\pi in}{2N} +O(N^{-2}).
\eqno (C.43)
$$
From (C.12) and (C.43) we obtain  (5.14). Proposition
5.2 is proved. It remains to prove (C.14).

{\it Proof of (C.14).} To simplify notations we will
drop super- and subscripts `0'. In the
proof we will neglect $O(N^{-1})$
terms. We will prove first that
$$
\tau'=\tilde\tau'+\hat\tau'
\eqno (C.44)
$$
and then we will check (C.14) at infinity. From (C.15),
$$
2(\tilde\tau'+\hat\tau')=\frac{\mu'-a_{11}'}{\mu-a_{11}}
-\frac{a'_{12}}{a_{12}}+\frac{a_{11}'a_{12}-a_{11}a_{12}'}{\mu
a_{12}}
=\frac{a_{12}(\mu\mu'-a_{11}a_{11}')+a_{12}'(a_{11}^2-\mu^2)}
{a_{12}\mu(\mu-a_{11})}.
$$
Since
$$
\mu^2=a_{11}^2+a_{12}a_{21},\qquad
2\mu\mu'=2a_{11}a_{11}'+a'_{12}a_{21}+a_{12}a'_{21},
$$
we obtain that
$$
4(\tilde\tau'+\hat\tau')=
\frac{a_{12}(a_{12}'a_{21}+a_{12}a_{21}')
-2a_{12}'a_{12}a_{21}}
{a_{12}\mu(\mu-a_{11})}=\frac{a_{12}a_{21}'-a_{12}'a_{21}}
{\mu(\mu-a_{11})},
$$
which coincides with $4\tau'$, see (5.5). This proves (C.44). 
As $z\to\infty$, $\Im z>0$,
$$
\tau(z)=O(z^{-2});\qquad \tilde\tau(z)=\frac{1}{2}\ln z-
\frac{1}{4}\ln R_n^0+O(z^{-2});
\qquad \hat\tau'(z)=-\frac{1}{2}z^{-1}+\hat r(z),
\quad \hat r(z)=O(z^{-3}),
$$
hence
$$
\hat\tau(z)=-\frac{1}{2}\ln z+\int_{\infty}^z\hat r(u)\,du.
$$
This implies that  
$$
\lim_{z\to\infty}\tau(z)=
\lim_{z\to\infty}\left[\tilde\tau(z)+\hat\tau(z)
+\frac{1}{4}\ln R_n^0\right]=0,\quad
\Im z>0.
$$
Equation (C.14) is proved.

\beginsection Appendix D. Proof of Lemma 6.2 \par

{\it Proof of (6.27).} Let us rewrite (6.24) as
$$
\hat\Psi_{\WKB}(z)=C_0\wt T^c(z)e^{-N\xi^c(z)\sg_3},
\eqno (D.1)
$$
where 
$$\eqalign{
\wt T^c(z)&\equiv T^c(z)
\pmatrix
1 & 1 \\
-1 & 1 
\endpmatrix
=\(\frac{ a_{12}^0(z)}{ \mu^c(z)}\)^{1/2}
\pmatrix
1 & 0 \\
-\frac{a^0_{11}(z)}{a^0_{12}(z)} &
\frac{\mu^c(z)}{a^0_{12}(z)}
\endpmatrix
\pmatrix
1 & 1 \\
-1 & 1 
\endpmatrix\cr
&=\({a_{12}^0(z)\over\mu^c(z)}\)^{1/2}
\pmatrix
1 & 1 \\
-{\mu^c(z)+a_{11}^0(z)\over a_{12}^0(z)} & {\mu^c(z)-a_{11}^0(z)\over
a_{12}^0(z)}  
\endpmatrix,\cr
&\cr}
\eqno (D.2) 
$$
and
$$
\xi^c(z)=\int_{z^N_0}^z \mu^c(u)\,du,
\qquad \mu^c(z)=\(U^0(z)\)^{1/2}.
\eqno (D.3)
$$
From (1.68),
$$\eqalign{
\mu^c(z)&=\(U^0(z)\)^{1/2}=\(-d(z)+N^{-1}U^1(z)\)^{1/2}=\mu(z)+N^{-1}
{1\over 2}\,{U^1(z)\over \mu(z)}+O(N^{-2}|z|^{-2}),\cr
U^1(z)&=(a_{11}^0)'(z) 
-a_{11}^0(z){(a_{12}^0)'(z)\over a_{12}^0(z)} \,,}
\eqno (D.4) 
$$
therefore we can replace $\mu^c(z)$ for $\mu(z)$ in formula (D.2) for
$\wt T^c(z)$, with an error term of the order of $N^{-1}|z|^{-1}$
which we will omit. This gives 
$$\eqalign{
\wt T^c(z)&
=\({a_{12}^0(z)\over\mu(z)}\)^{1/2}
\pmatrix
1 & 1 \\
-{\mu(z)+a_{11}^0(z)\over a_{12}^0(z)} & {\mu(z)-a_{11}^0(z)\over a_{12}^0(z)} 
\endpmatrix\cr
&=\({\mu(z)-a_{11}^0(z)\over\mu(z)}\)^{1/2}
\pmatrix
1 & {a_{12}^0(z)\over \mu(z)-a_{11}^0(z)} \\
-{a_{21}^0(z)\over \mu(z)-a_{11}^0(z)} & 1 
\endpmatrix
e^{- \tilde\tau(z)\sg_3}\cr
&=T_0(z)e^{- \tilde\tau(z)\sg_3}\,,}
\eqno (D.5) 
$$
where
$$
\tilde\tau(z)={1\over 2}\ln\[{\mu(z)-a_{11}^0(z)\over a_{12}^0(z)} \],
\eqno (D.6)
$$
hence by (D.1),
$$
\hat\Psi_{\WKB}(z)=C_0T_0(z)e^{-[N\xi^c(z)+\tilde\tau(z)]\sg_3}.
$$
From (D.3),
$$
\xi^c(z)=\int_{\infty}^z\mu^c(u)\,du+C^1,
\eqno (D.7)
$$
where
$$
C^1=\int_{z^N_0}^\infty\mu^c(u)\,du.
\eqno(D.8)
$$
From (D.4),
$$
\int_{\infty}^z\mu^c(u)\,du=\int_\infty^z\mu(u)\,du
+N^{-1}\hat\tau(z)+O(N^{-2}|z|^{-1})
=\xi(z)+N^{-1}\hat\tau(z)+O(N^{-2}|z|^{-1}), 
\eqno (D.9)
$$
where
$$
\hat\tau(z)=\frac{1}{2}\int_\infty^z\frac{U^1(u)}{\mu(u)}\,du.
\eqno (D.10)
$$
Thus, modulo terms of the order of $N^{-1}|z|^{-1}$,
$$
\hat\Psi_{\WKB}(z)=C_0 T_0(z)
e^{-\[N\xi(z)+\tilde\tau(z)+\hat\tau(z)+C^1\]\sg_3},
\eqno (D.11)
$$
From (5.5), (D.6,10) we obtain that
$$
\tilde\tau'(z)+\hat\tau'(z)=\tau'(z)
\eqno (D.12)
$$
(cf. Appendix C above).
In addition, as $z\to\infty$,
$$
\tilde\tau(z)=\frac{1}{2}\ln z
-\frac{1}{4}\ln R_n^0+O(z^{-2});\qquad
\hat\tau'(z)=-\frac{1}{2}z^{-1};\qquad \tau'(z)=O(z^{-3}).
$$
Hence, integrating (D.12) we obtain that
$$
\tilde\tau(z)+\hat\tau(z)=\tau(z)-\frac{1}{4}\ln R_n^0,
\eqno (D.13)
$$
where integral (D.10) is regularized as
$$
\hat\tau(z)=-\frac{1}{2}\ln z
+\frac{1}{2}\int_\infty^z\(\frac{U^1(u)}{\mu(u)}+\frac{1}{u}\)\,du.
$$
Thus, we obtain that, modulo terms of the order of $N^{-1}|z|^{-1}$, 
$$
\hat\Psi_{\WKB}(z)=C_0 T_0(z)
e^{-\[N\xi(z)+\tau(z)+C_1\]\sg_3}
=\Psi_{\WKB}(z),\qquad C_1=C^1-\frac{1}{4}\ln R_n^0,
$$
which was stated.

{\it Proof of (6.28).} Let $z\in\G_0^r$. For the sake of definiteness,
assume 
that $\Im z\ge 0$, so that
$$
\Psi_{\TP}(z)=
\tilde{C}_{1}W(z)Y_u(w(z)),
$$
where
$$
Y_u(z)=
\pmatrix
N^{1/6} & 0 \\
0 & N^{-1/6}
\endpmatrix
\pmatrix
y_0(N^{2/3}z) & y_{1}(N^{2/3}z) \\
y'_0(N^{2/3}z) & y'_{1}(N^{2/3}z) 
\endpmatrix.
\eqno (D.14)
$$
From Proposition 6.3,
$$\eqalign{
\pmatrix
y_0(N^{2/3}z)  \\
y'_0(N^{2/3}z)  
\endpmatrix
&=\vec{\Ai}(N^{2/3}z)\cr
&=\(1+O(N^{-1})\)\frac{1}{2\sqrt\pi}
\pmatrix
N^{-1/6}z^{-1/4} & 0 \\
0 & N^{1/6}z^{1/4}
\endpmatrix
\pmatrix
e^{-2Nz^{3/2}/3} \\
-e^{-2Nz^{3/2}/3}
\endpmatrix,\cr
& -\pi+\ep\le\arg z\le \pi-\ep,}
\eqno (D.15)
$$
and
$$\eqalign{
\pmatrix
y_1(N^{2/3}z)  \\
y'_1(N^{2/3}z)  
\endpmatrix
&=\pmatrix
e^{-\pi i/6} & 0 \\
0 & e^{-5\pi i/6}
\endpmatrix
\vec{\Ai}(N^{2/3}e^{-2\pi i/3}z)\cr
&=\(1+O(N^{-1})\)\frac{1}{2\sqrt\pi}
\pmatrix
N^{-1/6}z^{-1/4} & 0 \\
0 & -N^{1/6}z^{1/4}
\endpmatrix
\pmatrix
e^{2Nz^{3/2}/3} \\
-e^{2Nz^{3/2}/3}
\endpmatrix,\cr
& -\pi+\ep\le\arg z-\frac{2\pi}{3}\le \pi-\ep.}
\eqno (D.16)
$$
Thus,
$$\eqalign{
Y_u(z)&=
\(1+O(N^{-1})\)\frac{1}{2\sqrt\pi}
\pmatrix
z^{-1/4} & 0 \\
0 & z^{1/4}
\endpmatrix
\pmatrix
e^{-2Nz^{3/2}/3} & e^{2Nz^{3/2}/3} \\
-e^{-2Nz^{3/2}/3} & e^{2Nz^{3/2}/3} 
\endpmatrix,\cr
& -\frac{\pi}{3}+\ep\le\arg z\le \pi-\ep.}
\eqno (D.17)
$$
From (6.12),
$$
\frac{2w(z)^{3/2}}{3}=\int_{z^N_0}^z\mu^c(u)\,du=\xi^c(z),
\eqno (D.18)
$$
and from (6.10),
$$
w(z)=\frac{\mu^c(z)^2}{w'(z)^2},
\eqno (D.19)
$$
hence
$$\eqalign{
\Psi_{\TP}(z)&=
\tilde{C}_{1}W(z)\(1+O(N^{-1})\)\frac{1}{2\sqrt\pi}
\pmatrix
\frac{w'(z)^{1/2}}{\mu^c(z)^{1/2}} & 0 \\
0 & \frac{\mu^c(z)^{1/2}}{w'(z)^{1/2}}
\endpmatrix
\pmatrix
e^{-N\xi^c(z)} & e^{N\xi^c(z)} \\
-e^{-N\xi^c(z)} & e^{N\xi^c(z)} 
\endpmatrix,\cr
& -\frac{\pi}{3}+\ep\le\arg w(z)\le \pi-\ep.}
\eqno (D.20)
$$
Observe that if $z\in\G^0_r$ and $\Im z\ge 0$ then
$$
-\frac{\pi}{3}+\ep\le\arg w(z)\le \pi-\ep,
$$
hence (D.20) holds. From (1.79),
$$\eqalign{
W(z)
\pmatrix
\frac{w'(z)^{1/2}}{\mu^c(z)^{1/2}} & 0 \\
0 & \frac{\mu^c(z)^{1/2}}{w'(z)^{1/2}}
\endpmatrix
&=\({a_{12}^0(z)\over w'(z)}\)^{1/2}
\pmatrix
1 & 0 \\
-{a_{11}^0(z)\over a_{12}^0(z)} & {w'(z)\over a_{12}^0(z)}
\endpmatrix
\pmatrix
\frac{w'(z)^{1/2}}{\mu^c(z)^{1/2}} & 0 \\
0 & \frac{\mu^c(z)^{1/2}}{w'(z)^{1/2}}
\endpmatrix\cr
&=\(\frac{ a_{12}^0(z)}{ \mu^c(z)}\)^{1/2}
\pmatrix
1 & 0 \\
-\frac{a^0_{11}(z)}{a^0_{12}(z)} &
\frac{\mu^c(z)}{a^0_{12}(z)}
\endpmatrix=T^c(z),}
\eqno (D.21)
$$
hence from (D.20),
$$
\Psi_{\TP}(z)=
\(1+O(N^{-1})\)\frac{\tilde C_1}{2\sqrt\pi}
T^c(z)E(N\xi^c(z))=\(1+O(N^{-1})\)\hat\Psi_{\WKB}(z),
\eqno (D.22)
$$
if $C_0=\frac{\tilde C_1}{2\sqrt\pi}$. Lemma 6.2 is proven.

\beginsection  Appendix E. Proof of Lemma 4.2 \par

{\it Proof.} Let $R>0$ be a big  number,
independent on $N$, which will be chosen later.
Consider two cases: (1) $|z|\le RN^{-1/3},$
and (2) $|z|> RN^{-1/3}$. In the case (1) 
$\det \wt W(z)\not=0$ by (4.63).
Consider the case (2). First we notice that
by (4.48),
$$
4\z_0'(z)\z_0^2(z)=\frac{gz^2}{2}\sqrt{z_0^2-z^2}+O(N^{-2/3}),
\eqno (E.1)
$$
$$
z\in (-\Omega_{1})\cup \Omega^{0}\cup \Omega_{1}.
$$
Therefore by (4.67),
$$\eqalign{
&\wt a_{11}(z)=-\wt a_{22}(z)=O(N^{-1/3}z),\quad
\wt a_{12}(z)=\frac{gz^2}{2}\sqrt{z_0^2-z^2}+O(N^{-2/3}),\cr
&\wt
a_{21}(z)=-\frac{gz^2}{2}\sqrt{z_0^2-z^2}+O(N^{-2/3}).}
\eqno (E.2)
$$
By (4.4),
$$
\eqalign{
&a_{11}^0(z)=-a_{22}^0(z)=-\frac{gz^3}{2}+O(N^{-1/3}z),\quad
 a_{12}^0(z)=(R_n^0)^{1/2}gz^2+O(N^{-2/3}),\cr
& a_{21}^0(z)=-(R_n^0)^{1/2}gz^2+O(N^{-2/3}).}
\eqno (E.3)
$$
By (4.66) this gives the matrix elements $w_{ij}(z)$ of $W(z)$ as 
$$\eqalign{
&w_{11}(z)=\frac{gz^2}{2}\sqrt{z_0^2-z^2}+(R_n^0)^{1/2}gz^2
+O(N^{-2/3}),\cr
& w_{12}(z)=w_{21}(z)=\frac{gz^3}{2}+O(N^{-1/3}z),\cr
&w_{22}(z)=\frac{gz^2}{2}\sqrt{z_0^2-z^2}+(R_n^0)^{1/2}gz^2
+O(N^{-2/3}),}
\eqno (E.3)
$$
and these estimates are valied for all 
$z\in (-\Omega_{1})\cup \Omega^{0}\cup \Omega_{1}$.
Observe that by (1.45,53), 
$$
R_n^0=\frac{|t|}{2g}+O(N^{-1/3})=\frac{z_0^2}{4}+O(N^{-1/3}).
$$
Also, 
$$
N^{-1/3} = O(R^{-1}),
$$
for all $|z| > RN^{-1/3}$.
Hence
$$
W(z)=\frac{gz^2}{2}\left[\pmatrix
\sqrt{z_0^2-z^2}+z_0 &  z \\
z & \sqrt{z_0^2-z^2}+z_0 
\endpmatrix+O(R^{-1})\right],
\eqno (E.4)
$$
for all $z\in (-\Om_1)\cup\Om^0\cup\Om_1$ and $|z|>RN^{-1/3}$.
Since
$$
\det \pmatrix
\sqrt{z_0^2-z^2}+z_0 &  z \\
z & \sqrt{z_0^2-z^2}+z_0 
\endpmatrix=2\sqrt{z_0^2-z^2}\(\sqrt{z_0^2-z^2}+z_0\)\not=0,
\eqno (E.5)
$$
for $z\in (-\Om_1)\cup\Om^0\cup\Om_1$,
we obtain that $\det W(z)\not=0$ for sufficiently large $R$
if $z\in (-\Om_1)\cup\Om^0\cup\Om_1$ and $|z|>RN^{-1/3}$. Lemma 4.2
is proved.

\beginsection Appendix F. Alternative Forms for  $\Psi_{n}^{0}(z)$\par

Recall that each of the functions $\z(z)$ and
$\z_{0}(z)$  defines an analytic change of variable in the rectangle
$\Omega_{c} \equiv (-\Omega_{1})\cup \Omega^{0}\cup \Omega_{1}$,
and according to (4.84) we have that
$$
\z(z) = \z_{0}(z) + O(N^{-4/3}), \quad
z\in \Omega_{c}.
\eqno (F.1)
$$
Let us use the function $\z_{0}(z)$, instead of the
function $\z(z)$, in the 
right hand side of (5.10). This will lead us
to the  matrix function 
$$
\Psi^{0}_{\CP}(z)=
\left\{
\eqalign{
&\tilde{C}V(z)\Phi^u\(N^{1/3}\z_{0}(z)\),\qquad \Im z\ge 0,\cr 
&\tilde{C}V(z)\Phi^d\(N^{1/3}\z_{0}(z)\),\qquad \Im z\le 0.\cr}
\right. 
\eqno (F.2)
$$
It is clear that the functions
$\Psi^{0}_{\CP}(z)$ and $\Psi_{\CP}(z)$ have exactly the same Stokes matrices.
Taking also into account that, by (F.1),
$$
e^{-i\({4\over 3}N\z^{3}(z) + N^{1/3}y\z(z)\)\sigma_{3}}
e^{i\({4\over 3}N\z_{0}^{3}(z) + N^{1/3}y\z_{0}(z)\)\sigma_{3}}
= I + O(N^{-1/3}),
$$
we arrive to the asymptotic
formula,
$$
\Psi_{\CP}(z)[\Psi^{0}_{\CP}(z)]^{-1} =  I + O(N^{-1/3}),\quad
z\in \Omega^{0}.
\eqno (F.3)
$$
The last equation together with (8.21) yield the following
representation for the solution $\Psi_{n}(z)$ in the
domain $\Omega^{0}$,
$$
\Psi_n(z)=\(I+O(N^{-1/3})\)\Psi_{\CP}^0(z), \quad z \in \Omega^{0}.
\eqno (F.4)
$$
Of course, equation (F.4) has a bigger error term then 
a similar equation
$$
\Psi_n(z)=\(I+O(N^{-1})\)\Psi_{\CP}(z), \quad z \in \Omega^{0},
\eqno (F.5)
$$
which is given by Theorem 1.2. Sometimes, however,
a certain advantage can be achieved by using (F.4).
Indeed, the function $\z_{0}(z)$,
unlike the function $\z(z)$, is given by
an elementary explicit  formula (4.49). 
It is also worth noticing that estimate (F.4) can be 
improved as follows.

Consider the matrix ratio,
$$
R(z)\equiv \Psi_{\CP}(z)[\Psi^{0}_{\CP}(z)]^{-1}.
$$
The function $R(z)$ does not have a jump on the interval $[-d_{1}, d_{1}]$,
and hence it is analytic in $\Omega^{0}$. Therefore,
$$
R(z) = \int_{\partial \Omega^{0}}{{\Psi_{\CP}(u)[\Psi^{0}_{\CP}(u)]^{-1}}
\over {u-z}}{du\over{2\pi i}}, \quad z \in \Omega^{0}.
$$
At the same time, on the boundary of the domain $\Omega^{0}$
the function $\Psi_{\CP}(z)$ can be replaced by 
$\Psi_{\WKB}(z)$ with an error of the order of $N^{-1}$,
$$
\Psi_{\CP}(z) = \(I+O(N^{-1})\)\Psi_{\WKB}(z), \quad 
z \in \partial \Omega^{0},
\eqno (F.6)
$$
and therefore,
$$
R(z) = \int_{\partial \Omega^{0}}{\(I + O(N^{-1})\)
{\Psi_{\WKB}(u)[\Psi^{0}_{\CP}(u)]^{-1}}
\over {u-z}}{du\over{2\pi i}}.
\eqno (F.7)
$$
In addition, it follows from (F.6) and (F.3) that
$$
{\Psi_{\WKB}(u)[\Psi^{0}_{\CP}(u)]^{-1}} = I + O(N^{-1/3}), \quad 
z \in \partial \Omega^{0},
\eqno (F.8)
$$
and hence equation (F.7) can be rewritten as
$$
R(z) = \int_{\partial \Omega^{0}}{{\Psi_{\WKB}(u)[\Psi^{0}_{\CP}(u)]^{-1}}
\over {u-z}}{du\over{2\pi i}} + O\({1\over{N\dist(z; \partial \Omega^{0})}}\).
\eqno (F.9)
$$
The last equation, together with the standard arguments
based on the flexibility in the choice of the basic geometric
parameters $d_{1}$ and $d_{2}$, yields the following improvement of
estimate (F.4)
$$
\Psi_n(z)=\(I+O(N^{-1})\)\Psi_{\CP}^1(z), \quad z \in \Omega^{0},
\eqno (F.10)
$$
where 
$$
\Psi_{\CP}^1(z) = \Big\{\int_{\partial
\Omega^{0}}{{\Psi_{\WKB}(u)[\Psi^{0}_{\CP}(u)]^{-1}} 
\over {u-z}}{du\over{2\pi i}}\Big\}\Psi^{0}_{\CP}(z)
\eqno (F.11)
$$
 The only non-explicit element, apart of course from the
Painlev\'e functions, which is left in formulae (F.10) and
(F.11) is parameter $\alpha$ in definition (4.82) of the variable
$y$. To determine it one needs equation of periods (4.78). However,
the terms of order $N^{-1/3}$ can be extracted from (F.10)-(F.11)
in the very explicit form. Indeed, by a strightforward though 
a little bit involved calculations one can see that
$$
\Psi_{\WKB}(z)[\Psi^{0}_{\CP}(z)]^{-1} = 
I - N^{-1/3}{{ir_{1}(z)\over{\sqrt{z_{0}^{2}-z^{2}}}}}
\(z\sigma_{3}-2i(R_{n}^{0})^{1/2}\sigma_{2}\) + O(N^{-2/3}), \quad 
z \in \partial \Omega^{0},
\eqno (F.12)
$$
where
$$
r_{1}(z) = {iD\over{2\z_{\infty}(z)}}
+{1\over g}\int_{z}^{\infty}\(c_{3}
- {y^{2}c^{2}_{0}\over{u^{2}-z_{0}^{2}}}\)
{du\over{u^{2}\sqrt{u^{2}-z_{0}^{2}}}}
+iy^{2}{{\z_{1}(z)D_{1}(z)}\over{\z_{\infty}(z)}}.
$$ 
By a direct calculation (this time
much less involved) we check that $r_{1}(z)$
has no pole at $z=0$, i.e. it is an analytic
function in $\Omega^{0}$. Therefore, equation (F.11)
yields the estimate,
$$
\Psi_{\CP}^1(z) =
\(I - N^{-1/3}{{ir_{1}(z)\over{\sqrt{z_{0}^{2}-z^{2}}}}}
\(z\sigma_{3}-2i(R_{n}^{0})^{1/2}\sigma_{2}\) + O(N^{-2/3})\)
\Psi^{0}_{\CP}(z).
\eqno (F.13)
$$
 
Further improvement of (F.13) could be done. However,
when we go beyond the order $N^{-1/3}$ we have to
take into account the parameter $\alpha$ and hence
representation (F.10) will 
not give us any advantages comparing
with our basic formula (8.21). 

In principle, it is possible to avoid the use of
period equation (4.78). To this end, let us define the approximate solution
$\Psi^{0}_{n}(z)$ of the Riemann-Hilbert problem by relation
(8.1) where $\Psi_{\CP}(z)$ is replaced by $\Psi^{0}_{\CP}(z)$
and $\alpha$ in the definition of $y$ is put to zero. Then
for the ratio $X_{n}(z) \equiv \Psi_{n}(z)[\Psi^{0}_{n}(z)]^{-1} $ 
we will arrive to the Riemann-Hilbert problem whose 
jump matrix $G(z) \equiv [X_{n-}(z)]^{-1}X_{n+}(z)$ is given explicitely
in terms of $\Psi^{0}_{n}(z)$ and satisfies the uniform
estimates (cf. (8.11) and (8.12))
$$
||I-G(z)||_{L_{2}(\gamma)\cap L_{\infty}(\gamma)} = O(N^{-1/3}).
$$
This would mean the replacement of the final estimate (8.21)
by the estimate
$$
\Psi_n(z)=\(I+O\({1\over{N^{1/3}(1+|z|)}}\)\)\Psi_n^0(z).
\eqno (F.14)
$$
Theoretically, this estimate can be improved by iterating the
corresponding singular integral equation (cf. [DZ3]). In particular,
the first iteration leads to the function $\Psi^{1}_{\CP}(z)$ 
satisfying (F.13). The second iteration, however, looks
extremely cumbersome. In fact, our basic construction
involving the nontrivial change of variable function
$\z(z)$ can be thought of as a way to bring the second
iteration of the singular integral operator with the
weight $I- G(z)$ to a compact form.

Let us finally discuss the dependence of $\z_{0}(z)$ on the parameter $y$,
$
\z_{0}(z) \equiv \z_{0}(z;y).
$
To that end put
$$\eqalign{
M_{1}(z,\z_{0}(z;y),y)&\equiv
\Psi^{0}_{\CP}(z,\z_{0}(z;y),y)e^{\(N{{4i}\over{3}}\z^{3}_{0}(z;y)
+N^{1/3}iy\z_{0}(z;y) -i\gamma\)\sigma_{3}},\cr
z&\in\{\epsilon \leq \arg z \leq \pi - \epsilon\}\cup
\{-\pi +\epsilon \leq \arg z \leq - \epsilon\},}
\eqno (F.15)
$$
and
$$\eqalign{
M_{2}(z,\z_{0}(z;y),y)&\equiv
\Psi^{0}_{\CP}(z,\z_{0}(z;y),y)S^{-1}_{1,2}e^{\(N{{4i}\over{3}}\z^{3}_{0}(z;y)
+N^{1/3}iy\z_{0}(z;y) -i\gamma\)\sigma_{3}},\cr
z&\in\left\{
-{\pi \over 3}+\epsilon \leq \arg z \leq {\pi \over 3} - \epsilon,
\quad \pm \Im z \geq 0\right\}}
\eqno (F.16)
$$
[the matrices $S^{-1}_{1,2}$ are the Stokes matrices from (7.3)].
It is not difficult to see that
$$
M_{j}(z,\z_{0}(z;y),y) = \(I + O(N^{-1/3})\)M_{j}(z,\z_{0}(z;0),y).
\eqno (F.17)
$$
This implies that we can redefine $\Psi^0_{\CP}(z)$ as
$$\eqalign{
\Psi^{0}_{\CP}(z)&=M_{1}(z,\z_{0}(z;0),y)
e^{-\(N{{4i}\over{3}}\z^{3}_{0}(z;y)
+N^{1/3}iy\z_{0}(z;y) -i\gamma\)\sigma_{3}},\cr
z&\in\{\epsilon \leq \arg z \leq \pi - \epsilon\}\cup
\{-\pi +\epsilon \leq \arg z \leq - \epsilon\},}
\eqno (F.18)
$$
and
$$\eqalign{
\Psi^{0}_{\CP}(z)&=M_{2}(z,\z_{0}(z;0),y)
e^{-\(N{{4i}\over{3}}\z^{3}_{0}(z;y)
+N^{1/3}iy\z_{0}(z;y) -i\gamma\)\sigma_{3}}S_{1,2},\cr
z&\in\left\{
-{\pi \over 3}+\epsilon \leq \arg z \leq {\pi \over 3} - \epsilon,
\quad \pm \Im z \geq 0\right\},}
\eqno (F.19)
$$
and still have (F.4). Observe that by (4.49,44),
$$
\z_0(z;0)=\[\frac{3}{4}\int_0^z\frac{gu^2}{2}\sqrt{z_0^2-u^2}\,du\]^{1/3}.
\eqno (F.20)
$$
Formulae (F.18)-(F.20) are of the type that
would  appear if for the analysis of the
Riemann-Hilbert problem (1.35-38) we use the nonlinear steepest 
descent approach similar to the method used in
[BDJ] for the orthogonal polynomials
on the circle. The advantage of such scheme would be
the apearence of the Painlev\'e Riemann-Hilbert
problem in a completely deductive way, i.e. without use of
any prior information about the structure of the
asymptotics of the coefficients $R_{n}^{0}$. The price,
however, would be the much more subtle structure
of the relevant contour deformation
and much more complicated analysis of the all
error terms than in our approach. In our approach,
we take the full advantage of the prior heuristic
study of the Freud equation (1.24) which leads
us to the  choice of the approximate solution,
whose justification only needs a very ``light''
Riemann-Hilbert analysis. Moreover, we obtain the
error term up to the order $O(N^{-1})$. To get
this accuracy in the framework of
the [BDJ] scheme would in fact also require,
{\it a posterior}, a
WKB-type analysis of the associated Lax pair
(cf. [DZ3]). In our method, we use the WKB formulae
from the very beginning.

\beginsection References \par

\item {[ADMN]}
 G. Akemann, P. H. Damgaard, U. Magnea, and S. Nishigaki,
Multicritical microscopic spectral correlators of Hermitian and
complex matrices,
{\it Nucl. Phys. B}, {\bf 519}, 682--714 (1998).

\item {[BDJ]} J. Baik, P. Deift, and K. Johansson,
On the distribution of the length of the longest increasing
subsequence of random permutations,
{\it J. Am. Math. Soc.} {\bf 12}, 1119-1178 (1999). 

\item {[BDT]} R. Beals, P. A. Deift, and C. Tomei, Direct and inverse
scattering on the line. Mathematical Surveys and Monographs {\bf 28}.
AMS Providence, Rhode Island (1988).

\item {[BI]} P. M. Bleher and A. R. Its,
Semiclassical asymptotics of orthogonal polynomials, Riemann-Hilbert
problem, and universality in the matrix model. 
{\it Ann. Math.} {\bf 150}, 185-266 (1999).

\item {[BIZ]} D. Bessis, C. Itzykson, and J.-B. Zuber,
Quantum field theory techniques in graphical enumeration,
{\it Adv. Applied Math.} {\bf 1}, 109--157 (1980).

\item {[Ble1]} P. M. Bleher, 
Quasi-classical expansion and the problem of quantum chaos,
{\it Lect. Notes Math.} {\bf 1469}, Eds. J. Lindenstrauss and
V. D. Milman, 60--89 (1991).

\item {[Ble2]}
Semiclassical quantization rules near
separatrices, {\it Commun. Math. Phys.}, 1994, {\bf 165},
3, 621--640. 

\item {[BPS]} A. Boutet de Monvel, L. A. Pastur, and M. Shcherbina,
On the statistical mechanics approach in the random matrix
theory. Integrated density of states,
{\it J. Statist. Phys.} {\bf 79}, 585--611 (1995).

\item {[BB]} M. J. Bowick and E. Br\'ezin,
Universal scaling of the tail of the density of eigenvalues in random
matrix models, {\it Phys. Letts.} {\bf B268}, 21--28 (1991).

\item {[BIPZ]} E. Br\'ezin, C. Itzykson, G. Parisi, and J. B. Zuber,
Planar diagrams,
{\it Commun. Math. Phys.} {\bf 59}, 35--51 (1978).

\item {[BK]} E. Br\'ezin and V. A. Kazakov,
Exactly solvable field theories of closed strings,
{\it Phys. Lett. B} {\bf 236}, 144--150 (1990). 

\item {[BZ]} E. Br\'ezin and A. Zee,
Universality of the correlations between eigenvalues of large random
matrices, {\it Nuclear Physics B} {\bf 402}, 613--627 (1993).


\item {[CG]} K. Clancey and I. Gohberg, Factorization of matrix
functions and singular integral operators. Operator Theory. {\bf 3},
Birkh{\"{a}}user Verlag Basel (1981).


\item {[CCM]} G. M. Cicuta, L. Molinari, and Montaldi,
Large $N$ phase transition in low dimensions,
{\it Mod. Phys. Lett.} {\bf A1}, 125, 1986.

\item {[CM]} \v C. Crnkovi\' c and G. Moore,
Multicritical multi-cut matrix models, 
{\it Phys. Lett. B} {\bf 257}, (1991).

\item {[DIZ]} P. Deift, A.  Its, and X. Zhou, A Riemann-Hilbert
 approach to asymptotic problems arising in the theory
of random matrix models, and also in the theory of integrable
statistical mechanics, {\it Ann. of Math.} {\bf 146}, 149-235 (1997).

\item{[DKM]} P. Deift, T. Kriecherbauer, K. T-R. McLaughlin,
New results on the equilibrium measure for logarithmic
potentials in the presence of an external field,
{\it J. Appr. Theory} {\bf 95}, 399-475 (1998). 

\item{[DKMVZ1]} P. Deift, T. Kriecherbauer, K. T-R. McLaughlin,
S. Venakides, and X. Zhou,
Uniform asymptotics for polynomials orthogonal with respect
to varying exponential weights and applications to universality
questions in random matrix theory,
{\it Commun. Pure Appl. Math.}, {\bf 52}, 1335-1425 (1999). 

\item{[DKMVZ2]} P. Deift, T. Kriecherbauer, K. T-R. McLaughlin,
S. Venakides, and X. Zhou,
Strong asymptotics of orthogonal polynomials with respect
to exponential weights,
{\it Commun. Pure Appl. Math.}, {\bf 52}, 1491-1552 (1999). 

\item {[DZ1]} P.  Deift and X. Zhou, A steepest descent method for
oscillatory Riemann-Hilbert problems. Asymptotics for the MKdV equation,
{\it Ann. of Math.} {\bf 137}, 295--368 (1995).

\item {[DZ2]} P.  Deift and X. Zhou, Asymptotics for the
Painlev{\'{e}} II equation, {\it Commun. Pure Appl. Math.} {\bf 48},
277--337 (1995). 

\item {[DZ3]} P. Deift and X. Zhou, Long-time asymptotics for
integrable systems. Higher order theory,
{\it Commun. Math. Phys.} {\bf 165},
175--191 (1994). 

\item {[DSS]} M. R. Douglas, N. Seiberg, and S. H. Shenker,
Flow and instability in quantum gravity,
{\it Phys. Lett. B} {\bf 244}, 381--386 (1990).  

\item {[DS]} M. R. Douglas and S. H. Shenker,
Strings in less than one dimension,
{\it Nucl. Phys. B} {\bf 335}, 635--654 (1990).

\item {[Dys]} F. J. Dyson,
Correlation between the eigenvalues of a random matrix,
{\it Commun. Math. Phys.} {\bf 19}, 235--250 (1970).

\item {[FI]} A. S. Fokas and A. R. Its, The isomonodromy method and the
Painlev{\'{e}} equations, {\it in the book: Important Developments in Soliton
Theory}, A. S. Fokas, V. E. Zakharov (eds.), Berlin, Heidelberg, New York,
Springer (1993).

\item {[FN]} H. Flaschka and A. Newell, Monodromy and spectral preserving deformations, {\it Commun. Math. Phys.} {\bf 76}, 67--116 (1980).

\item {[FIK1]} A. S. Fokas, A. R. Its, and A. V. Kitaev,
Isomonodromic approach in the theory of two-dimensional quantum
gravity, {\it Usp. Matem. Nauk } {\bf 45}, 6,  135--136 (1990) (in Russian).

\item {[FIK2]} A. S. Fokas, A. R. Its, and A. V. Kitaev,
Discrete Painlev\'e equations and their appearance in quantum gravity,
{\it Commun. Math. Phys.} {\bf 142}, 313--344 (1991).

\item {[FIK3]} A. S. Fokas, A. R. Its, and A. V. Kitaev, The matrix model
of the two-dimensional quantum gravity and isomonodromic solutions
of the discrete Painlev{\'{e}} equations;
 A. V. Kitaev,
Calculations of nonperturbation parameter in matrix model $\Phi^{4}$;
A. R. Its and A. V. Kitaev, Continuous limit for hermitian
matrix model $\Phi^{6}$, {\it in the book:  Zap. Nauch.
Semin. LOMI } {\bf 187}, 12 (1991) (in Russian).

\item {[FIK4]} A. S. Fokas, A. R. Its, and A. V. Kitaev,
The isomonodromy approach to matrix models in 2D quantum gravity,
{\it Commun. Math. Phys.} {\bf 147}, 395--430 (1992).

\item {[For]} P. J. Forrester,
The spectrum edge of random matrix ensembles, {\it Nucl. Phys.} {\bf
B402}, 709--728 (1993).

\item {[Fre]} G. Freud,
On the coefficients in the recursion formulae of orthogonal
polynomials, {\it Proc. Royal Irish Acad.} {\bf 76A}, 1--6 (1976).


\item {[G]} R. Garnier, Sur les \'equationes diff\'erentielles du 
troisi\`me ordre dont l'int\'egrale g\'en\'erale est uniforme et sur une
classe d'\`equationnes nouvelles d'ordre sup\'erieur dont l'int\'egrale
g\'en\'erale a ses points critique fixes.
{\it Ann.\ Sci.\ Ec.\ Norm.\ Sup\'er.} {\bf 29}, 1-126  (1912).

\item {[GM]} D. J. Gross and A. A. Migdal,
Nonperturbative two-dimensional quantum gravity,
{\it Phys. Rev. Lett.} {\bf 64}, 127--130 (1990).

\item {[HM]} S. P. Hastings and J. B. McLeod,
A boundary value problem associated with the
second Painlev\'e transcendent and the Korteweg de 
Vries equation. {\it Arch. Rational Mech.
Anal.} {\bf 73}, 31-51 (1980).

\item {[I]} A. R. Its, Connection Formulae for the Painleve Transcendents,
   in the book: {\it The Stokes Phenomenon and Hilbert's 16th Problem},
   B. L. J. Braaksma, G. K. Immink, and M. van der Put, eds., 
   World Scientific, Singapore, pp. 139-165 (1996).

\item {[IK]} A. R. Its, A. A. Kapaev, 
The nonlinear steepest descent approach to the 
asymptotics of the second
Painleve transcendent in the complex domain,
nlin.SI/0108054

\item {[IN]} A. R. Its and V. Yu. Novokshenov,
The isomonodromic deformation method in the theory of Painlev\'e
equations. {\it Lecture Notes in Math.} {\bf 1191}. Springer-Verlag,
Berlin -- New York, 1986.

\item {[IZ]} C. Itzykson and J.-B. Zuber,
The planar approximation. II,
{\it J. Math. Phys.}, {\bf 21}, 411--421 (1980).


\item {[JMU]} M. Jimbo, T. Miwa, and K. Ueno, Monodromy preserving deformation
of linear ordinary differential equations with rational 
coefficients I, {\it Physica D2}, 306--352 (1981). 

\item {[Kap1]} A. A. Kapaev, Global asymptotics of the second Painlev\'e transcendent.
{\it Physics Letters A.}, {\bf 167},  356-362 (1992).

\item {[Kap2]} A. A. Kapaev The essential singularity 
of the Painlev\'e function of the
second kind and nonlinear Stokes phenomenon,
{\it Zapiski Nauchn.\ Seminarov
LOMI} {\bf 187},  139-170 (1991).

\item {[Kap3]} A. A. Kapaev,
WKB method for $2\times 2$ systems of linear ordinary differential
equations with rational coefficients, Preprint \# 96-6, 
Indiana University -- Purdue University at Indianapolis, 1996.

\item{[Kit1]} A. V. Kitaev,
Turning points of linear systems and double asymptotics of the
Painlev\'e transcendents,
 {\it in the book :  Zap. Nauch.
Semin. LOMI } {\bf 187}, 12, 53 (1991) (in Russian).

\item {[Kit2]} A. V. Kitaev, Elliptic asymptotics of the 
first and the second Painlev\`e transcendents,
{\it Russian Math. Surveys.} {\bf 49:1}, 81 - 150 (1994). 

\item {[KM]} A. B. J. Kuijlaars and K. T-R McLauglin, Generic behavior
of the density of states in random matrix theory and equilibrium
problems in the presence of real analytic
external field, {\it Commun. Pure Appl. Math.} {\bf 53}, 736-785 (2000).

\item {[LiS]} G. Litvinchuk and T. Spitkovskii, Factorization of measurable
matrix functions. Birk\-h{\"{a}}user Verlag, Basel, 51 (1987).

\item {[LS]} D. S. Lubinsky and E. B. Saff,
Strong asymptotics for extremal polynomials associated with weights on
$\R$, {\it Lect. Notes Math.} {\bf 1305} (1988).

\item {[Lub]} D. S. Lubinsky,
Strong asymptotics of extremal errors and polynomials 
associated with Erd\H os--type weights,
Pitman Research Notes in Mathematics {\bf 202}, Longman Scientific and
Technical, Harlow, UK, 1989.

\item {[Meh]} M. L. Mehta,
{\it Random Matrices}, 2nd ed., Academic Press, Boston, 1991. 

\item {[Moo]} G. Moore,
Matrix models of 2D gravity and isomonodromic deformations, {\it
Progr. Theor. Phys. Suppl.} {\bf 102}, 255--285 (1990).

\item {[Nev]} P. Nevai,
G\'eza Freud, orthogonal polynomials and Christoffel functions. A case
study, {\it J. Approx. Theory} {\bf 48}, 3--167 (1986).

\item {[Nov]} S. P. Novikov, Quantization of finite-gap potentials and
nonlinear quasiclassical approximation in nonperturbative string
theory, {\it Funkt. Analiz i Ego Prilozh.} {\bf 24}, 4, 43-53 (1990).

\item {[Novok]} Yu. V. Novokshenov, The Boutroux ansatz for the second
Painlev\'e equation in the complex domain,
{\it Izv. Acad. Nauk SSSR Ser. Mat.} {\bf 54} 1229-1251 (1990)

\item {[PS]} L. Pastur and M. Shcherbina,
Universality of the local eigenvalue statistics for a class of unitary
invariant random matrix ensembles,
{\it Journ. Statist. Phys.} {\bf 86}, 109--147 (1997).

\item {[PeS]} V. Periwal and D. Shevitz,
Exactly solvable unitary matrix models: multicritical potentials and
correlations, {\it Nucl. Phys. B} {\bf 333}, 731--746 (1990).

\item {[Sib]} Y. Sibuya,
{\it Linear differential equations in the complex domain: problems of
analytic continuation.} Transl. Math. Monogr., {\bf 82}, AMS,
Providence, RI, 1990.

\item {[ST]}  E. B. Saff and V. Totik,
{\it Logarithmic potentials with external fields},
 Springer-Verlag, New York, 1997

\item {[Sze]} G. Szeg\"o,
{\it Orthogonal Polynomials}, 3rd ed., AMS, Providence, 1967.

\item {[TW1]} C. A. Tracy and H. Widom,
Introduction to random matrices, In: {\it Geometric and quantum
aspects of integrable systems (Scheveningen, 1992),} 103--130, Lecture
Notes in Phys. {\bf 424}, Springer, Berlin, 1993.

\item {[TW2]} C. A. Tracy and H. Widom,
Level-spacing distribution and the Airy kernel, {\it
Commun. Math. Phys.} {\bf 159}, 151--174 (1994).

\item {[Zh]} X. Zhou, Riemann-Hilbert problem and inverse scattering,
{\it SIAM J. Math. Anal.} {\bf 20}, 966--986 (1989).

\bye